\documentclass[12pt,a4paper]{iopart} 
\pdfoutput=1
\usepackage{iopams}
\usepackage{graphicx,psfrag,bbm,latexsym,color,dcolumn,bm,dsfont,bbm,
  color,mathrsfs,bbold,latexsym,amsfonts,amssymb,xfrac,cite}

\newcommand{\ua}{\uparrow} \newcommand{\da}{\downarrow}

\newcommand{\mediac}[1]{\left\langle #1 \right\rangle_{N}^{(c)}}

\newcommand{\arc}[1]{\langle\!\langle #1 \rangle\!\rangle}
\newcommand{\EE}{\mbox{$\mathsf E$}} 
\def\1{ \mathit{1}\! } %
\def\mod{ \mathop{\rm mod} } %
\def\Arg{ \mathop{\rm Arg} } %
\def\I{\mathrm{i}} %
\def\E{\mathrm{e}} %
\def\D{\mathrm{d}} %
\definecolor{rosso}{rgb}{0.667,0,0.118}

\begin{document}
\title[A perturbative probabilistic approach to quantum many-body
systems] {A perturbative probabilistic approach to quantum many-body
  systems}

\author{Andrea Di Stefano$^{1,2}$, Massimo Ostilli$^{1,3}$ and Carlo
  Presilla$^{1,4}$}

\address{$^1$\ Dipartimento di Fisica, Universit\`a di Roma ``La
  Sapienza'', Piazzale A. Moro 2, Roma 00185, Italy} \address{$^2$\
  Fakult\"at f\"ur Mathematik, Universit\"at Bielefeld, Postfach 110
  131, 33501 Bielefeld, Germany} \address{$^3$\ Cooperative
  Association for Internet Data Analysis (CAIDA), University of
  California, San Diego (UCSD), 9500 Gilman Drive, La Jolla,
  California 92093, USA} \address{$^4$\ Istituto Nazionale di Fisica
  Nucleare (INFN), Sezione di Roma 1, Roma 00185, Italy}

\date{\today}

\begin{abstract}
  In the probabilistic approach to quantum many-body systems, the
  ground-state energy is the solution of a nonlinear scalar equation
  written either as a cumulant expansion or as an expectation with
  respect to a probability distribution of the potential and hopping
  (amplitude and phase) values recorded during an infinitely lengthy
  evolution.  We introduce a perturbative expansion of this
  probability distribution which conserves, at any order, a
  multinomial-like structure, typical of uncorrelated systems, but
  includes, order by order, the statistical correlations provided by
  the cumulant expansion.  The proposed perturbative scheme is
  successfully tested in the case of pseudo-spin 1/2
  hard-core boson Hubbard models also when affected by a phase problem
  due to an applied magnetic field.
\end{abstract}


\maketitle

\section{Introduction}
\label{introduction}

A multitude of evolution problems, including quantum many-body
systems, can be cast in the form of a linear flow, namely a system of
linear differential equations with respect to a parameter, the time,
which can be real or imaginary.  The solution of a linear flow, with a
real or an imaginary time, admits an exact probabilistic
representation, namely a Feynman-Kac--like formula, in terms of a
proper collection of independent Poisson processes
\cite{DAJLS,DJS,BPDAJL1,BPDAJL2,BPDAJL3}.  For a lattice system, the
Poisson processes are associated with the links of the lattice and the
probabilistic representation leads to an optimal algorithm
\cite{BPDAJL1,BPDAJL2,BPDAJL3} which coincides with the Green function
quantum Monte Carlo method in the limit when the latter becomes exact
\cite{SC}.  The algorithm can be rigorously generalized \cite{OP3} to
include fluctuation control techniques, like reconfigurations and
importance sampling, and allows the exact simulation of time-dependent
correlation functions for system not affected by the so called sign
problem \cite{CK}.

In the limit of an infinitely long imaginary time, the above exact
probabilistic representation has been developed to yield
semi-analytical results.  In fact, for an arbitrary many-body system
we are able to relate the energy of its ground state to the unique
solution of a nonlinear scalar equation \cite{OP1,OP2,OP4,OP5}.  This
equation can be written in terms of a series involving the cumulants
of integers, the multiplicities $N_V$, $N_T$ and $N_\lambda$, which
count how many times the potential, hopping and phase variables take
the values $V$, $T$ and $\lambda$ during an infinitely long evolution
of the system.  The potential variables are, in some chosen base, the
diagonal matrix elements of the Hamiltonian of the system, whereas the
hopping and phase variables are related to the amplitude and phase of
the off-diagonal matrix elements.  Alternatively, the equation for the
ground-state energy can be written in terms of an expectation
involving the probability distribution of the multiplicities $N_V$,
$N_T$ and $N_\lambda$.

The two ways of writing the equation for $E_0$, cumulant expansion or
probabilistic expectation, correspond to two different approaches to
evaluate the ground state of a many-body system.  In the former case,
we can imagine measuring the exact cumulants of the system up to some
finite (small) order, inserting them in a corresponding truncated
equation and solving it obtaining an approximation to $E_0$ \cite{OP4}.
In the latter case, we have the possibility to make some guess on the
probability distribution of the multiplicities $N_V$, $N_T$ and
$N_\lambda$ bypassing the microscopic connection between
configurations of the system and values of the variables $V$, $T$ and
$\lambda$. The simplest guess is to neglect any correlation among the
multiplicities and assume that they are multinomially distributed.
There is a class of systems for which, in the thermodynamic limit, a
multinomial distribution exactly applies.  The class includes the
uniformly fully connected models, namely a collection of states all
connected with equal hopping coefficients and in the presence of a
potential operator with arbitrary levels and degeneracies, and the
random potential systems, in which the hopping operator is generic and
arbitrary potential levels are assigned randomly to the states with
arbitrary probabilities.  For this class of models we have found a
zero-temperature universal thermodynamic limit displaying a quantum
phase transition \cite{OP5}.

Both the two approaches described above have limitations. The more
severe drawback of the cumulant expansion is that truncating the
series corresponds to introducing a rather artificial equivalent
system with zero cumulants of large order. Since the determination of
$E_0$ is a problem which involves large fluctuations \cite{OP1}, we
expect, and find indeed, some nonphysical behavior of the solutions
corresponding to large interaction energies.  On the other hand, we
expect that the multinomial probability distribution used in the
second approach may give quantitatively inaccurate results for most of
the systems, i.e. when the correlations among the potential, hopping
and phase multiplicities cannot be neglected.

In the present paper we introduce a perturbative scheme merging the
merits of the cumulant expansion with those of the expectation taken
from an uncorrelated multinomial distribution.  The idea is to develop
a perturbative expansion of the probability distribution of the
multiplicities $N_V$, $N_T$ and $N_\lambda$ which conserves, at any
order, a multinomial-like structure.  Order by order, we add
correlations among the multiplicities in such a way as to modify the
cumulants of the multinomial-like distribution and make them to
coincide with those measured in the system up to the order
considered. As a result, we gain better and better approximations to
the real probability distribution. At the first order, we have a
probability distribution which contain infinitely many cumulants and
the first one is exact; at the second order, we have a probability
distribution with infinitely many cumulants and the first two are
exact; and so on. In this way, we expect to obtain, even at very small
perturbative orders, accurate results for the ground-state energy of
both weakly- and strongly-interacting many-body systems.  We have
checked our perturbative scheme in the case of pseudo-spin
1/2 hard core boson Hubbard models in one- and
two-dimensional lattices.  In particular, we have considered the case
of a ring threaded by a magnetic flux, a model which is affected by a
phase problem.  It is remarkable that, already at the second
perturbative order, we find a ground-state energy which compares
rather well with the exact value of $E_0$.

The main advantage of our method lies in its semi-analytical
character.  Once all the cumulants up to some order $k$ are measured,
via a \textit{una tantum} simulation, the perturbative probabilistic
distribution built from these input data provides, within an
approximation with improves with $k$, the ground state energy $E_0$ as
a function of the Hamiltonian parameters (e.g. interaction and hopping
amplitudes).  In contrast, in a standard Monte Carlo method any
different choice of the Hamiltonian parameters requires a distinct
simulation.  Furthermore, the cumulants are easily measured also in
the case of fermions (bosons in the presence of magnetic fields). The
sign (phase) problem that occurs in this case remains confined in the
expression of the perturbative probability distribution and can, in
principle, be addressed analytically.

There is another useful result provided by the semi-analytical
character of our approach.  Assuming a Hamiltonian $\hat{H}(\xi)$
function of the parameter $\xi$, we are able to evaluate the
derivatives of the ground-state energy $E_0(\xi)$ with respect to
$\xi$. This allows the determination of arbitrary ground-state
correlation functions via the Hellman-Feynman theorem
\begin{eqnarray*}
  \frac{\partial E_0(\xi)}{\partial\xi} = 
  \langle E_0(\xi), \frac{\partial\hat{H}(\xi)}{\partial\xi} E_0(\xi) \rangle,
\end{eqnarray*}
where we have assumed a normalized ground state $\langle
E_0(\xi),E_0(\xi) \rangle=1$. In fact, the quantum expectation of an
arbitrary observable $\hat{O}$ in the ground state of the Hamiltonian
$\hat{H}$ can be obtained by evaluating the ground-state energy
$E_0(\xi)$ of the Hamiltonian $\hat{H}(\xi) = \hat{H}+\xi\hat{O}$ and
taking the derivative $\left.\partial_\xi E_0(\xi)\right|_{\xi=0}$.

The paper is organized as follows.  In section
\ref{probabilistic.approach} we review the probabilistic approach to
quantum many-body systems.  The equation for the ground-state energy
is written as an expansion over the cumulants of the potential,
hopping and phase multiplicities in section
\ref{cumulant.expansion} and as an expectation with respect to the
probability distribution of the same variables in section
\ref{multinomial.probability.density}.  In the latter section the
case of an uncorrelated multinomial distribution is described in
detail.  In section \ref{multinomial.perturbative.scheme} we introduce
the probabilistic perturbative scheme bringing together the merits of
the multinomial probability distribution with the statistical details
provided by the cumulant expansion.  The parameters defining the
perturbative probability distribution of the potential, hopping and
phase multiplicities are explicitly discussed, up to the third order,
in sections \ref{evaluation.arc} to \ref{equation.p3}.  This
is a technical part which could be skipped in a first reading.  Some
considerations on the higher perturbative orders are given in
section \ref{considerations.on.higher.orders}.  The equation for
the evaluation of $E_0$ resulting from the above perturbative scheme
is discussed in section \ref{E0.evaluation}. Sections
\ref{numerical.results} and \ref{magnetic.field} deal with some
numerical results. In particular, in section \ref{magnetic.field} we
discuss how the determination of the ground-state energy in the
presence of a phase problem is handled by our approach.  Concluding
remarks are drawn in section \ref{conclusions}.  Two appendices close
the paper. In \ref{calculation.p2} we review the methods for solving
the nonsymmetric algebraic Riccati equation which appears in
determining the perturbative parameters at the second order.
\ref{equation.p4} summarizes the evaluation of the perturbative
parameters at the fourth order.

\section{Probabilistic approach to quantum many-body systems}
\label{probabilistic.approach}
In this section we review the probabilistic approach to quantum
many-body systems developed in \cite{OP1,OP2,OP4,OP5}. Consider a
system of particles represented by a Hamiltonian operator $\hat{H}$
acting on a $M$ dimensional space of states labeled by configuration
indices $\boldsymbol{n}$. As an example, think about spinless
particles undergoing a simple exclusion dynamics in a lattice with
configurations given by the site occupations
$\boldsymbol{n}=(0,1,1,1,0,0,1,\dots)$.  In the chosen
$\bm{n}$-representation, we separate, as usual,
$\hat{H}=\hat{K}+\hat{V}$, where $\hat{V}=\mathrm{diag}(\hat{H})$ is
called the potential operator and has non vanishing matrix elements
\begin{eqnarray}
  V_{\bm{n},\bm{n}} = V_{\bm{n}}.
  \label{V}
\end{eqnarray}
The hopping operator $\hat{K}$ is defined by the matrix elements
\begin{eqnarray}
  K_{\bm{n},\bm{n}'} = - \lambda_{\bm{n},\bm{n}'}~\eta_{\bm{n},\bm{n}'},
  \qquad
  \eta_{\bm{n},\bm{n}'}>0, 
  \qquad
  |\lambda_{\bm{n},\bm{n}'}|=0,1,
  \label{K}
\end{eqnarray}
conventionally written in terms of links $\lambda_{\bm{n},\bm{n}'}$
and positive strengths $\eta_{\bm{n},\bm{n}'}$.  The matrix with
elements $|\lambda_{\bm{n},\bm{n}'}|$ forms the so called adjacency
matrix and establishes whether two configurations $\bm{n}$ and
$\bm{n}'$ are first neighbors with respect to $\hat{K}$ or not.  The
phase change (sign in the case of fermions) registered by connecting
two first neighbor configurations $\bm{n}$ and $\bm{n}'$ is given by
$\Arg{\lambda_{\bm{n},\bm{n}'}}$.  Note that $\sum_{\bm{n}'}
|\lambda_{\bm{n},\bm{n}'}|$ is the number of configurations which may
be connected to the configuration $\bm{n}$.

\begin{figure}[t]
  \centering {% Picture saved by xtexcad 2.4
    \unitlength=0.700000pt
    \begin{picture}(430.00,50.00)(0.00,-10.00) \textcolor{rosso}{
        \put(0.00,10.00){\line(1,0){232.00}}
        \put(410.00,15.00){\line(0,-1){10.00}}
        \put(380.00,25.00){\line(0,-1){30.00}}
        \put(310.00,15.00){\line(0,-1){10.00}}
        \put(140.00,15.00){\line(0,-1){10.00}}
        \put(50.00,15.00){\line(0,-1){10.00}}
        \put(0.00,25.00){\line(0,-1){30.00}}
        \put(240.00,18.00){\line(-1,-1){16.00}}
        \put(245.00,18.00){\line(-1,-1){16.00}}
        \put(237.00,10.00){\vector(1,0){245.00}} }
      \put(410.00,30.00){\makebox(0.00,0.00)[tl]{$s_{N_t+1}$}}
      \put(380.00,45.00){\makebox(0.00,0.00)[tc]{$t$}}
      \put(310.00,30.00){\makebox(0.00,0.00)[tl]{$s_{N_t}$}}
      \put(140.00,30.00){\makebox(0.00,0.00)[tl]{$s_2$}}
      \put(50.00,30.00){\makebox(0.00,0.00)[tl]{$s_1$}}
      \put(0.00,45.00){\makebox(0.00,0.00)[tc]{$0$}}
      \put(340.00,0.00){\makebox(0.00,0.00)[tl]{$\bm{n}_{N_t}$}}
      \put(250.00,0.00){\makebox(0.00,0.00)[tl]{$\bm{n}_{N_t-1}$}}
      \put(175.00,0.00){\makebox(0.00,0.00)[tl]{$\bm{n}_2$}}
      \put(90.00,0.00){\makebox(0.00,0.00)[tl]{$\bm{n}_1$}}
      \put(20.00,0.00){\makebox(0.00,0.00)[tl]{$\bm{n}_0$}}
    \end{picture}}
  \caption{Random walk with $N_t$ jumps in the time interval $[0,t)$:
    scheme of the visited configurations and of the corresponding jump
    times.}
  \label{fig1}
\end{figure}
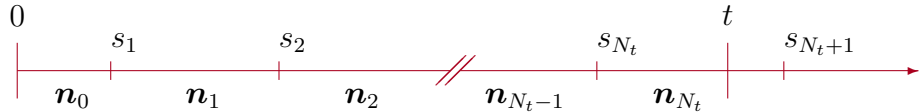

The evolution of the system from an initial state $\psi_0$ is earned
by solving the Schr\"odinger equation. After a time $t$ the system is
found in the state $\psi(t)$ which in the chosen
$\bm{n}$-representation has components (we put $\hbar=1$)
\begin{eqnarray}
  \langle \bm{n} , \psi(t) \rangle 
  = \sum_{\bm{n}_0} 
  \langle \bm{n} , \E^{-\I \hat{H}t}  \bm{n}_0 \rangle 
  \langle \bm{n}_0 , \psi_0 \rangle.
\end{eqnarray}    
This evolution, in the same way as for any linear flow, admits an exact
probabilistic representation \cite{DAJLS,DJS,BPDAJL1,BPDAJL2,BPDAJL3}.
In a one-to-one correspondence with the links, we introduce $M^2$
independent Poisson processes $\{ N^t_{\bm{n},\bm{n}'} \}$ with rates
$\{\rho_{\bm{n},\bm{n}'}\}$. We recall that for these processes the
probability to jump $k$ times in the time interval $[t,t+s)$ is
\cite{Kingman}
\begin{eqnarray}
  \mathrm{prob}(N^{t+s}_{\bm{n},\bm{n}'} - N^t_{\bm{n},\bm{n}'}=k)
  = \frac{(\rho_{\bm{n},\bm{n}'} s)^k}{k!} e^{-\rho_{\bm{n},\bm{n}'}s},
  \qquad k=0,1,2,\dots.
\end{eqnarray}
We establish that each time a Poisson process $N^t_{\bm{n},\bm{n}'}$
jumps, the configuration of the system changes from $\bm{n}$ to
$\bm{n}'$ if $|\lambda_{\bm{n},\bm{n}'}|=1$, otherwise it remains
$\bm{n}$.  Arranging the jumps according to the times,
$s_1<s_2<\dots<s_{N_t}<t<s_{N_t+1}$, at which they take place in the
interval $[0,t)$ we define a random walk in the configuration space of
the system $\bm{n}_0 \to \bm{n}_1 \to \bm{n}_2 \to \dots
\to\bm{n}_{N_t}$ generated by the above rule from a chosen initial
configuration $\bm{n}_0$, see figure \ref{fig1}.  
It is simple to prove that the fundamental
matrix elements of the evolution operator $\langle \bm{n} , \E^{-\I
  \hat{H}t} \bm{n}_0 \rangle$ can be written as an expectation over
the above $M^2$ independent Poisson processes\footnotemark[1]
\footnotetext[1]{%
For $t=0$ we have $ \EE \left( \delta_{ \bm{n} ,
      \bm{n}_{N_0}} {\cal M}^0_{\bm{n}_0} \right) = \langle \bm{n},
  \E^{-\I \hat{H}0} \bm{n}_0 \rangle$ and for $t>0$ we calculate \newline
  $ \D
  \EE ( \delta_{ \bm{n} , \bm{n}_{N_t}} {\cal M}^t_{\bm{n}_0} ) = \EE
  ( \delta_{ \bm{n} , \bm{n}_{N_{t+\D{t}}}} {\cal
    M}^{t+\D{t}}_{\bm{n}_0} ) - \EE( \delta_{ \bm{n} , \bm{n}_{N_t}}
  {\cal M}^t_{\bm{n}_0} ) = - \I\sum_{\bm{n}'} \langle
  \bm{n},\hat{H}\bm{n}' \rangle \EE( \delta_{ \bm{n}' , \bm{n}_{N_t}}
  {\cal M}^t_{\bm{n}_0} ) \D{t} +\mathcal{O}(\D{t^2})$.  The claim follows
  from the uniqueness of the solution of the system of ordinary
  differential equations $\D \langle \bm{n},\E^{-\I \hat{H}t} \bm{n}_0
  \rangle / \D{t} = - \I \sum_{\bm{n}'} \langle \bm{n}, \hat{H}
  \bm{n}' \rangle \langle \bm{n}', \E^{-\I \hat{H}t} \bm{n}_0 \rangle
  $.}
\begin{eqnarray}
  \langle \bm{n} , \E^{-\I \hat{H}t} \bm{n}_0 \rangle =  
  \EE \left( \delta_{ \bm{n} , \bm{n}_{N_t}} {\cal M}^t_{\bm{n}_0} \right),
  \label{pr1}
\end{eqnarray}
\begin{eqnarray}
  {\cal M}^t_{\bm{n}_0} =
  \E^{\sum_{\bm{n},\bm{n}'} \rho_{\bm{n},\bm{n}'}t}
  \left( \prod_{k=1}^{N_{t}} \I \lambda_k \eta_k \rho_k^{-1}
    \E^{-\I V_{k-1}(s_{k}-s_{k-1})} \right)  
  \E^{- \I V_{N_{t}}(t-s_{N_{t}})},
  \label{pr2}
\end{eqnarray}
where we put $s_0=0$ and introduced the shorthand
\begin{eqnarray}
  V_k=V_{\bm{n}_k},  \qquad k=0,1,\dots,N_t,
  \label{V_k}
  \\
  \lambda_k = \lambda_{\bm{n}_{k-1},\bm{n}_k},  \qquad k=1,\dots,N_t,
  \label{lambda_k}
  \\
  \eta_k=\eta_{\bm{n}_{k-1},\bm{n}_k}, \quad 
  \rho_k = \rho_{\bm{n}_{k-1},\bm{n}_k}, \qquad k=1,\dots,N_t.
  \label{etarho_k}
\end{eqnarray}
The rates $\{\rho_{\bm{n},\bm{n}'}\}$ of the Poisson processes are
completely arbitrary, in fact it is easy to check that $\EE( \D {\cal
  M}^t_{\bm{n}_0} /\D{\rho_{\bm{n},\bm{n}'}} )=0$.  Here, for
simplicity, we take $\rho_{\bm{n},\bm{n}'}=\rho$ uniform, whereas
other choices, e.g.  $\rho_{\bm{n},\bm{n}'}=\eta_{\bm{n},\bm{n}'}$,
allow to define optimal Monte Carlo numerical algorithms
\cite{BPDAJL1,OP3}.  The above probabilistic representation holds also
for non-autonomous systems with a time dependent potential
$V_{\boldsymbol{n}}(t)$.  In this case the time-ordered quantum
evolution operator $\mathsf{T}\exp( -\I \int_0^t \hat{H}(u)\D{u})$ has
matrix elements given by (\ref{pr1}-\ref{pr2}) with $V_k \to
\int_{s_k}^{s_{k+1}}V_k(u)\D{u}$.  The representation holds also at
imaginary times $t\to -\I t$ with the substitutions $\lambda_k \to -\I
\lambda_k$ and $V_k \to -\I V_k$.

A convenient way to study the properties of the ground state of a
particle system is to consider its evolution for a long imaginary
time.  Starting from an arbitrary configuration $\bm{n}_0$ the system
finally relaxes into the ground state, assumed not to be orthogonal to
$\bm{n}_0$. In this way we can evaluate any ground-state correlation
function via time asymptotic probabilistic expressions.  For instance,
the ground state energy $E_0$ is given by
\begin{eqnarray}
  E_0 = \lim_{t\to\infty} 
  - \partial_t \log \sum_{\bm{n}}  
  \langle \bm{n} , \E^{-\hat{H}t}  \bm{n}_0 \rangle
  = \lim_{t \to \infty}
  - \partial_t \log \EE (\mathcal{M}_{\bm{n}_0}^{t}),
  \label{pre0}
\end{eqnarray}    
where $\mathcal{M}_{\bm{n}_0}^{t}$ is the imaginary time variant of
(\ref{pr2}). It is of fundamental importance that for the expectation
$\EE(\mathcal{M}_{\bm{n}_0}^{t})$ we can find, at large times, an
analytical result. In the following we outline how this is obtained.

First, we decompose $\EE (\mathcal{M}_{\bm{n}_0}^{t})$ in a series of
canonical expectations over random walks with a fixed number of jumps
\begin{eqnarray}
  \EE \left( {\cal M}^{t}_{\bm{n}_0} \right) =
  \sum_{N=0}^{\infty}
  \EE \left( {\cal M}^{t}_{\bm{n}_0},N_t=N \right) 
  \label{prcandec}
\end{eqnarray}
and evaluate each term of this series by integrating over all possible
jump times.  The probability to have $N$ Poisson processes jumping in
the interval $[0,t)$ with the $k$th process jumping in the interval
$[s_k,s_k+\D{s_k})$ amounts to
\begin{eqnarray}
  \prod_{k=1}^N \E^{-\sum_{\bm{n},\bm{n}'} \rho_{\bm{n},\bm{n}'}(s_k-s_{k-1})}
  \rho_k \D{s_k}.
\end{eqnarray}
With a simple calculation we then obtain
\begin{eqnarray}
  \EE \left( {\cal M}^{t}_{\bm{n}_0},N_t=N \right) 
  =\sum_{r\in\Omega_N}
  \mathcal{W}_N^{(r)}(t) \prod_{k=1}^N \lambda_k^{(r)} \eta_k^{(r)},
  \label{prcan0}
\end{eqnarray}
where $\Omega_N=\Omega_N(\bm{n}_0)$ is the set of all possible random
walks with $N$ jumps branching from $\bm{n}_0$.  The contribution of
the $r$th random walk $\bm{n}_0 \to \bm{n}_1^{(r)} \to \dots \to
\bm{n}_N^{(r)}$ includes the factor
$\mathcal{W}_N^{(r)}(t)=\mathcal{L}^{-1}[\tilde{\mathcal{W}}_N^{(r)}(z)](t)$,
namely the inverse Laplace transform of
\begin{eqnarray}
  \tilde{\mathcal{W}}_N^{(r)}(z) = 
  \prod_{k=0}^N \frac{1}{z+V_k^{(r)}}.
  \label{laplacetrasform}
\end{eqnarray}
From equation (\ref{prcan0}) it is evident that only the random walks
with $|\lambda_k^{(r)}|=1$, i.e. $\bm{n}_{k-1}^{(r)} \neq
\bm{n}_k^{(r)}$, for $k=1,\dots,N$, contribute to the sum.  Thus we
can rewrite the sum over $\Omega_N$ as a probabilistic expectation
over these effective random walks.  According to the choice
$\rho_{\bm{n},\bm{n}'}=\rho$ uniform, the effective random walks
correspond to a Markov chain with transition matrix
$P_{\bm{n},\bm{n}'} = |\lambda_{\bm{n},\bm{n}'}| /\sum_{\bm{n}''}
|\lambda_{\bm{n},\bm{n}''}|$ \cite{Bremaud}.  The probability that we
must associate with the $r$th element of the set $\Omega_N$ is,
therefore,
\begin{eqnarray}
  p_N^{(r)} =
  \prod_{k=1}^N 
  \frac{|\lambda_{\bm{n}_{k-1}^{(r)},\bm{n}_k^{(r)}}|}
  {\sum_{\bm{n}} |\lambda_{\bm{n}_{k-1}^{(r)},\bm{n}}|}.
  \label{prob1}
\end{eqnarray}
Note that $p_N^{(r)}=0$ for non effective random walks and
$\sum_{r\in\Omega_N} p_N^{(r)} =1$.  Multiplying and dividing 
each addend of equation (\ref{prcan0}) by
$p_N^{(r)}$, we rewrite the canonical expectations as
\begin{eqnarray}
  \EE \left( {\cal M}^{t}_{\bm{n}_0},N_t=N \right) 
  = \sum_{r\in\Omega_N} p_N^{(r)}
  \mathcal{L}^{-1}\left[\prod_{k=0}^N \frac{1}{z+V_k^{(r)}}\right](t)
  \prod_{k=1}^N T_k^{(r)}
  \prod_{k=1}^N \lambda_k^{(r)},
  \label{prcan1}
\end{eqnarray}
where
\begin{eqnarray}
  T_k = \eta_{\bm{n}_{k-1},\bm{n}_k}
  \sum_{\bm{n}} |\lambda_{\bm{n}_{k-1},\bm{n}}|, \qquad k=1,\dots,N.
  \label{T_k}
\end{eqnarray}

Equation (\ref{prcan1}) shows that $\EE \left( {\cal
    M}^{t}_{\bm{n}_0},N_t=N \right)$ is the average of a quantity
which does not rely on the detailed sequence of the configurations
visited during the time $t$. It depends just on the multiplicities, or
numbers of occurrences, of the potential, hopping and phase variables,
$V$, $T$ and $\lambda$, defined by (\ref{V_k}), (\ref{T_k}) and
(\ref{lambda_k}), respectively.  For a random walk with $N$ jumps,
these multiplicities are explicitly defined as
\begin{eqnarray}
  N_V &= \sum_{k=0}^{N}\delta_{V,V_k}, \qquad V \in \mathscr{V},
  \label{N_V}
  \\
  N_T &= \sum_{k=1}^{N}\delta_{T,T_k}, \qquad T \in \mathscr{T},
  \label{N_T}
  \\
  N_{\lambda} &= \sum_{k=1}^{N}\delta_{\lambda,\lambda_{k}},
  \qquad \lambda \in \mathscr{L},
  \label{N_lambda}
\end{eqnarray}
where $\mathscr{V}$, $\mathscr{T}$ and $\mathscr{L}$ are the sets of
all possible values assumed by (\ref{V_k}), (\ref{T_k}) and
(\ref{lambda_k}) during a random walk with infinitely many jumps. Note
that $0 \notin \mathscr{L}$ as jumps between configurations $\bm{n}$
and $\bm{n}'$ with $\lambda_{\bm{n},\bm{n}'}=0$ have zero probability
to be realized.  Since any configuration can be obtained from any
other one by a finite number of jumps, i.e. the Markov chain we are
considering is irreducible, the elements in the sets $\mathscr{V}$,
$\mathscr{T}$ and $\mathscr{L}$ do not depend on the initial
configuration $\bm{n}_0$.  Let us indicate the multiplicities $N_V$,
$N_T$ and $N_\lambda$ collectively by a vector with as many components
as the elements in the set
$\mathscr{H}=\mathscr{V}\cup\mathscr{T}\cup\mathscr{L}$,
\begin{eqnarray}
  \bm{\mu}^\mathrm{T} =
  (\ldots N_V \ldots; \ldots N_T \ldots; \ldots N_\lambda \ldots).
\end{eqnarray}
If we split the set $\Omega_N$ into subsets of random walks with equal
values of $\bm{\mu}$, we conclude that
\begin{eqnarray}\fl
  \EE \left( {\cal M}^{t}_{\bm{n}_0},N_t=N \right) =
  \sum_{\bm{\mu}} \mathcal{P}_{N}(\bm{\mu}) ~
  \mathcal{L}^{-1}\left[\prod_{V\in\mathscr{V}} (z+V)^{-N_V}\right](t)
  \prod_{T \in \mathscr{T}} T^{N_T}\prod_{\lambda \in \mathscr{L}} 
  {\lambda}^{N_{\lambda}},
  \label{prcan2}
\end{eqnarray}
where
\begin{eqnarray}
  \mathcal{P}_{N}(\bm{\mu}) =
  \sum_{r \in \Omega_{N}} 
  \left(
    \prod_{k=1}^N 
    \frac{|\lambda_{\bm{n}_{k-1}^{(r)},\bm{n}_k^{(r)}}|}
    {\sum_{\bm{n}} 
      |\lambda_{\bm{n}_{k-1}^{(r)},\bm{n}}|}
  \right)
  ~\delta_{\bm{\mu}^{(r)},\bm{\mu}}
  \label{prob2}
\end{eqnarray}
is the probability to have random walks with multiplicities $\bm{\mu}$
after $N$ jumps from the configuration $\bm{n}_0$.  Note that
$\mathcal{P}_{N}(\bm{\mu})=0$ unless $\bm{\mu}$ satisfies the
following three constraints
\begin{eqnarray}
  \sum_{V\in\mathscr{V}} N_V=N+1, \qquad
  \sum_{T\in\mathscr{T}} N_T=N, \qquad
  \sum_{\lambda\in\mathscr{L}} N_\lambda=N.
  \label{constraintsmu}
\end{eqnarray}

As a second step, we observe that when the imaginary time $t$ becomes
large the full expectation (\ref{prcandec}) takes exponentially
leading contributions from terms with $N \sim t$.  Thus, we can
replace all previous results with their $N\to\infty$ asymptotic
expressions. Due to the ergodicity of the underlying Markov chain, the
probability (\ref{prob2}) looses memory of the initial configuration
$\bm{n}_0$.  We can evaluate the inverse Laplace transform which
appears in (\ref{prcan2}) by a saddle point technique in the complex
plane. In the same equation we can also substitute the sum over the
multiplicities by an integral over $\bm{\mu}$ and avoid distinguishing
the normalizations $N$ and $N+1$ in (\ref{constraintsmu}).  As a final
result we have the following asymptotic logarithm equality
\begin{eqnarray}
  \label{ceint}
  \EE \left( {\cal M}^{t}_{\bm{n}_0},N_t=N \right) \simeq  
  \int \D{(N\bm{\nu})} ~\mathcal{P}_N(N\bm{\nu})~ 
  \frac{ \E^{x_{0}t + N\left( \bm{\nu},\bm{u} \right) } }
  { \sqrt {2\pi N (\bm{\nu},\bm{w}) }} ,
\end{eqnarray}
where we have introduced the frequencies $\bm{\nu}= \bm{\mu}/N$, which
have a finite limit for $N\to\infty$, the vectors
\begin{eqnarray}
  \bm{u}^\mathrm{T} &=& 
  (\ldots -\log(x_{0}+V) \ldots;
  \ldots \log T \ldots; \ldots \log \lambda \ldots),
  \\
  \bm{w}^\mathrm{T} &=& 
  (\ldots (x_{0}+V)^{-2} \ldots; \ldots 0 \ldots; \ldots 0 \ldots),
\end{eqnarray}
and the scalar product $(\bm{a},\bm{b})=\sum_{\alpha\in\mathscr{H}}
a_\alpha b_\alpha$.  The quantity $x_0$, which is the real
saddle-point in the complex contour used to evaluate the Laplace
antitransform, is the unique solution of the scalar equation
\begin{eqnarray}
  \sum_{V \in \mathscr{V}}\frac{\nu_{V}}{x_{0}+V}=\frac{t}{N}, 
  \qquad x_{0}>-V_{\mathrm{min}}.
  \label{spx0}
\end{eqnarray}
Note that this equation has a regular scaling behavior for
$t,N\to\infty$ with $N\sim t$.

The integral over the frequencies $\bm{\nu}$ in equation (\ref{ceint})
is easily performed by a saddle-point method whenever the asymptotic
probability density $\mathcal{P}_{N}(N\boldsymbol{\nu})$ is known.  We
cannot hope to evaluate this probability from the microscopic
definition (\ref{prob2}) except for very particular models. For
general systems, two different strategies have been considered, see
\cite{OP4} and \cite{OP5}.  We can relate
$\mathcal{P}_{N}(N\boldsymbol{\nu})$ to proper statistical moments of
the system under consideration, measure or calculate some of these
moments and, lastly, obtain partial information about $E_0$.  Or we
can postulate some expressions for
$\mathcal{P}_{N}(N\boldsymbol{\nu})$ and see which kind of systems are
described, exactly or approximately, by this guess.

\subsection{Cumulant expansion}
\label{cumulant.expansion}
If we rewrite the probability density $\mathcal{P}_N(N\bm{\nu})$ in
terms of its Fourier transform ${\tilde{\mathcal{P}}}_N(\bm{q})$
\begin{eqnarray}
  \label{FOURIER}
  \mathcal{P}_N(N\bm{\nu}) = (2\pi)^{-|\mathscr{H}|}
  \int \D{\bm{q}} ~
  \tilde{\mathcal{P}}_N(\bm{q})
  \E^{-\I (\bm{q},N\bm{\nu})},
\end{eqnarray}
we can associate $\log{\tilde{\mathcal{P}}}_N(\bm{q})$ with the
cumulants, or connected correlation functions, of the multiplicities
$N\bm{\nu}$ sampled with respect to the measure
$\mathcal{P}_N(N\bm{\nu})$.  Indicating with $\mediac{N\nu_{\alpha_1}
  \dots N\nu_{\alpha_k}}$ the component $\alpha_1 \dots \alpha_k$ of
the cumulant of order $k$, we have the well known relation
\cite{Shiryayev}
\begin{eqnarray}
  \label{LOGP}
  \log {\tilde{\mathcal{P}}}_N(\bm{q}) &=
  \sum_{k=1}^{\infty} \frac{1}{k!}
  \langle
  (N\boldsymbol{\nu},\I\bm{q})^k
  \rangle_N^{(c)}
  \nonumber \\ &=
  \sum_{k=1}^{\infty} \frac{\I^k}{k!}
  \sum_{\alpha_1\in\mathscr{H}} \dots \sum_{\alpha_k\in\mathscr{H}}
  \mediac{N\nu_{\alpha_1} \dots N\nu_{\alpha_k}}
  q_{\alpha_1} \dots q_{\alpha_k}.
\end{eqnarray}
At this point, it is simple to calculate the canonical expectations
$\EE \left( {\cal M}^{t}_{\bm{n}_0},N_t=N \right)$ performing the
integrals over the $2|\mathscr{H}|$ variables $\bm{\nu}$ and $\bm{q}$
by a saddle-point approximation which is asymptotically exact for
$N\to\infty$.  The last step is to evaluate $E_0$ by resumming the
series (\ref{prcandec}).  By virtue of the limit $t\to\infty$ to be
taken at the end, we still have an exact result if we replace the
series over $N$ by an integral and estimate this integral at its
maximum.  There is one important point to observe.  Independently
of their order $k$, the cumulants $\mediac{N\nu_{\alpha_1} \dots
  N\nu_{\alpha_k}}$ diverge as $N$ for $N\to\infty$.  Thus we
introduce the asymptotic rescaled cumulants
\begin{eqnarray}
  \Sigma^{(k)}_{\alpha_1 \ldots \alpha_k} =
  \lim_{N\to \infty}
  \frac{1}{N} \mediac{N_{\alpha_1} \ldots N_{\alpha_k}},
  \label{arc}
\end{eqnarray}
$\bm{\Sigma}^{(k)}$ in a compact notation. Existence and finiteness of
these limits are ensured by the finite correlation length $N_c$ which
characterizes the correlations functions, connected or not, of the the
multiplicities \cite{OP4}.  We conclude that $E_0$ is the unique
solution of the scalar equation
\begin{eqnarray}
  \sum_{k=1}^{\infty}
  \frac{1}{k!}
  \sum_{\alpha_1\in\mathscr{H}}
  \ldots
  \sum_{\alpha_k\in\mathscr{H}}
  \Sigma_{\alpha_1 \ldots \alpha_k}^{(k)}
  u_{\alpha_1}(E_{0}) \dots u_{\alpha_k}(E_{0}) =0, 
  \qquad E_{0} \leq V_\mathrm{min},
  \label{e0cumulants}
\end{eqnarray}
where
\begin{eqnarray}
  \boldsymbol{u}^\mathrm{T}(E_{0}) =
  (\ldots -\log(-E_0+V) \ldots;
  \ldots \log T \ldots; \ldots \log \lambda \ldots).
  \label{ue0}
\end{eqnarray}
Equation (\ref{e0cumulants}) is exact and the uniqueness of its
solution is ensured by the constraint $E_{0} \leq V_\mathrm{min}$
which stems from the Laplace transform causality condition
(\ref{spx0}).

To find the exact ground-state energy $E_0$ from equation
(\ref{e0cumulants}) we have to know the cumulants $\bm{\Sigma}^{(k)}$
at any order $k$. Of course, for a general system this is not
conceivable. However, up to some small order $k$ and even for systems
of relatively large size, the cumulants can be measured by reliable
statistical simulations \cite{OP4}.  With these input data, we can
truncate the series in (\ref{e0cumulants}) at some order and obtain an
approximation to $E_0$.  Independently of the truncation order, there
is an important feature to note.  Suppose that we change the
Hamiltonian $\hat{H}$ leaving unaltered the adjacency matrix
$|\lambda_{\bm{n},\bm{n}'}|$ and the number of elements in the sets
$\mathscr{V}$, $\mathscr{T}$ and $\mathscr{L}$.  The asymptotic
rescaled cumulants $\bm{\Sigma}^{(k)}$ are unaffected by this change
and the only modifications to equation (\ref{e0cumulants}) are encoded
analytically by $\bm{u}(E_0)$.  Therefore, the same input data
$\bm{\Sigma}^{(k)}$, $k=1,\dots,k_\mathrm{max}$, can be used to find
the ground-state energy of parametric Hamiltonians as a function of
their parameters. The simplest example is to consider
$\hat{H}(\gamma)=\hat{K}+\gamma\hat{V}$ and evaluate the function
$E_0(\gamma)$.

At the lowest truncation order $k_\mathrm{max}=1$, equation
(\ref{e0cumulants}) reads
\begin{eqnarray}\fl
  \sum_{V\in\mathscr{V}} \Sigma^{(1)}_V \log(-E_0+V) =
  \sum_{T\in\mathscr{T}} \Sigma^{(1)}_T \log T +
  \sum_{\lambda\in\mathscr{L}} \Sigma^{(1)}_\lambda \log \lambda,
  \qquad E_{0} \leq V_\mathrm{min}.
\end{eqnarray}
In general this equation must be solved numerically.  Conversely, for
$\hat{V}=0$ we have $\mathscr{V}=\{0\}$ and $\Sigma^{(1)}_{V=0}=1$
which allow us to find the analytical solution
\begin{eqnarray}
  E_0^{(0)} = - \left( \prod_{T\in\mathscr{T}} T^{\Sigma^{(1)}_T} \right)
  \left( \prod_{\lambda\in\mathscr{L}} \lambda^{\Sigma^{(1)}_\lambda} \right).  
  \label{e00order1}
\end{eqnarray}
This represents the lowest order approximation to the ground-state
energy of the hopping operator $\hat{K}$.

The cumulant expansion described above has been applied to study
Hubbard models in a two dimensional lattice \cite{OP4}.  By including
cumulants up to order 4, the results compare rather well with the
exact ground-state energy (determined numerically in other ways) at
least in the case of hard core bosons, i.e. $\mathscr{L}=\{1\}$, and
for interaction energies not too large with respect to the hopping
term.  The results, however, are disappointing for large interaction
energies, in fact in this limit $E_0$ diverges.  Moreover, in the case
of fermions when $\mathscr{L}=\{-1,1\}$ pointless complex solutions
can be found for $E_0$ as shown, for instance, for $\hat{V}=0$ and at
the lowest order by equation (\ref{e00order1}).  These problems stem
form the fact that truncating at some order $k_\mathrm{max}$ equation
(\ref{e0cumulants}) is equivalent to consider an artificial
probability density $\mathcal{P}_N(N\bm{\nu})$ with zero cumulants at
any order $k>k_\mathrm{max}$.

\subsection{Multinomial probability density}
\label{multinomial.probability.density}
In view of the definition (\ref{arc}) of the asymptotic rescaled
cumulants, equation (\ref{e0cumulants}) can be compactly rewritten as
\begin{eqnarray}
  \lim_{N\to\infty} \frac{1}{N}
  \sum_{k=1}^{\infty} \frac{1}{k!}
  \langle
  (N\boldsymbol{\nu},\boldsymbol{u}(E_0))^k
  \rangle_N^{(c)} = 0,
  \qquad E_0 \leq V_{\mathrm{min}}.
\end{eqnarray}
The series sums up to $\log {\tilde{\mathcal{P}}}_N(-\I \bm{u}(E_0))$
hence we can state that $E_0$ is the unique solution of the exact
equation
\begin{eqnarray}
  \label{E0probden}
  \lim_{N\to\infty} \frac{1}{N} \log
  \int \D{(N\boldsymbol{\nu})} ~
  \mathcal{P}_{N}(N\boldsymbol{\nu}) ~
  \E^{(N\boldsymbol{\nu},\boldsymbol{u}(E_{0}))} =0, 
  \qquad 
  E_{0}\leq V_{\mathrm{min}},
\end{eqnarray}
with $\bm{u}(E_0)$ given by (\ref{ue0}).  Equation (\ref{E0probden})
makes crystal clear that the knowledge of $E_0$ stems from that of
$\mathcal{P}_{N}(N\boldsymbol{\nu})$.

The random walks in the configuration space definitely induce
correlations among the multiplicities $\bm{\mu}$ due to the dependence
of the potential, hopping and phase variables on the visited
configurations.  If we could neglect these correlations, for $N$
sufficiently large the multiplicities of each set $\mathscr{V}$,
$\mathscr{T}$ and $\mathscr{L}$ would be equivalent to multinomial
trials processes with success probabilities
\begin{eqnarray}
  p_V &= \lim_{N\to\infty} \frac{1}{N+1} 
  \sum_{k=0}^{N}\delta_{V,V_k}=\Sigma^{(1)}_V, \qquad V \in \mathscr{V},
  \label{p_V}
  \\
  p_T &= \lim_{N\to\infty} \frac{1}{N}
  \sum_{k=1}^{N}\delta_{T,T_k}=\Sigma^{(1)}_T, \qquad T \in \mathscr{T},
  \label{p_T}
  \\
  p_{\lambda} &= \lim_{N\to\infty} \frac{1}{N}
  \sum_{k=1}^{N}\delta_{\lambda,\lambda_{k}}=\Sigma^{(1)}_\lambda,
  \qquad \lambda \in \mathscr{L}.
  \label{p_lambda}
\end{eqnarray}
Note that we made use of the ergodic properties of the underlying
Markov chain.  In that case, the probability
$\mathcal{P}_{N}(\boldsymbol{\mu})$ would be a product of multinomial
distributions
\begin{eqnarray}
  \label{p_mult}
  \mathcal{P}_{N}(\boldsymbol{\mu}) =
  (N+1)!\prod_{V\in\mathscr{V}}\frac{p_{V}^{N_{V}}}{N_{V}!}
  ~~
  N!\prod_{T\in\mathscr{T}}\frac{p_{T}^{N_{T}}}{N_{T}!}
  ~~
  N!\prod_{\lambda\in\mathscr{L}}\frac{p_{\lambda}^{N_{\lambda}}}{N_{\lambda}!},
\end{eqnarray}
where the multiplicities $N_V$, $N_T$ and $N_\lambda$ are integers
which satisfy the constraints (\ref{constraintsmu}).

Equation (\ref{p_mult}) greatly simplifies for $N$ large.  By using
Stirling's approximation for the factorials and explicitly taking into
account the constraints (\ref{constraintsmu}), we have the following
asymptotic equality for the associated probability density.
\begin{eqnarray}\fl
  \mathcal{P}_{N}(N\boldsymbol{\nu}) \simeq \exp[N \omega(\bm{\nu})]~ 
  \delta\left(\sum_{V\in\mathscr{V}} N \nu_V -N \right) 
  \delta\left(\sum_{T\in\mathscr{T}} N \nu_T -N \right)
  \delta\left(\sum_{\lambda\in\mathscr{L}} N \nu_\lambda -N \right),
\end{eqnarray}
where
\begin{eqnarray}
  \omega(\bm{\nu})=
  \sum_{\alpha\in\mathscr{H}}
  \nu_{\alpha}\log\left(\frac{p_{\alpha}}{\nu_{\alpha}}\right)
  \label{omega.multinomial}
\end{eqnarray}
and $\bm{p}^T=(\dots p_V \dots;\dots p_T \dots;\dots p_\lambda \dots)$
is a vector collecting the success probabilities (\ref{p_V}),
(\ref{p_T}) and (\ref{p_lambda}).  To find $E_0$ it remains to
calculate the integral of equation (\ref{E0probden}) over the
frequencies $\bm{\nu}$. The integration can be performed by steepest
descent as detailed in the following section. The result is that $E_0$
is the solution of
\begin{eqnarray}
  \label{E0multinomial}
  \sum_{V\in\mathscr{V}} 
  \frac{p_{V}}{-E_{0}+V} = 
  \frac{1}
  {\left( \sum_{T\in\mathscr{T}} p_{T} T \right)
    \left( \sum_{\lambda\in\mathscr{L}} p_{\lambda} \lambda \right)},
  \qquad E_0 \leq V_\mathrm{min}.
\end{eqnarray}
For $\hat{V}=0$, i.e. $\mathscr{V}=\{0\}$ and $p_{V=0}=1$, equation
(\ref{E0multinomial}) can be solved analytically and we obtain the
value
\begin{eqnarray}
  E_0^{(0)} = - \left( \sum_{T\in\mathscr{T}} p_{T} T \right)
  \left( \sum_{\lambda\in\mathscr{L}} p_{\lambda} \lambda \right)  
  \label{E00multinomial}
\end{eqnarray}
for the ground-state energy of the hopping operator $\hat{K}$.  For
$\hat{V}\neq 0$ equation (\ref{E0multinomial}) is straightforwardly
solved numerically by bisection method.

The uncorrelated multinomial probability density considered here has
the great advantage that cumulants of any order are included in the
determination of $E_0$. This eliminates the artifacts obtained by
truncating the cumulant expansion (\ref{e0cumulants}), namely the
wrong behavior of $E_0$ at large interaction energies and its complex
value in the case of fermions.  In fact, for large interaction
strength equation (\ref{E0multinomial}) admits the asymptotic finite
solution
\begin{eqnarray}
  E_0 = V_\mathrm{min} + p_{V_\mathrm{min}} E_0^{(0)}.
  \label{E0Vmultinomial}
\end{eqnarray}
Moreover, for intermediate interaction strengths the solution of
equation (\ref{E0multinomial}) varies monotonously between the two
limits (\ref{E00multinomial}) and (\ref{E0Vmultinomial}), as expected.
In the case of fermions, the $\hat{V}=0$ solution
(\ref{E00multinomial}) is real and negative, as must be.  In fact,
among the links $\lambda_{\bm{n},\bm{n}'}\neq 0$ those with a positive
sign are the majority, so that
\begin{eqnarray}
  \sum_{\lambda\in\mathscr{L}} p_{\lambda} \lambda = 
  \Sigma_{\lambda=1}^{(1)} - \Sigma_{\lambda=-1}^{(1)} >0.
\end{eqnarray}

The main drawback of the present approach has already been mentioned.
The correlations among the multiplicities $\bm{\mu}$ are neglected so
that the comparison of $E_0$, solution of equation
(\ref{E0multinomial}), with the ground-state energy of a system of
particles can be quantitatively poor. Whenever these correlations are
absent, as in the case of the uniformly fully connected models and of
the random potential systems considered in \cite{OP5}, the approach
provides exact results together with the possibility to study the
appearance of quantum phase transitions in the thermodynamic limit.

\section{Multinomial perturbative scheme}
\label{multinomial.perturbative.scheme}
Here we propose a probabilistic perturbative scheme with the aim of
merging the merits of the multinomial probability density of section
\ref{multinomial.probability.density} with the statistical details
provided by the cumulant expansion of section
\ref{cumulant.expansion}.  Basically the idea is as follows. We
consider an asymptotic probability density which has the same
structure of a multinomial density but with parameters $p_{\alpha}$,
$\alpha\in\mathscr{H}$, that are functions of the frequencies
$\boldsymbol{\nu}$,
\begin{eqnarray}
  \label{OMEGA_pert}
  \omega\left(\boldsymbol{\nu}\right)=
  \lim_{N\rightarrow\infty}\frac{1}{N}\log\mathcal{P}_{N}
  \left(N\boldsymbol{\nu}\right) =
  \sum_{\alpha\in\mathscr{H}}\nu_{\alpha}\log\left(\frac{p_{\alpha}
      \left(\boldsymbol{\nu}\right)}{\nu_{\alpha}}\right),
\end{eqnarray}
and write these functions as power series of the form
\begin{eqnarray}
  \label{p_alpha}
  p_{\alpha}\left(\boldsymbol{\nu}\right)=
  \sum_{n=0}^{\infty} \frac{1}{n!}
  \sum_{\beta_1\in\mathscr{H}}\dots\sum_{\beta_n\in\mathscr{H}}
  p^{(n+1)}_{\alpha\beta_1\dots\beta_n}
  \left(\nu_{\beta_1}-p^{(1)}_{\beta_1}\right)
  \dots
  \left(\nu_{\beta_n}-p^{(1)}_{\beta_n}\right).
\end{eqnarray}
The scalars $p^{(k)}_{\alpha_1\ldots\alpha_{k}}$, with
$\alpha_i\in\mathscr{H}$ for $i=1,\dots,k$, in a compact notation
$\boldsymbol{p}^{(k)}$, are the perturbative parameters at order
$k$. Note that we identify the perturbative order $k$ with the rank of
the tensor $\boldsymbol{p}^{(k)}$ not the index $n$ of the series
(\ref{p_alpha}), i.e. $k=n+1=1,2,\dots$.  At the lowest perturbative
order $k=1$ we have
$p_{\alpha}\left(\boldsymbol{\nu}\right)=p^{(1)}_\alpha$,
$\alpha\in\mathscr{H}$, constant as in the strict multinomial case.

We determine the perturbative parameters $\boldsymbol{p}^{(k)}$ as
follows.  First, we assume that for $k\geq 2$ they are symmetric under
the exchange of any two of their indices $\alpha_1, \ldots, \alpha_k$
\begin{eqnarray}
  \label{PPAR_SIM}
  p_{\dots\alpha_i\dots\alpha_j\dots}^{(k)} = 
  p_{\dots\alpha_j\dots\alpha_i\dots}^{(k)} .
\end{eqnarray}
Second, in order to maintain as much as possible the structure of a
multinomial, we ask that, for any value of $\boldsymbol{\nu}$, the
functions $p_{\alpha}\left(\boldsymbol{\nu}\right)$ are normalized in
each set $\mathscr{V},\mathscr{T},\mathscr{L}$, i.e.
\begin{eqnarray}
  \label{Pa_eq1}
  \sum_{\alpha\in\mathscr{A}}p_{\alpha}(\boldsymbol{\nu})=1,
  \qquad \mathscr{A}=\mathscr{V},\mathscr{T},\mathscr{L}.
\end{eqnarray}
Third, we require that the asymptotic rescaled cumulants of order $k$
evaluated from (\ref{OMEGA_pert}--\ref{p_alpha}), hereafter indicated
by $\arc{\nu_{\alpha_1}\ldots\nu_{\alpha_k}}$, coincide with those
effectively owned by the system, $\boldsymbol{\Sigma}^{(k)}$,
\begin{eqnarray}
  \label{SYSTEM_PAR}
  \arc{\nu_{\alpha_1}\ldots\nu_{\alpha_k}}
  (\boldsymbol{p}^{(1)},\boldsymbol{p}^{(2)},\ldots,\boldsymbol{p}^{(k)}) = 
  \Sigma_{\alpha_1\ldots\alpha_{k}}^{(k)}.
\end{eqnarray}
In this expression we have anticipated that
$\arc{\nu_{\alpha_1}\ldots\nu_{\alpha_k}}$ depends only on the
parameters $\boldsymbol{p}^{(j)}$ with $j\leq k$.  This property
implies that we can first find $\boldsymbol{p}^{(1)}$ by solving the
system of $|\mathscr{H}|$ equations
\begin{eqnarray}
  \label{SYSTEM_P1}
  \arc{\nu_{\alpha_1}}(\boldsymbol{p}^{(1)}) = \Sigma_{\alpha_1}^{(1)},
  \qquad \alpha_1\in\mathscr{H}, 
\end{eqnarray}
next find $\boldsymbol{p}^{(2)}$ by solving the system of
$|\mathscr{H}|^2$ equations
\begin{eqnarray}
  \label{SYSTEM_P2}
  \arc{\nu_{\alpha_1}\nu_{\alpha_2}}(\boldsymbol{p}^{(1)}, 
  \boldsymbol{p}^{(2)}) = \Sigma_{\alpha_1\alpha_2}^{(2)},
  \qquad \alpha_1,\alpha_2\in\mathscr{H},
\end{eqnarray}
next find $\boldsymbol{p}^{(3)}$ by solving the system of
$|\mathscr{H}|^3$ equations
\begin{eqnarray}
  \label{SYSTEM_P3}
  \arc{\nu_{\alpha_1}\nu_{\alpha_2}\nu_{\alpha_3}}(\boldsymbol{p}^{(1)}, 
  \boldsymbol{p}^{(2)},\boldsymbol{p}^{(3)}) = 
  \Sigma_{\alpha_1\alpha_2\alpha_3}^{(3)},
  \qquad \alpha_1,\alpha_2,\alpha_3\in\mathscr{H},
\end{eqnarray}
and so on up to a chosen maximum order $k_{\mathrm{max}}$ which
corresponds to truncating the series (\ref{p_alpha}) at the term
$n=k_{\mathrm{max}}-1$ included.

To be precise, the probability density (\ref{OMEGA_pert}) is well
defined only if each $p_\alpha(\bm{\nu})$, $\alpha\in\mathscr{H}$, is
a non negative function of the vector $\bm{\nu}$ with components
$\nu_\alpha \geq 0$ varying in the unit simplex
$\sum_{\alpha\in\mathscr{A}} \nu_\alpha=1$ for
$\mathscr{A}=\mathscr{V},\mathscr{T},\mathscr{L}$.
\textit{A priori} we don't know whether this condition can be met by
introducing the perturbative parameters $\bm{p}^{(k)}$ as described
above.  We will then proceed heuristically.  Whenever an effective
solution of equation (\ref{E0probden}) can be found by using the
probability density (\ref{OMEGA_pert}--\ref{p_alpha}), that will be
the signal that the proposed perturbative scheme is meaningful.

In the following sections, we will first evaluate, up to the third
order, the asymptotic rescaled cumulants associated with the probability
density (\ref{OMEGA_pert}--\ref{p_alpha}), then we will discuss the
solution of the systems of equations (\ref{SYSTEM_P1}),
(\ref{SYSTEM_P2}) and (\ref{SYSTEM_P3}).  Finally, we will make some
comments on the higher perturbative orders.

\subsection{Evaluation of the asymptotic rescaled cumulants}
\label{evaluation.arc}
The definition of the asymptotic rescaled cumulants of order
$k=1,2,\dots$ is
\begin{eqnarray}
  \arc{\nu_{\alpha_{1}}\ldots\nu_{\alpha_{k}}}=
  \lim_{N\rightarrow\infty}
  \frac{1}{N}
  \left\langle N\nu_{\alpha_{1}}\dots 
    N\nu_{\alpha_{k}}\right\rangle_{N}^{\left(c\right)},
  \qquad \alpha_1,\dots,\alpha_k\in\mathscr{H},
\end{eqnarray}
where $\left\langle N\nu_{\alpha_{1}}\dots N\nu_{\alpha_{k}}
\right\rangle_{N}^{\left(c\right)}$ are the cumulants, or connected
correlation functions, of the multiplicities $N_\alpha=N\nu_\alpha$
sampled with respect to the $N$-jumps probability density
$\mathcal{P}_{N}(N\boldsymbol{\nu})$.  In turn, as standard, the
cumulants are obtained from the generating function associated with
$\mathcal{P}_{N}(N\boldsymbol{\nu})$
\begin{eqnarray}
  \label{asym_cum}
  \langle N\nu_{\alpha_{1}} \ldots N\nu_{\alpha_{k}}\rangle^{(c)}_N = 
  \left.\frac{\partial^{k}\log Z_{N}(\boldsymbol{J})}
    {\partial J_{\alpha_1}\ldots\partial J_{\alpha_k}} 
  \right|_{\boldsymbol{J}=\boldsymbol{0}},
\end{eqnarray}
\begin{eqnarray}
  Z_{N}(\boldsymbol{J}) = \int \mathrm{d}(N\boldsymbol{\nu})
  \mathcal{P}_N(N\boldsymbol{\nu})
  \mathrm{e}^{\left(\boldsymbol{J},N\boldsymbol{\nu}\right)}.
\end{eqnarray}
Assuming that for $N\to \infty$ the logarithm of
$\mathcal{P}_{N}(N\boldsymbol{\nu})$ is given by
(\ref{OMEGA_pert}--\ref{p_alpha}), for $N$ large we have, up to an
inessential constant,
\begin{eqnarray}
  \label{Z_N}
  Z_{N}(\boldsymbol{J}) &=& \int \mathrm{d}(N\boldsymbol{\nu})
  \mathrm{e}^{N \omega(\boldsymbol{\nu}) + 
    \left(\boldsymbol{J},N\boldsymbol{\nu}\right)}
  \prod_{\mathscr{A}=\mathscr{V},\mathscr{T},\mathscr{L}}
  \delta\left(\sum_{\alpha\in\mathscr{A}}N\nu_{\alpha}-N\right),
\end{eqnarray}
where we have explicitly taken into account the fact that the
multiplicities of each set $\mathscr{V},\mathscr{T},\mathscr{L}$ must
sum to $N$.  Using the Fourier integral representation of the Dirac
$\delta$, we rewrite the generating function, up to a constant, as
\begin{eqnarray}
  \label{Z_pert}
  Z_{N}(\boldsymbol{J}) &=&  \int 
  \prod_{\mathscr{A}=
    \mathscr{V},\mathscr{T},\mathscr{L} }\mathrm{d}k_{\mathscr{A}}
  \prod_{\alpha\in\mathscr{H}}\mathrm{d}\nu_{\alpha}
  \mathrm{e}^{N\phi(\boldsymbol{\nu},\boldsymbol{k},\boldsymbol{J})},
\end{eqnarray}
where
\begin{eqnarray}\fl
  \label{phi_pert}
  \phi(\boldsymbol{\nu},\boldsymbol{k},\boldsymbol{J}) = 
  \sum_{\alpha\in\mathscr{H}} \nu_{\alpha}
  \left[  
    \log\left(\frac{p_{\alpha}\left(\boldsymbol{\nu}\right)}{\nu_{\alpha}}
    \right) + J_{\alpha} \right] + 
  \sum_{\mathscr{A}=\mathscr{V},\mathscr{T},\mathscr{L}} 
  \I k_{\mathscr{A}}\left(\sum_{\alpha\in\mathscr{A}}\nu_{\alpha}-1\right).
\end{eqnarray}
In the above expressions we used the compact notation
$\boldsymbol{k}=\left(k_{\mathscr{V}},k_{\mathscr{T}},k_{\mathscr{L}}\right)$
and $\boldsymbol{J}=\left(\ldots J_{V} \ldots;\ldots J_{T}
  \ldots;\ldots J_{\lambda} \ldots \right)$ with $V\in\mathscr{V}$,
$T\in\mathscr{T}$ and $\lambda\in\mathscr{L}$.  Evaluating the
integrals in equation (\ref{Z_pert}) by the saddle-point method, we
get the $N\to\infty$ asymptotic logarithm equality
\begin{eqnarray}
  Z_{N}(\boldsymbol{J}) \simeq 
  \mathrm{e}^{N\phi_{\mathrm{sp}}(\boldsymbol{J})},
  \qquad 
  \phi_{\mathrm{sp}}(\boldsymbol{J})=
  \phi\left(\boldsymbol{\nu}^{\mathrm{sp}}(\boldsymbol{J}),
    \boldsymbol{k}^{\mathrm{sp}}(\boldsymbol{J}),\boldsymbol{J}\right),
\end{eqnarray}
where $(\boldsymbol{\nu}^{\mathrm{sp}},\boldsymbol{k}^{\mathrm{sp}})$
is the saddle point (actually, the maximum point in the case of real
variables) of $\phi$, i.e. the solution of the following system of equations
\begin{eqnarray}
  \label{spe_pert_a}
  &\frac{\partial\phi}{\partial\nu_{\alpha}}=0, 
  \qquad \alpha\in\mathscr{H},
  \\
  \label{spe_pert_b}
  &\frac{\partial\phi}{\partial k_{\mathscr{A}}}=0, 
  \qquad \mathscr{A}=\mathscr{V},\mathscr{T},\mathscr{L}.
\end{eqnarray}
By differentiating equation (\ref{phi_pert}), the saddle-point
equations can be written as
\begin{eqnarray}
  \label{spe_pert2_a}
  &\log\left( \frac{ \nu_{\alpha}^{\mathrm{sp}}}
    {p_{\alpha}(\boldsymbol{\nu}^{\mathrm{sp}}) }\right) =
  J_{\alpha} + \sum_{\beta\in\mathscr{H}}
  \frac{p'_{\alpha\beta}\left(\boldsymbol{\nu}^{\mathrm{sp}}\right)}
  {p_{\beta}\left(\boldsymbol{\nu}^{\mathrm{sp}}\right)}
  {\nu_{\beta}^{\mathrm{sp}}}
  + \I k_{\mathscr{A}_{\alpha}}^{\mathrm{sp}} - 1, 
  \qquad \alpha\in\mathscr{H},
  \\
  \label{spe_pert2_b}
  &\sum_{\alpha\in\mathscr{A}}\nu_{\alpha}^{\mathrm{sp}} = 1,
  \qquad \mathscr{A}=\mathscr{V},\mathscr{T},\mathscr{L},
\end{eqnarray}
where we have defined
\begin{eqnarray}
  \label{p'}
  p'_{\alpha\beta}\left(\boldsymbol{\nu}\right) = 
  \frac{\partial p_{\alpha}\left(\boldsymbol{\nu}\right)}{\partial\nu_{\beta}}
  = p_{\alpha\beta}^{(2)} + 
  \sum_{\gamma\in\mathscr{H}}
  p_{\alpha\beta\gamma}^{(3)}\left(\nu_{\gamma}-p_{\gamma}^{(1)}\right)+
  \ldots
\end{eqnarray}
and
\begin{eqnarray}
  \mathscr{A}_{\alpha}=
  \left\{ \begin{array}{c}
      \mathscr{V},\qquad \alpha\in\mathscr{V},\\
      \mathscr{T},\qquad \alpha\in\mathscr{T},\\
      \mathscr{L},\qquad \alpha\in\mathscr{L}.\end{array}
  \right. 
\end{eqnarray}
Note that, due to the properties assumed for the perturbative
parameters, $p'_{\alpha\beta}\left(\boldsymbol{\nu}\right)$ is
symmetric under the exchange of the indices $\alpha,\beta$.  From
equation (\ref{spe_pert2_a}) we find
\begin{eqnarray}
  \nu_{\alpha}^{\mathrm{sp}} =
  p_{\alpha}\left(\boldsymbol{\nu}^{\mathrm{sp}}\right) 
  \mathrm{e}^{ 
    J_{\alpha} + 
    \sum_{\beta\in\mathscr{H}}
    \frac{p'_{\alpha\beta}\left(\boldsymbol{\nu}^{\mathrm{sp}}\right)}
    {p_{\beta}\left(\boldsymbol{\nu}^{\mathrm{sp}}\right)}
    {\nu_{\beta}^{\mathrm{sp}}} 
    + \I k_{\mathscr{A}_{\alpha}}^{\mathrm{sp}} - 1 },
  \qquad \alpha\in\mathscr{H},
  \label{nuspe_pert3}
\end{eqnarray}
which inserted into equation (\ref{spe_pert2_b}) provides
\begin{eqnarray}
  \label{kspe_pert}
  \mathrm{e}^{ \I k_{\mathscr{A}}^{\mathrm{sp}} - 1 } = 
  \frac{1}{\sum_{\alpha\in\mathscr{A}}
    p_{\alpha}\left(\boldsymbol{\nu}^{\mathrm{sp}}\right) 
    \mathrm{e}^{ 
      J_{\alpha} + 
      \sum_{\beta\in\mathscr{H}}
      \frac{p'_{\alpha\beta}\left(\boldsymbol{\nu}^{\mathrm{sp}}\right)}
      {p_{\beta}\left(\boldsymbol{\nu}^{\mathrm{sp}}\right)}
      {\nu_{\beta}^{\mathrm{sp}}}}}, 
  \qquad \mathscr{A}=\mathscr{V},\mathscr{T},\mathscr{L}.
\end{eqnarray}
By using equation (\ref{kspe_pert}) we conclude that the saddle-point
frequencies (\ref{nuspe_pert3}) are the solution of the system of
nonlinear equations
\begin{eqnarray}
  \label{nuspe_pert}
  \nu_{\alpha}^{\mathrm{sp}} =
  \frac{ p_{\alpha}\left(\boldsymbol{\nu}^{\mathrm{sp}}\right) 
    \mathrm{e}^{ 
      J_{\alpha} + 
      \sum_{\beta\in\mathscr{H}}
      \frac{p'_{\alpha\beta}\left(\boldsymbol{\nu}^{\mathrm{sp}}\right)}
      {p_{\beta}\left(\boldsymbol{\nu}^{\mathrm{sp}}\right)} 
      \nu_{\beta}^{\mathrm{sp}} }}
  {\sum_{\alpha'\in\mathscr{A}_{\alpha}}
    p_{\alpha'}\left(\boldsymbol{\nu}^{\mathrm{sp}}\right) 
    \mathrm{e}^{ 
      J_{\alpha'} + 
      \sum_{\beta\in\mathscr{H}}
      \frac{p'_{\alpha'\beta}\left(\boldsymbol{\nu}^{\mathrm{sp}}\right)}
      {p_{\beta}\left(\boldsymbol{\nu}^{\mathrm{sp}}\right)}
      \nu_{\beta}^{\mathrm{sp}} }},
  \qquad \alpha\in\mathscr{H}.
\end{eqnarray}
Note that both $\boldsymbol{\nu}^{\mathrm{sp}}$ and
$\boldsymbol{k}^{\mathrm{sp}}$ are functions of the source
$\boldsymbol{J}$.  It follows that the function
$\phi\left(\boldsymbol{\nu},\boldsymbol{k},\boldsymbol{J}\right)$
evaluated at the saddle point
$(\boldsymbol{\nu}^{\mathrm{sp}}(\boldsymbol{J}),
\boldsymbol{k}^{\mathrm{sp}}(\boldsymbol{J}))$ is
\begin{eqnarray}
  \label{phitilde_pert} 
  \phi_{\mathrm{sp}}\left(\boldsymbol{J}\right)  &=&  
  \sum_{\alpha\in\mathscr{H}} \nu_{\alpha}^{\mathrm{sp}}
  \left[
    \log\left( \frac{p_{\alpha}\left(\boldsymbol{\nu}^{\mathrm{sp}} \right)}
      {\nu_{\alpha}^{\mathrm{sp}}} \right) + J_{\alpha} \right]
  \nonumber \\ &=&
  \sum_{\mathscr{A}=\mathscr{V},\mathscr{T},\mathscr{L}} \!\!\!\!
  \log\left( \sum_{\alpha\in\mathscr{A}}
    p_{\alpha}\left(\boldsymbol{\nu}^{\mathrm{sp}}\right)
    \mathrm{e}^{J_{\alpha}+\sum_{\beta\in\mathscr{H}}
      \frac{p'_{\alpha\beta}\left(\boldsymbol{\nu}^{\mathrm{sp}}\right)}
      {p_{\beta}\left(\boldsymbol{\nu}^{\mathrm{sp}}\right)} 
      \nu_{\beta}^{\mathrm{sp}}} \right) 
  \nonumber \\ &&\qquad - 
  \sum_{\alpha\in\mathscr{H}}\sum_{\beta\in\mathscr{H}}
  \frac{p'_{\alpha\beta}\left(\boldsymbol{\nu}^{\mathrm{sp}}\right)}
  {p_{\beta}\left(\boldsymbol{\nu}^{\mathrm{sp}}\right)}
  \nu_{\alpha}^{\mathrm{sp}}\nu_{\beta}^{\mathrm{sp}}.
\end{eqnarray}

We are now ready to evaluate the asymptotic rescaled cumulants
associated with the probability density
(\ref{OMEGA_pert}--\ref{p_alpha}) by using the formula
\begin{eqnarray}
  \label{arc-der}
  \arc{\nu_{\alpha_{1}}\ldots\nu_{\alpha_{k}}} = 
  \left.
    \frac{ \partial^{k}\phi_{\mathrm{sp}}(\boldsymbol{J})}
    { \partial J_{\alpha_1} \ldots \partial J_{\alpha_k} }
  \right|_{\boldsymbol{J}=\boldsymbol{0}}.
\end{eqnarray}

\subsection{Derivatives of $\phi_{\mathrm{sp}}(\boldsymbol{J})$}
\label{phitilde.derivatives}
In this section we supply the derivatives of
$\phi_{\mathrm{sp}}(\boldsymbol{J})$ with respect to the source
$\boldsymbol{J}$ up to the third order.  Note that
$\phi_{\mathrm{sp}}$ depends on $\boldsymbol{J}$ explicitly and
through the saddle-point frequencies.

The first-order derivative of $\phi_{\mathrm{sp}}(\boldsymbol{J})$
with respect to the source component $J_{\alpha_1}$,
$\alpha_1\in\mathscr{H}$, is
\begin{eqnarray*}\fl
  \frac{\partial\phi_{\mathrm{sp}}(\boldsymbol{J})}{\partial J_{\alpha_1}} 
  =
  \sum_{\mathscr{A}=\mathscr{V},\mathscr{T},\mathscr{L}}
  \sum_{\alpha\in\mathscr{A}}\nu_{\alpha}^{\mathrm{sp}}
  \left(
    \frac{1}{p_{\alpha}(\boldsymbol{\nu}^{\mathrm{sp}})}
    \frac{\partial p_{\alpha}(\boldsymbol{\nu}^{\mathrm{sp}})}
    {\partial J_{\alpha_1}} + \delta_{\alpha\alpha_1} 
  \right) -
  \sum_{\alpha\in\mathscr{H}}\sum_{\beta\in\mathscr{H}}
  \frac{\partial\nu_{\alpha}^{\mathrm{sp}}}{\partial J_{\alpha_{1}}}
  \frac{p'_{\alpha\beta}\left(\boldsymbol{\nu}^{\mathrm{sp}}\right)}
  {p_{\beta}\left(\boldsymbol{\nu}^{\mathrm{sp}}\right)}
  \nu_{\beta}^{\mathrm{sp}}.
\end{eqnarray*}
Since
\begin{eqnarray}
  \label{p_J1}
  \frac{\partial p_{\alpha}(\boldsymbol{\nu}^{\mathrm{sp}})}
  {\partial J_{\alpha_1}} = 
  \sum_{\beta\in\mathscr{H}}
  p'_{\alpha\beta}\left(\boldsymbol{\nu}^{\mathrm{sp}}\right)
  \frac{\partial\nu_{\beta}^{\mathrm{sp}}}{\partial J_{\alpha_1}}
\end{eqnarray}
and $p'_{\alpha\beta}$ is symmetric, the previous expression reduces
to
\begin{eqnarray}
  \label{d1_phi}
  \frac{\partial\phi_{\mathrm{sp}}(\boldsymbol{J})}{\partial J_{\alpha_1}} =
  \nu_{\alpha_{1}}^{\mathrm{sp}}.
\end{eqnarray}

The second-order derivative of $\phi_{\mathrm{sp}}(\boldsymbol{J})$
with respect to the source components $J_{\alpha_1}$ and
$J_{\alpha_2}$, $\alpha_1,\alpha_2\in\mathscr{H}$, is
\begin{eqnarray*}\fl
  \frac{\partial^{2}\phi_{\mathrm{sp}}(\boldsymbol{J})}
  {\partial J_{\alpha_1} \partial J_{\alpha_2}} &=&
  \frac{\partial\nu_{\alpha_{1}}^{\mathrm{sp}}}
  {\partial J_{\alpha_2}}
  \\ \fl &=&
  \nu_{\alpha_{1}}^{\mathrm{sp}}\delta_{\alpha_{1}\alpha_{2}} - 
  \nu_{\alpha_{1}}^{\mathrm{sp}}\nu_{\alpha_{2}}^{\mathrm{sp}}
  \chi_{\alpha_{1}\alpha_2} + 
  \sum_{\alpha\in\mathscr{H}}\left(
    \nu_{\alpha_{1}}^{\mathrm{sp}}\delta_{\alpha_1\alpha} - 
    \nu_{\alpha_{1}}^{\mathrm{sp}}\nu_{\alpha}^{\mathrm{sp}}
    \chi_{\alpha_{1}\alpha} \right)
  \nonumber \\ \fl &&\qquad\times
  \left[
    \frac{1}{p_{\alpha}(\boldsymbol{\nu}^{\mathrm{sp}})}
    \frac{\partial p_{\alpha}(\boldsymbol{\nu}^{\mathrm{sp}})}
    {\partial J_{\alpha_2}} +
    \sum_{\beta\in\mathscr{H}}\frac{\partial}{\partial J_{\alpha_2}}
    \left(\frac{p'_{\alpha\beta}\left(\boldsymbol{\nu}^{\mathrm{sp}}\right)}
      {p_{\beta}\left(\boldsymbol{\nu}^{\mathrm{sp}}\right)}
      \nu_{\beta}^{\mathrm{sp}}\right)
  \right],
\end{eqnarray*}
where
\begin{eqnarray}
  \label{chi}
  \chi_{\alpha \beta} = \chi_{\beta\alpha} = 
  \left\{ 
    \begin{array}{ll}
      1,\qquad & \alpha\in\mathscr{A}_\beta,\\
      0,\qquad & \mbox{otherwise}.
    \end{array}
  \right. 
\end{eqnarray}
Recalling the definition (\ref{p'}) and using equation (\ref{p_J1}),
we find
\begin{eqnarray}
  \label{d2_phi}\fl
  \frac{\partial^{2}\phi_{\mathrm{sp}}(\boldsymbol{J})}
  {\partial J_{\alpha_1} \partial J_{\alpha_2}} =
  \nu_{\alpha_{1}}^{\mathrm{sp}}\delta_{\alpha_{1}\alpha_2} - 
  \nu_{\alpha_{1}}^{\mathrm{sp}}\nu_{\alpha_{2}}^{\mathrm{sp}}
  \chi_{\alpha_1\alpha_2}
  \nonumber \\  + 
  \sum_{\alpha\in\mathscr{H}} \sum_{\beta\in\mathscr{H}}
  \left(
    \nu_{\alpha_{1}}^{\mathrm{sp}}\delta_{\alpha_1\alpha} - 
    \nu_{\alpha_{1}}^{\mathrm{sp}}\nu_{\alpha}^{\mathrm{sp}}
    \chi_{\alpha_1\alpha} \right)
  M_{\alpha\beta}\left(\boldsymbol{\nu}^{\mathrm{sp}}\right)
  \frac{\partial\nu_{\beta}^{\mathrm{sp}}}
  {\partial J_{\alpha_2}},
\end{eqnarray}
where we have introduced the symmetric matrix
\begin{eqnarray}\fl
  \label{Md3_phi}
  M_{\alpha\beta}\left(\boldsymbol{\nu}^{\mathrm{sp}}\right) =
  \frac{p'_{\alpha\beta}(\boldsymbol{\nu}^{\mathrm{sp}})}
  {p_{\alpha}(\boldsymbol{\nu}^{\mathrm{sp}})} + 
  \frac{p'_{\alpha\beta}(\boldsymbol{\nu}^{\mathrm{sp}})}
  {p_{\beta}(\boldsymbol{\nu}^{\mathrm{sp}})}
  + \sum_{\gamma\in\mathscr{H}}
  \left(
    \frac{p''_{\alpha\beta\gamma}\left(\boldsymbol{\nu}^{\mathrm{sp}}\right)}
    {p_{\gamma}\left(\boldsymbol{\nu}^{\mathrm{sp}}\right)}
    - \frac{p'_{\alpha\gamma}\left(\boldsymbol{\nu}^{\mathrm{sp}}\right)
      p'_{\gamma\beta}\left(\boldsymbol{\nu}^{\mathrm{sp}}\right)}
    {p_{\gamma}\left(\boldsymbol{\nu}^{\mathrm{sp}}\right)^{2}} 
  \right) \nu_{\gamma}^{\mathrm{sp}}
\end{eqnarray}
and defined
\begin{eqnarray}
  \label{p''}
  p''_{\alpha\beta\gamma}(\boldsymbol{\nu}) = 
  \frac{\partial^{2}p_{\alpha}(\boldsymbol{\nu})}
  {\partial\nu_{\beta} \partial\nu_{\gamma}}
  =
  p_{\alpha\beta\gamma}^{(3)} + 
  \sum_{\delta\in\mathscr{H}} p_{\alpha\beta\gamma\delta}^{(4)}
  \left(\nu_{\delta}-p_{\delta}^{(1)}\right)+\ldots
\end{eqnarray}
Note that, due to the properties assumed for the perturbative
parameters, $p''_{\alpha\beta\gamma}(\boldsymbol{\nu})$ is symmetric
under the exchange of any two of its indices $\alpha,\beta,\gamma$.

The third-order derivative of $\phi_{\mathrm{sp}}(\boldsymbol{J})$
with respect to the source components $J_{\alpha_1}$, $J_{\alpha_2}$
and $J_{\alpha_3}$, $\alpha_1,\alpha_2,\alpha_3\in\mathscr{H}$, is
\begin{eqnarray}\fl
  \label{d3_phi}
  \frac{\partial^{3}\phi_{\mathrm{sp}}(\boldsymbol{J})}
  {\partial J_{\alpha_1} \partial J_{\alpha_2}\partial J_{\alpha_3}} &=&
  \frac{\partial^{2}\nu_{\alpha_{1}}^{\mathrm{sp}}}
  {\partial J_{\alpha_2}\partial J_{\alpha_3}} 
  \nonumber \\ \fl &=& 
  \frac{\partial\nu_{\alpha_{1}}^{\mathrm{sp}}}{\partial J_{\alpha_3}}
  \delta_{\alpha_1\alpha_2} - 
  \left(
    \frac{\partial\nu_{\alpha_{1}}^{\mathrm{sp}}}{\partial J_{\alpha_3}}
    \nu_{\alpha_2}^{\mathrm{sp}} +
    \nu_{\alpha_{1}}^{\mathrm{sp}}
    \frac{\partial\nu_{\alpha_2}^{\mathrm{sp}}}{\partial J_{\alpha_3}}
  \right)
  \chi_{\alpha_1\alpha_2}
  \nonumber \\ \fl &&+
  \sum_{\alpha\in\mathscr{H}}\sum_{\beta\in\mathscr{H}}
  \left(
    \frac{\partial\nu_{\alpha_{1}}^{\mathrm{sp}}}{\partial J_{\alpha_3}}
    \delta_{\alpha_1\alpha} - 
    \left(
      \frac{\partial\nu_{\alpha_{1}}^{\mathrm{sp}}}{\partial J_{\alpha_3}}
      \nu_{\alpha}^{\mathrm{sp}} +
      \nu_{\alpha_{1}}^{\mathrm{sp}}
      \frac{\partial\nu_{\alpha}^{\mathrm{sp}}}{\partial J_{\alpha_3}}
    \right)
    \chi_{\alpha_1\alpha}
  \right)
  M_{\alpha\beta}\left(\boldsymbol{\nu}^{\mathrm{sp}}\right)
  \frac{\partial\nu_{\beta}^{\mathrm{sp}}}
  {\partial J_{\alpha_2}}
  \nonumber \\ \fl &&+ 
  \sum_{\alpha\in\mathscr{H}}\sum_{\beta\in\mathscr{H}}
  \left( \nu_{\alpha_{1}}^{\mathrm{sp}}\delta_{\alpha_1\alpha} - 
    \nu_{\alpha_{1}}^{\mathrm{sp}}\nu_{\alpha}^{\mathrm{sp}}
    \chi_{\alpha_1\alpha} \right)
  \nonumber \\ \fl &&\qquad\times 
  \left(
    M_{\alpha\beta}\left(\boldsymbol{\nu}^{\mathrm{sp}}\right)
    \frac{\partial^{2}\nu_{\beta}^{\mathrm{sp}}}
    {\partial J_{\alpha_2}\partial J_{\alpha_3}} +
    \frac{\partial M_{\alpha\beta}\left(\boldsymbol{\nu}^{\mathrm{sp}}\right)}
    {\partial J_{\alpha_3}}
    \frac{\partial\nu_{\beta}^{\mathrm{sp}}}
    {\partial J_{\alpha_2}}
  \right),
\end{eqnarray} 
with
\begin{eqnarray}\fl
  \frac{\partial M_{\alpha\beta}\left(\boldsymbol{\nu}^{\mathrm{sp}}\right)}
  {\partial J_{\alpha_3}} &=&
  \sum_{\gamma\in\mathscr{H}}\left[
    \vphantom{\sum_{\beta\in\mathscr{H}}}
    \frac{p''_{\alpha\beta\gamma}\left(\boldsymbol{\nu}^{\mathrm{sp}}\right)}
    {p_{\alpha}\left(\boldsymbol{\nu}^{\mathrm{sp}}\right)} + 
    \frac{p''_{\alpha\beta\gamma}\left(\boldsymbol{\nu}^{\mathrm{sp}}\right)}
    {p_{\beta}\left(\boldsymbol{\nu}^{\mathrm{sp}}\right)} + 
    \frac{p''_{\alpha\beta\gamma}\left(\boldsymbol{\nu}^{\mathrm{sp}}\right)}
    {p_{\gamma}\left(\boldsymbol{\nu}^{\mathrm{sp}}\right)}
  \right.
  \nonumber \\ \fl &&-
  \frac{p'_{\alpha\beta}\left(\boldsymbol{\nu}^{\mathrm{sp}}\right)
    p'_{\alpha\gamma}\left(\boldsymbol{\nu}^{\mathrm{sp}}\right)}
  {p_{\alpha}\left(\boldsymbol{\nu}^{\mathrm{sp}}\right)^2} -
  \frac{p'_{\alpha\beta}\left(\boldsymbol{\nu}^{\mathrm{sp}}\right)
    p'_{\beta\gamma}\left(\boldsymbol{\nu}^{\mathrm{sp}}\right)}
  {p_{\beta}\left(\boldsymbol{\nu}^{\mathrm{sp}}\right)^2} - 
  \frac{p'_{\alpha\gamma}\left(\boldsymbol{\nu}^{\mathrm{sp}}\right)
    p'_{\gamma\beta}\left(\boldsymbol{\nu}^{\mathrm{sp}}\right)}
  {p_{\gamma}\left(\boldsymbol{\nu}^{\mathrm{sp}}\right)^2} 
  \nonumber \\ \fl &&+
  \sum_{\delta\in\mathscr{H}} \left(
    \frac{p'''_{\alpha\beta\gamma\delta}
      \left(\boldsymbol{\nu}^{\mathrm{sp}}\right)} 
    {p_{\delta}\left(\boldsymbol{\nu}^{\mathrm{sp}}\right)} -  
    \frac{p''_{\alpha\beta\delta}\left(\boldsymbol{\nu}^{\mathrm{sp}}\right)
      p'_{\delta\gamma}\left(\boldsymbol{\nu}^{\mathrm{sp}}\right)}
    {p_{\delta}\left(\boldsymbol{\nu}^{\mathrm{sp}}\right)^2} 
  \right.
  \nonumber \\ \fl &&\qquad-
  \frac{p''_{\alpha\delta\gamma}\left(\boldsymbol{\nu}^{\mathrm{sp}}\right)
    p'_{\delta\beta}\left(\boldsymbol{\nu}^{\mathrm{sp}}\right)}
  {p_{\delta}\left(\boldsymbol{\nu}^{\mathrm{sp}}\right)^2} -
  \frac{p''_{\delta\beta\gamma}\left(\boldsymbol{\nu}^{\mathrm{sp}}\right)
    p'_{\delta\alpha}\left(\boldsymbol{\nu}^{\mathrm{sp}}\right)}
  {p_{\delta}\left(\boldsymbol{\nu}^{\mathrm{sp}}\right)^2} 
  \nonumber \\ \fl &&\qquad+ \left.\left.
      2\frac{p'_{\alpha\delta}\left(\boldsymbol{\nu}^{\mathrm{sp}}\right)
        p'_{\beta\delta}\left(\boldsymbol{\nu}^{\mathrm{sp}}\right)      
        p'_{\gamma\delta}\left(\boldsymbol{\nu}^{\mathrm{sp}}\right)}
      {p_{\delta}\left(\boldsymbol{\nu}^{\mathrm{sp}}\right)^3}
    \right) \nu_{\delta}^{\mathrm{sp}} 
    \vphantom{\sum_{\delta\in\mathscr{H}}}\right]
  \frac{\partial\nu_{\gamma}^{\mathrm{sp}}}{\partial J_{\alpha_3}},
\end{eqnarray}
where we have defined
\begin{eqnarray}
  \label{p'''}
  p'''_{\alpha\beta\gamma\delta}(\boldsymbol{\nu}) = 
  \frac{\partial^{3}p_{\alpha}(\boldsymbol{\nu})}
  {\partial\nu_{\beta} \partial\nu_{\gamma} \partial\nu_{\delta}}
  =
  p_{\alpha\beta\gamma\delta}^{(4)} + 
  \sum_{\varepsilon\in\mathscr{H}}
  p_{\alpha\beta\gamma\delta\varepsilon}^{(5)}
  \left(\nu_{\varepsilon}-p_{\varepsilon}^{(1)}\right)+\ldots
\end{eqnarray}

\subsection{Equations for the perturbative parameters: first order}
\label{equation.p1}
The perturbative parameters $\boldsymbol{p}^{(1)}$ are determined by
the system of equations (\ref{SYSTEM_P1}) which, according to
(\ref{arc-der}) and (\ref{d1_phi}), reads
\begin{eqnarray}
  \label{EQ_spe}
  \left.\nu_{\alpha_1}^{\mathrm{sp}}\right|_{\boldsymbol{J}=\boldsymbol{0}} = 
  \Sigma_{\alpha_1}^{(1)},
  \qquad \alpha_{1}\in\mathscr{H}.
\end{eqnarray} 
Using equation (\ref{nuspe_pert}), we explicitly have
\begin{eqnarray}
  \label{EQ_p1}
  \frac{ p_{\alpha_{1}}(\boldsymbol{\Sigma}^{(1)}) 
    \mathrm{e}^{ 
      \sum_{\beta\in\mathscr{H}}
      \frac{p'_{\alpha_{1}\beta}(\boldsymbol{\Sigma}^{(1)})}
      {p_{\beta}(\boldsymbol{\Sigma}^{(1)})}
      {\Sigma_{\beta}^{(1)}} } }
  { \sum_{\alpha\in\mathscr{A}_{\alpha_{1}}}
    p_{\alpha}(\boldsymbol{\Sigma}^{(1)}) 
    \mathrm{e}^{ 
      \sum_{\beta\in\mathscr{H}}
      \frac{p'_{\alpha\beta}(\boldsymbol{\Sigma}^{(1)})}
      {p_{\beta}(\boldsymbol{\Sigma}^{(1)})}
      {\Sigma_{\beta}^{(1)}} } }
  = \Sigma_{\alpha_1}^{(1)},
  \qquad \alpha_{1}\in\mathscr{H}.
\end{eqnarray} 
A solution of this system of nonlinear equations is
\begin{eqnarray}
  \label{SOL_p1}
  p_{\alpha_{1}}^{(1)}=\Sigma_{\alpha_{1}}^{(1)}, 
  \qquad\qquad\alpha_{1}\in\mathscr{H}. 
\end{eqnarray}
To prove it, we start to observe that if
$\boldsymbol{p}^{(1)}=\boldsymbol{\Sigma}^{(1)}$ then for any
$\alpha,\beta\in\mathscr{H}$
\begin{eqnarray}
  \label{psigma}
  p_{\alpha}(\boldsymbol{\Sigma}^{(1)}) = p_{\alpha}^{(1)}, 
\end{eqnarray}
and
\begin{eqnarray}
  \label{p1sigma}
  p'_{\alpha\beta}(\boldsymbol{\Sigma}^{(1)}) = p_{\alpha\beta}^{(2)}.
\end{eqnarray}
The position $\boldsymbol{p}^{(1)}=\boldsymbol{\Sigma}^{(1)}$ has
another consequence which is pivotal also in determining the
perturbative parameteres of higher order.  In fact, due to the
constraints (\ref{Pa_eq1}), the analogous normalization conditions
valid for $\boldsymbol{\Sigma}^{(1)}$ and the sysmmetry properties
(\ref{PPAR_SIM}), we have
\begin{eqnarray}
  \label{PPAR_SUM}
  \sum_{\alpha_i\in\mathscr{A}} p_{\alpha_1\dots\alpha_k}^{(k)} = 0,
  \qquad k\geq 2,
  \qquad 1\leq i \leq k,
  \qquad \mathscr{A}=\mathscr{V},\mathscr{T},\mathscr{L}.
\end{eqnarray}
In the present case, the above sum rules allow to write
$\sum_{\beta\in\mathscr{H}} p^{(2)}_{\alpha\beta}=0$, which, together
with (\ref{psigma}) and (\ref{p1sigma}), permits reducing the
lhs of equation (\ref{EQ_p1}) to $\Sigma_{\alpha_{1}}^{(1)}$.

Equation (\ref{SOL_p1}) is not the unique solution of the system
(\ref{EQ_p1}).  However, besides being the natural solution for which
the perturbed probability density (\ref{OMEGA_pert}-\ref{p_alpha})
reduces, for $k_\mathrm{max}=1$, to the uncorrelated multinomial
(\ref{omega.multinomial}), it allows us to establish the sum rules
(\ref{PPAR_SUM}). We shall show that these, in turn, entail the
asymptotic rescaled cumulants
$\arc{\nu_{\alpha_1}\ldots\nu_{\alpha_k}}$ of arbitrary order $k$ to
depend only by the parameters $\boldsymbol{p}^{(j)}$ with $j\leq k$.
In this way, the parameters
$\boldsymbol{p}^{(1)},\boldsymbol{p}^{(2)},\boldsymbol{p}^{(3)},\dots$
we find do not depend on the order $k_\mathrm{max}$ at which we decide
to truncate the series (\ref{p_alpha}).

\subsection{Equations for the perturbative parameters: second order}
\label{equation.p2}
The perturbative parameters $\boldsymbol{p}^{(2)}$ are determined by
the system of equations (\ref{SYSTEM_P2}) which, according to
(\ref{arc-der}) and (\ref{d2_phi}), reads
\begin{eqnarray*}\fl
  \Sigma_{\alpha_1\alpha_2}^{(2,0)} + 
  \sum_{\alpha\in\mathscr{H}}\sum_{\beta\in\mathscr{H}}
  \Sigma_{\alpha_1\alpha}^{(2,0)}
  \left[
    \frac{p_{\alpha\beta}^{(2)}}{\Sigma_{\alpha}^{(1)}} + 
    \frac{p_{\alpha\beta}^{(2)}}{\Sigma_{\beta}^{(1)}} + 
    \sum_{\gamma\in\mathscr{H}}
    \left(
      \frac{p_{\alpha\beta\gamma}^{(3)}}{\Sigma_{\gamma}^{(1)}} -
      \frac{p_{\alpha\gamma}^{(2)}p_{\gamma\beta}^{(2)}}
      {{\Sigma_{\gamma}^{(1)}}^2}
    \right) \Sigma_{\gamma}^{(1)}
  \right]
  \Sigma_{\beta\alpha_2}^{(2)} 
  = \Sigma_{\alpha_1\alpha_2}^{(2)}. 
\end{eqnarray*} 
In writing this expression we have used the results (\ref{EQ_spe}),
(\ref{SOL_p1}), (\ref{psigma}) and (\ref{p1sigma}) of the previous
section as well as the fact that, since
$\boldsymbol{p}^{(1)}=\boldsymbol{\Sigma}^{(1)}$, for any
$\alpha,\beta,\gamma \in \mathscr{H}$ it is
\begin{eqnarray}
  \label{p2sigma}
  p''_{\alpha\beta\gamma}\left(\boldsymbol{\Sigma}^{(1)}\right) = 
  p_{\alpha\beta\gamma}^{(3)}.
\end{eqnarray}
We have also defined
\begin{eqnarray}
  \label{Sigma20}
  \Sigma_{\alpha\beta}^{(2,0)} =
  \Sigma_{\alpha}^{(1)}\delta_{\alpha\beta} - 
  \Sigma_{\alpha}^{(1)}\Sigma_{\beta}^{(1)}
  \chi_{\alpha\beta},
  \qquad \alpha,\beta\in\mathscr{H},
\end{eqnarray}
$\boldsymbol{\Sigma}^{(2,0)}$ in a compact notation, which is the
second cumulant of an uncorrelated multinomial probability density
with parameters $\boldsymbol{\Sigma}^{(1)}$.  According to equation
(\ref{PPAR_SUM}), we have $\sum_{\gamma\in\mathscr{H}}
p_{\alpha\beta\gamma}^{(3)}=0$ so that the above system of equations
becomes
\begin{eqnarray}
  \label{SOL_p2}
  \sum_{\alpha\in\mathscr{H}}\sum_{\beta\in\mathscr{H}}
  \Sigma_{\alpha_1\alpha}^{(2,0)}
  \left[
    \frac{p_{\alpha\beta}^{(2)}}{\Sigma_{\alpha}^{(1)}} + 
    \frac{p_{\alpha\beta}^{(2)}}{\Sigma_{\beta}^{(1)}} - 
    \sum_{\gamma\in\mathscr{H}}
    \frac{p_{\alpha\gamma}^{(2)}p_{\gamma\beta}^{(2)}}{\Sigma_{\gamma}^{(1)}} 
  \right]
  \Sigma_{\beta\alpha_2}^{(2)} 
  = \Sigma_{\alpha_1\alpha_2}^{(2)} - \Sigma_{\alpha_1\alpha_2}^{(2,0)}. 
\end{eqnarray}
For $\alpha_1,\alpha_2\in\mathscr{H}$, this is a nonlinear system of
$m^2$, $m=|\mathscr{H}|$, equations.  On the other hand,
$\boldsymbol{p}^{(2)}$, recalling that it satisfies the sum rules
(\ref{PPAR_SUM}), has $\hat{m}^2$,
$\hat{m}=(|\mathscr{V}|-1)+(|\mathscr{T}|-1)+(|\mathscr{L}|-1)$,
independent components.  Therefore, the system of equations
(\ref{SOL_p2}) is overdetermined and we have to lower its dimension to
find $\boldsymbol{p}^{(2)}$. Note that also the $m \times m$ matrices
$\boldsymbol{\Sigma}^{(2)}$ and $\boldsymbol{\Sigma}^{(2,0)}$ are
singular and their rank is $\hat{m}^2$.

Let us introduce the $\hat{m}$-dimensional index set
$\hat{\mathscr{H}} = \mathscr{H} \setminus \{V_*,T_*,\lambda_*\}$,
where $V_*$, $T_*$ and $\lambda_*$ are three arbitrarily chosen
elements of the sets $\mathscr{V}$, $\mathscr{T}$ and $\mathscr{L}$,
respectively.  Let $\hat{\boldsymbol{p}}^{(1)}$ be the vector with
components $\hat{p}^{(1)}_\alpha=p^{(1)}_\alpha$,
$\alpha\in\hat{\mathscr{H}}$, and $\hat{\boldsymbol{p}}^{(2)}$ the
matrix with components
$\hat{p}^{(2)}_{\alpha\beta}=p^{(2)}_{\alpha\beta}$,
$\alpha,\beta\in\hat{\mathscr{H}}$.  Now consider the $\hat{m}^2$
equations (\ref{SOL_p2}) with $\alpha_1,\alpha_2\in\hat{\mathscr{H}}$.
Observing that
$\sum_{\alpha\in\mathscr{H}}=\sum_{\alpha\in\hat{\mathscr{H}}}
+\sum_{\alpha\in\{V_*,T_*,\lambda_*\}}$ and using the sum rules
(\ref{PPAR_SUM}), we rewrite the first term in the lhs of equation
(\ref{SOL_p2}) as
\begin{eqnarray*}\fl
  \sum_{\alpha\in\mathscr{H}}\sum_{\beta\in\mathscr{H}}
  \Sigma_{{\alpha}_{1}\alpha}^{(2,0)}
  \frac{p_{\alpha\beta}^{(2)}}{\Sigma_{\alpha}^{(1)}}
  \Sigma_{\beta{\alpha}_{2}}^{(2)} &=&
  \sum_{{\alpha}\in\hat{\mathscr{H}}}\sum_{{\beta}\in\hat{\mathscr{H}}}
  \left(
    \frac{\Sigma_{{\alpha}_{1}{\alpha}}^{(2,0)}}{\Sigma_{{\alpha}}^{(1)}} - 
    \frac{\Sigma_{{\alpha}_{1}\alpha_{*}}^{(2,0)}}{\Sigma_{\alpha_{*}}^{(1)}}
  \right)
  \hat{p}_{{\alpha}{\beta}}^{(2)}
  \left(
    \Sigma_{{\beta}{\alpha}_{2}}^{(2)} - 
    \Sigma_{\beta_{*}\alpha_{2}}^{(2)}
  \right),
\end{eqnarray*}
where $\alpha_*$, $\beta_*$ are the components eliminated from the
sets $\mathscr{A}_\alpha$, $\mathscr{A}_\beta$, i.e.
\begin{eqnarray}
  \alpha_{*} =
  \left\{ \begin{array}{ll}
      V_*, \qquad &\alpha\in\mathscr{V}, \\
      T_*, \qquad &\alpha\in\mathscr{T}, \\
      \lambda_*, \qquad &\alpha\in\mathscr{L}.
    \end{array}\right. 
\end{eqnarray}
Similarly, the second term in the lhs of (\ref{SOL_p2}) becomes
\begin{eqnarray*}\fl
  \sum_{\alpha\in\mathscr{H}}\sum_{\beta\in\mathscr{H}}
  \Sigma^{(2,0)}_{{\alpha}_{1}\alpha} 
  \frac{p_{\alpha\beta}^{(2)}}{\Sigma_{\beta}^{(1)}}
  \Sigma_{\beta{\alpha}_{2}}^{(2)} &=&
  \sum_{{\alpha}\in\hat{\mathscr{H}}}\sum_{{\beta}\in\hat{\mathscr{H}}}
  \left(
    \Sigma_{{\alpha}_{1}{\alpha}}^{(2,0)} - 
    \Sigma_{{\alpha}_{1}\alpha_{*}}^{(2,0)}
  \right)
  \hat{p}_{{\alpha}{\beta}}^{(2)}
  \left(
    \frac{\Sigma_{{\beta}{\alpha}_{2}}^{(2)}}{\Sigma_{{\beta}}^{(1)}} - 
    \frac{\Sigma_{\beta_{*}\alpha_{2}}^{(2)}}{\Sigma_{\beta_{*}}^{(1)}}
  \right),
\end{eqnarray*}
whereas the third term gives
\begin{eqnarray*}\fl
  \sum_{\alpha\in\mathscr{H}}\sum_{\beta\in\mathscr{H}}
  \sum_{\gamma\in\mathscr{H}}
  \Sigma^{(2,0)}_{{\alpha}_{1}\alpha}
  \frac{p_{\alpha\gamma}^{(2)}p_{\gamma\beta}^{(2)}}{\Sigma_{\gamma}^{(1)}}
  \Sigma_{\beta{\alpha}_{2}}^{(2)} 
  \nonumber \\ \fl \qquad=
  \sum_{{\alpha}\in\hat{\mathscr{H}}}\sum_{{\beta}\in\hat{\mathscr{H}}}
  \sum_{{\gamma}\in\hat{\mathscr{H}}}\sum_{{\delta}\in\hat{\mathscr{H}}} 
  \left(
    \Sigma_{\alpha_{1}\alpha}^{(2,0)} - 
    \Sigma_{\alpha_{1}\alpha_{*}}^{(2,0)}
  \right)
  \hat{p}_{{\alpha}{\gamma}}^{(2)}
  \left(
    \frac{\delta_{\gamma\delta}}{\Sigma_{{\gamma}}^{(1)}} + 
    \frac{\chi_{\gamma\delta}}{\Sigma_{\gamma_{*}}^{(1)}}
  \right)
  \hat{p}_{{\delta}{\beta}}^{(2)}
  \left(
    \Sigma_{{\beta}{\alpha}_{2}}^{(2)} - 
    \Sigma_{\beta_{*}\alpha_{2}}^{(2)}
  \right).
\end{eqnarray*}
We conclude that $\hat{\boldsymbol{p}}^{(2)}$ is the solution of the
nonlinear matrix equation
\begin{eqnarray}
  \label{SOL_p2M}
  \overline{\boldsymbol{\Sigma}}^{(2,0)}
  \hat{\boldsymbol{p}}^{(2)}
  \widetilde{\boldsymbol{\Sigma}}^{(2)} +
  \widetilde{\boldsymbol{\Sigma}}^{(2,0)}
  \hat{\boldsymbol{p}}^{(2)}
  \overline{\boldsymbol{\Sigma}}^{(2)} -
  \widetilde{\boldsymbol{\Sigma}}^{(2,0)} 
  \hat{\boldsymbol{p}}^{(2)}
  \boldsymbol{\Gamma}
  \hat{\boldsymbol{p}}^{(2)}
  \widetilde{\boldsymbol{\Sigma}}^{(2)} =
  \hat{\boldsymbol{\Sigma}}^{(2)} - \hat{\boldsymbol{\Sigma}}^{(2,0)}.
\end{eqnarray}
where $\hat{\boldsymbol{\Sigma}}^{(2)}$ and
$\hat{\boldsymbol{\Sigma}}^{(2,0)}$ are the reduced versions of the
matrices $\boldsymbol{\Sigma}^{(2)}$ and $\boldsymbol{\Sigma}^{(2,0)}$
and we have introduced the matrices
$\widetilde{\boldsymbol{\Sigma}}^{(2)}$,
$\overline{\boldsymbol{\Sigma}}^{(2)}$,
$\widetilde{\boldsymbol{\Sigma}}^{(2,0)}$,
$\overline{\boldsymbol{\Sigma}}^{(2,0)}$ and $\boldsymbol{\Gamma}$
with components $\alpha,\beta\in\hat{\mathscr{H}}$ given by
\begin{eqnarray}
  \label{Sigma2.tilde}
  \widetilde{\Sigma}_{{\alpha}{\beta}}^{(2)} =
  \Sigma_{{\alpha}{\beta}}^{(2)} - \Sigma_{\alpha_{*}{\beta}}^{(2)},
  \\
  \label{Sigma2.bar}
  \overline{\Sigma}_{{\alpha}{\beta}}^{(2)} = 
  \frac{\Sigma_{{\alpha}{\beta}}^{(2)}}{\Sigma_{{\alpha}}^{(1)}} - 
  \frac{\Sigma_{\alpha_{*}{\beta}}^{(2)}}{\Sigma_{\alpha_{*}}^{(1)}},
  \\
  \label{Sigma20.tilde}
  \widetilde{\Sigma}_{{\alpha}{\beta}}^{(2,0)} =
  \Sigma_{{\alpha}{\beta}}^{(2,0)} - \Sigma_{{\alpha}\beta_{*}}^{(2,0)},
  \\
  \label{Sigma20.bar}
  \overline{\Sigma}_{{\alpha}{\beta}}^{(2,0)} = 
  \frac{\Sigma_{{\alpha}{\beta}}^{(2,0)}}{\Sigma_{{\beta}}^{(1)}} - 
  \frac{\Sigma_{{\alpha}\beta_{*}}^{(2,0)}}{\Sigma_{\beta_{*}}^{(1)}},
  \\
  \label{Gamma}
  \Gamma_{\alpha\beta} =
  \frac{\delta_{\alpha\beta}}{\Sigma_{{\alpha}}^{(1)}} + 
  \frac{\chi_{\alpha\beta}}{\Sigma_{\alpha_{*}}^{(1)}} .
\end{eqnarray}
Since $\widetilde{\boldsymbol{\Sigma}}^{(2)}$ and
$\widetilde{\boldsymbol{\Sigma}}^{(2,0)}$ are nonsingular, equation
(\ref{SOL_p2M}) can be rewritten as
\begin{eqnarray}\fl
  \label{SOL_p2M_fin}
  \left.\widetilde{\boldsymbol{\Sigma}}^{(2,0)}\right.^{-1}
  \overline{\boldsymbol{\Sigma}}^{(2,0)}
  \hat{\boldsymbol{p}}^{(2)} +
  \hat{\boldsymbol{p}}^{(2)}
  \overline{\boldsymbol{\Sigma}}^{(2)} 
  \left.\widetilde{\boldsymbol{\Sigma}}^{(2)}\right.^{-1} -
  \hat{\boldsymbol{p}}^{(2)} 
  \boldsymbol{\Gamma}
  \hat{\boldsymbol{p}}^{(2)} =
  \left.\widetilde{\boldsymbol{\Sigma}}^{(2,0)}\right.^{-1}
  \left(
    \hat{\boldsymbol{\Sigma}}^{(2)} - \hat{\boldsymbol{\Sigma}}^{(2,0)}
  \right)
  \left.\widetilde{\boldsymbol{\Sigma}}^{(2)}\right.^{-1}.
\end{eqnarray}
This is a nonsymmetric algebraic Riccati equation (NARE), which can be
solved numerically by matrix factorization (Sch\"ur method)
\cite{Laub} or iterative methods \cite{BIMP}. For details we refer to
\ref{calculation.p2}.  Once $\hat{\boldsymbol{p}}^{(2)}$ has been
found, the complete set of second-order perturbative parameters
$\boldsymbol{p}^{(2)}$ is obtained using the sum rules
(\ref{PPAR_SUM}) for $k=2$.

\subsection{Equations for the perturbative parameters: third order}
\label{equation.p3}
The perturbative parameters $\boldsymbol{p}^{(3)}$ are determined by
the system of equations (\ref{SYSTEM_P3}) which, according to
(\ref{arc-der}) and (\ref{d3_phi}), reads
\begin{eqnarray} \fl
  \label{SOL_p3_0}
  \Sigma_{\alpha_{1}\alpha_{2}\alpha_{3}}^{(3,0)} +
  \sum_{\alpha\in\mathscr{H}} \sum_{\beta\in\mathscr{H}} 
  \Sigma_{\alpha_{1}\alpha\alpha_{3}}^{(3,0)}
  \left(
    \frac{p_{\alpha\beta}^{(2)}}{\Sigma_{\alpha}^{(1)}} + 
    \frac{p_{\alpha\beta}^{(2)}}{\Sigma_{\beta}^{(1)}} + 
    \sum_{\gamma\in\mathscr{H}}
    \left(      
      \frac{p_{\alpha\beta\gamma}^{(3)}}{\Sigma_{\gamma}^{(1)}} -      
      \frac{p_{\alpha\gamma}^{(2)}p_{\gamma\beta}^{(2)}}
      {{\Sigma_{\gamma}^{(1)}}^2}
    \right) 
    \Sigma_{\gamma}^{(1)}
  \right)
  \Sigma_{\beta\alpha_{2}}^{(2)}
  \nonumber \\ +
  \sum_{\alpha\in\mathscr{H}} \sum_{\beta\in\mathscr{H}}
  \Sigma_{\alpha_{1}\alpha}^{(2,0)} 
  \left(
    \frac{p_{\alpha\beta}^{(2)}}{\Sigma_{\alpha}^{(1)}} + 
    \frac{p_{\alpha\beta}^{(2)}}{\Sigma_{\beta}^{(1)}} + 
    \sum_{\gamma\in\mathscr{H}}
    \left(      
      \frac{p_{\alpha\beta\gamma}^{(3)}}{\Sigma_{\gamma}^{(1)}} -      
      \frac{p_{\alpha\gamma}^{(2)}p_{\gamma\beta}^{(2)}}
      {{\Sigma_{\gamma}^{(1)}}^2}
    \right) 
    \Sigma_{\gamma}^{(1)}
  \right)
  \Sigma_{\beta\alpha_{2}\alpha_{3}}^{(3)}
  \nonumber \\ + 
  \sum_{\alpha\in\mathscr{H}} \sum_{\beta\in\mathscr{H}}
  \Sigma_{\alpha_{1}\alpha}^{(2,0)}
  \sum_{\gamma\in\mathscr{H}}
  \left[
    \frac{p_{\alpha\beta\gamma}^{(3)}}{\Sigma_{\alpha}^{(1)}} +
    \frac{p_{\alpha\beta\gamma}^{(3)}}{\Sigma_{\beta}^{(1)}} + 
    \frac{p_{\alpha\beta\gamma}^{(3)}}{\Sigma_{\gamma}^{(1)}} -
    \frac{p_{\beta\alpha}^{(2)}p_{\alpha\gamma}^{(2)}}
    {{\Sigma_{\alpha}^{(1)}}^2} -
    \frac{p_{\alpha\beta}^{(2)}p_{\beta\gamma}^{(2)}}
    {{\Sigma_{\beta}^{(1)}}^2} - 
    \frac{p_{\alpha\gamma}^{(2)}p_{\gamma\beta}^{(2)}}
    {{\Sigma_{\gamma}^{(1)}}^2}
  \right.
  \nonumber \\ +
  \sum_{\delta\in\mathscr{H}}
  \left(
    \frac{p_{\alpha\beta\gamma\delta}^{(4)}}{\Sigma_{\delta}^{(1)}} -
    \frac{p_{\alpha\beta\delta}^{(3)}p_{\delta\gamma}^{(2)}}
    {{\Sigma_{\delta}^{(1)}}^{2}} -
    \frac{p_{\alpha\delta\gamma}^{(3)}p_{\delta\beta}^{(2)}}
    {{\Sigma_{\delta}^{(1)}}^{2}} -
    \frac{p_{\delta\beta\gamma}^{(3)}p_{\delta\alpha}^{(2)}}
    {{\Sigma_{\delta}^{(1)}}^{2}}
  \right.
  \nonumber \\ + 
  \left.\left.
      2 \frac{p_{\alpha\delta}^{(2)} p_{\beta\delta}^{(2)}
        p_{\gamma\delta}^{(2)}}
      {{\Sigma_{\delta}^{(1)}}^3}
    \right)
    \Sigma_{\delta}^{(1)}
  \right]
  \Sigma_{\beta\alpha_{2}}^{(2)}
  \Sigma_{\gamma\alpha_{3}}^{(2)} = 
  \Sigma_{\alpha_{1}\alpha_{2}\alpha_{3}}^{(3)}.
\end{eqnarray}
In writing this expression we have used the results (\ref{EQ_spe}),
(\ref{SOL_p1}), (\ref{psigma}), (\ref{p1sigma}) and (\ref{p2sigma}) of
the previous sections as well as the fact that, since
$\boldsymbol{p}^{(1)}=\boldsymbol{\Sigma}^{(1)}$, for any
$\alpha,\beta,\gamma,\delta \in \mathscr{H}$ it is
\begin{eqnarray}
  \label{p3sigma}
  p'''_{\alpha\beta\gamma\delta}\left(\boldsymbol{\Sigma}^{(1)}\right) = 
  p_{\alpha\beta\gamma\delta}^{(4)}.
\end{eqnarray}
We have also defined
\begin{eqnarray}
  \label{Sigma30}
  \Sigma^{(3,0)}_{\alpha\beta\gamma} = 
  \Sigma^{(2)}_{\alpha\gamma} \delta_{\alpha\beta} - 
  \left(
    \Sigma^{(2)}_{\alpha\gamma} \Sigma^{(1)}_{\beta} +
    \Sigma^{(1)}_{\alpha}\Sigma^{(2)}_{\beta\gamma}
  \right) \chi_{\alpha\beta},
\end{eqnarray}
$\boldsymbol{\Sigma}^{(3,0)}$ in a compact notation, which is,
formally, the third cumulant of an uncorrelated multinomial
probability density with the first two cumulants equal to
$\boldsymbol{\Sigma}^{(1)}$ and $\boldsymbol{\Sigma}^{(2)}$,
respectively.  According to equation (\ref{PPAR_SUM}), in the first
two lines of (\ref{SOL_p3_0}) we have $\sum_{\gamma\in\mathscr{H}}
p_{\alpha\beta\gamma}^{(3)}=0$ and in the fourth line
$\sum_{\delta\in\mathscr{H}} p_{\alpha\beta\gamma\delta}^{(4)}=0$, so
that equation (\ref{SOL_p3_0}) becomes
\begin{eqnarray} \fl
  \label{SOL_p3}
  \sum_{\alpha\in\mathscr{H}}
  \sum_{\beta\in\mathscr{H}}
  \sum_{\gamma\in\mathscr{H}}
  \left(
    \Lambda_{{\alpha}_{1}\alpha}^{(2,0)}
    \Sigma_{\beta\alpha_2}^{(2)}
    \Sigma_{\gamma\alpha_3}^{(2)} +
    \Sigma_{\alpha_1\alpha}^{(2,0)}
    \Lambda_{\beta\alpha_2}^{(2)}
    \Sigma_{\gamma\alpha_3}^{(2)} +
    \Sigma_{\alpha_1\alpha}^{(2,0)}
    \Sigma_{\beta\alpha_2}^{(2)}
    \Lambda_{\gamma\alpha_3}^{(2)} 
  \right) p_{\alpha\beta\gamma}^{(3)} 
  \nonumber \\ =
  \Delta_{\alpha_1\alpha_2\alpha_3}, 
\end{eqnarray}
where we have introduced the matrices $\boldsymbol{\Lambda}^{(2)}$ and
$\boldsymbol{\Lambda}^{(2,0)}$ with components
$\alpha,\beta\in\mathscr{H}$
\begin{eqnarray}
  \label{Lambda2}
  \Lambda_{\alpha\beta}^{(2)} =
  \frac{1}{\Sigma_{\alpha}^{(1)}}
  \left(
    \Sigma_{\alpha\beta}^{(2)} - 
    \sum_{\gamma\in\mathscr{H}}
    p^{(2)}_{\alpha\gamma}\Sigma_{\gamma\beta}^{(2)}
  \right),
\end{eqnarray}
\begin{eqnarray}
  \label{Lambda20}
  \Lambda_{\alpha\beta}^{(2,0)} =
  \left(
    \Sigma_{\alpha\beta}^{(2,0)} - 
    \sum_{\gamma\in\mathscr{H}}
    \Sigma_{\alpha\gamma}^{(2,0)}p^{(2)}_{\gamma\beta}
  \right)
  \frac{1}{\Sigma_{\beta}^{(1)}},
\end{eqnarray} 
and the tensor $\boldsymbol{\Delta}$ with components
$\alpha_1,\alpha_2,\alpha_3\in\mathscr{H}$
\begin{eqnarray}\fl
  \label{Delta3}
  \Delta_{\alpha_1\alpha_2\alpha_3} =
  \Sigma_{\alpha_1\alpha_2\alpha_3}^{(3)} -
  \Sigma_{\alpha_1\alpha_2\alpha_3}^{(3,0)}
  \nonumber \\ -
  \sum_{\alpha\in\mathscr{H}}\sum_{\beta\in\mathscr{H}} 
  \Sigma_{\alpha_1\alpha\alpha_3}^{(3,0)}
  \left(
    \frac{p_{\alpha\beta}^{(2)}}{\Sigma_{\alpha}^{(1)}} + 
    \frac{p_{\alpha\beta}^{(2)}}{\Sigma_{\beta}^{(1)}} - 
    \sum_{\gamma\in\mathscr{H}}
    \frac{p_{\alpha\gamma}^{(2)}p_{\gamma\beta}^{(2)}}
    {\Sigma_{\gamma}^{(1)}}
  \right)
  \Sigma_{\beta\alpha_2}^{(2)}
  \nonumber \\ -
  \sum_{\alpha\in\mathscr{H}}\sum_{\beta\in\mathscr{H}} 
  \Sigma_{\alpha_1\alpha}^{(2,0)}
  \left(
    \frac{p_{\alpha\beta}^{(2)}}{\Sigma_{\alpha}^{(1)}} + 
    \frac{p_{\alpha\beta}^{(2)}}{\Sigma_{\beta}^{(1)}} - 
    \sum_{\gamma\in\mathscr{H}}
    \frac{p_{\alpha\gamma}^{(2)}p_{\gamma\beta}^{(2)}}
    {\Sigma_{\gamma}^{(1)}}
  \right)
  \Sigma_{\beta\alpha_2\alpha_3}^{(3)}
  \nonumber \\ +
  \sum_{\alpha\in\mathscr{H}}
  \sum_{\beta\in\mathscr{H}}
  \sum_{\gamma\in\mathscr{H}}
  \Sigma_{\alpha_{1}\alpha}^{(2,0)}
  \left(
    \frac{p_{\alpha\beta}^{(2)}p_{\alpha\gamma}^{(2)}}
    {{\Sigma_{\alpha}^{(1)}}^2} +
    \frac{p_{\alpha\beta}^{(2)}p_{\beta\gamma}^{(2)}}
    {{\Sigma_{\beta}^{(1)}}^2} +
    \frac{p_{\alpha\gamma}^{(2)}p_{\gamma\beta}^{(2)}}
    {{\Sigma_{\gamma}^{(1)}}^2} 
  \right.
  \nonumber \\ -
  \left.
    2\sum_{\delta\in\mathscr{H}}
    \frac{p_{\alpha\delta}^{(2)}
      p_{\beta\delta}^{(2)}
      p_{\gamma\delta}^{(2)}}
    {{\Sigma_{\delta}^{(1)}}^{2}}
  \right)
  \Sigma_{\beta\alpha_{2}}^{(2)}
  \Sigma_{\gamma\alpha_{3}}^{(2)}.
\end{eqnarray}
Note that, once $\boldsymbol{p}^{(1)}$ and $\boldsymbol{p}^{(2)}$ have
been determined, this tensor can be considered known.

Equation (\ref{SOL_p3}) is, for
$\alpha_1,\alpha_2,\alpha_3\in\mathscr{H}$, an overdetermined linear
system of $m^3$ equations in the unknown $\boldsymbol{p}^{(3)}$ which,
according to the sum rules (\ref{PPAR_SUM}), has $\hat{m}^3$
independent components.  To find $\boldsymbol{p}^{(3)}$ we proceed as
in the previous section.  Let $\hat{\boldsymbol{p}}^{(3)}$ be the
tensor with components
$\hat{p}^{(3)}_{\alpha\beta\gamma}=p^{(3)}_{\alpha\beta\gamma}$,
$\alpha,\beta,\gamma\in\hat{\mathscr{H}}$, and consider the
$\hat{m}^3$ equations (\ref{SOL_p3}) with
$\alpha_1,\alpha_2,\alpha_3\in\hat{\mathscr{H}}$.  Observing that
$\sum_{\alpha\in\mathscr{H}}=\sum_{\alpha\in\hat{\mathscr{H}}}
+\sum_{\alpha\in\{V_*,T_*,\lambda_*\}}$ and using the sum rules
(\ref{PPAR_SUM}), we rewrite equation (\ref{SOL_p3}) as
\begin{eqnarray}\fl
  \label{SOL_p3_rid}
  \sum_{\alpha\in\hat{\mathscr{H}}}
  \sum_{\beta\in\hat{\mathscr{H}}}
  \sum_{\gamma\in\hat{\mathscr{H}}}
  \left(
    \widetilde{\Lambda}_{{\alpha}_{1}\alpha}^{(2,0)}
    \widetilde{\Sigma}_{\beta\alpha_2}^{(2)}
    \widetilde{\Sigma}_{\gamma\alpha_3}^{(2)} +
    \widetilde{\Sigma}_{\alpha_1\alpha}^{(2,0)}
    \widetilde{\Lambda}_{\beta\alpha_2}^{(2)}
    \widetilde{\Sigma}_{\gamma\alpha_3}^{(2)} +
    \widetilde{\Sigma}_{\alpha_1\alpha}^{(2,0)}
    \widetilde{\Sigma}_{\beta\alpha_2}^{(2)}
    \widetilde{\Lambda}_{\gamma\alpha_3}^{(2)} 
  \right)     \hat{p}_{\alpha\beta\gamma}^{(3)}
  \nonumber \\ =
  \Delta_{\alpha_1\alpha_2\alpha_3}, 
\end{eqnarray}
where we have introduced the matrices
$\widetilde{\boldsymbol{\Lambda}}^{(2)}$ and
$\widetilde{\boldsymbol{\Lambda}}^{(2,0)}$ with components
$\alpha,\beta\in\hat{\mathscr{H}}$
\begin{eqnarray}
  \label{Lambda2.tilde}
  \widetilde{\Lambda}_{\alpha\beta}^{(2)} =
  \Lambda_{\alpha\beta}^{(2)} - \Lambda_{\alpha_*\beta}^{(2)},
  \\
  \label{Lambda20.tilde}
  \widetilde{\Lambda}_{\alpha\beta}^{(2,0)} =
  \Lambda_{\alpha\beta}^{(2,0)} - \Lambda_{\alpha\beta_*}^{(2,0)},
\end{eqnarray}
whereas the matrices $\widetilde{\boldsymbol{\Sigma}}^{(2)}$ and
$\widetilde{\boldsymbol{\Sigma}}^{(2,0)}$ are defined by
(\ref{Sigma2.tilde}) and (\ref{Sigma20.tilde}).

The system of equations (\ref{SOL_p3_rid}) with
$\alpha_1,\alpha_2,\alpha_3\in\hat{\mathscr{H}}$ is a linear tensorial
equation in the unknown $\hat{\boldsymbol{p}}^{(3)}$.  It can be
solved by vectorization, i.e. by introducing a bijective map between
the set $\hat{\mathscr{H}}^3$ and the integers
$\{1,2,\dots,\hat{m}^3\}$.  Let $n(\alpha): \hat{\mathscr{H}}\mapsto
\{1,2,\dots,\hat{m} \}$ be some ordering of the elements
$\alpha\in\hat{\mathscr{H}}$ and $n^{-1}:
\{1,2,\dots,\hat{m}\}\mapsto\hat{\mathscr{H}}$ its inverse.  We define
the integer map $i(\alpha,\beta,\gamma): \hat{\mathscr{H}}^3 \mapsto
\{1,2,\dots,\hat{m}^3\}$ by
\begin{eqnarray}
  i(\alpha,\beta,\gamma) = 
  (n(\alpha)-1)\hat{m}^2 + (n(\beta)-1)\hat{m} + n(\gamma),
\end{eqnarray}
so that the inverse triplet $(\alpha(i),\beta(i),\gamma(i))$ is given
by
\begin{eqnarray}
  \alpha(i) &=& n^{-1}\left(
    \left\lfloor \frac{i-0^+}{\hat{m}^2} \right\rfloor + 1
  \right),
  \\
  \beta(i) &=& n^{-1}\left(
    \left\lfloor 
      \frac{i-(n(\alpha(i))-1)\hat{m}^2-0^+}{\hat{m}} 
    \right\rfloor + 1
  \right),
  \\
  \gamma(i) &=& n^{-1}\left(
    i - (n(\alpha(i))-1)\hat{m}^2 - (n(\beta(i))-1)\hat{m}
  \right),
\end{eqnarray}
where $0^+$ is a positive infinitesimal.  Instead of equation
(\ref{SOL_p3_rid}) with
$\alpha_1,\alpha_2,\alpha_3\in\hat{\mathscr{H}}$, we thus obtain the
equivalent system
\begin{eqnarray}
  \label{SOL_p3_rid_vec}
  \sum_{j=1}^{\hat{m}^3}
  Q_{ij} ~ \hat{p}^{(3)}_{j} =
  \Delta_{i},
  \qquad i=1,2,\dots,\hat{m}^3,
\end{eqnarray}
where we have defined
\begin{eqnarray}\fl
  Q_{ij} =
  \widetilde{\Lambda}^{(2,0)}_{\alpha(i)\alpha(j)}
  \widetilde{\Sigma}^{(2)}_{\beta(j)\beta(i)}
  \widetilde{\Sigma}^{(2)}_{\gamma(j)\gamma(i)} + 
  \widetilde{\Sigma}^{(2,0)}_{\alpha(i)\alpha(j)}
  \widetilde{\Lambda}^{(2)}_{\beta(j)\beta(i)}
  \widetilde{\Sigma}^{(2)}_{\gamma(j)\gamma(i)} + 
  \widetilde{\Sigma}^{(2,0)}_{\alpha(i)\alpha(j)}
  \widetilde{\Sigma}^{(2)}_{\beta(j)\beta(i)}
  \widetilde{\Lambda}^{(2)}_{\gamma(j)\gamma(i)},
\end{eqnarray} 
as well as
$\hat{p}^{(3)}_{j}=\hat{p}^{(3)}_{\alpha(j)\beta(j)\gamma(j)}$ and
$\Delta_{i}=\Delta_{\alpha(i)\beta(i)\gamma(i)}$.  Equation
(\ref{SOL_p3_rid_vec}) is a linear matrix equation which can be solved
by standard methods, e.g. LU-factorization \cite{NR}.

Once $\hat{\boldsymbol{p}}^{(3)}$ has been found, the complete set of
third-order perturbative parameters $\boldsymbol{p}^{(3)}$ is obtained
using the sum rules (\ref{PPAR_SUM}) for $k=3$.

\subsection{Some considerations on higher orders}
\label{considerations.on.higher.orders}
In the previous sections we have evaluated the perturbative
parameters $\boldsymbol{p}^{(k)}$ for the first three perturbative
orders $k=1,2,3$.  In this section we will show that for any $k\geq 3$
the equations which determine $\boldsymbol{p}^{(k)}$ are (i)
linear tensorial equations (ii) depending only on the
parameters $\boldsymbol{p}^{(j)}$ with $j\leq k$.
This ensures that the perturbative parameters are, in principle,
solvable at all orders and that their value is independent of the
integer $k_\mathrm{max}-1$ at which we decide to truncate the series
(\ref{p_alpha}).  The evaluation of $\boldsymbol{p}^{(4)}$ is outlined
in \ref{equation.p4}.

To demonstrate the properties (i) and (ii), first of
all let us point up why they hold in the analyzed case $k=3$.  The
tensor $p^{(3)}_{\alpha_1\alpha_2\alpha_3}$ is determined by the
system of equations (\ref{SYSTEM_P3}) the structure of which is
established, see equation (\ref{arc-der}), by the third-order
derivatives $\partial^3 \phi_{\mathrm{sp}}(\boldsymbol{J})/
\partial J_{\alpha_1}\partial J_{\alpha_2}\partial J_{\alpha_3}$
evaluated at $\boldsymbol{J}=\boldsymbol{0}$.  According to equation
(\ref{d3_phi}), these derivatives contain rational combinations of the
functions $\boldsymbol{p}(\boldsymbol{\nu})$ and their derivatives
$\boldsymbol{p}'(\boldsymbol{\nu})$,
$\boldsymbol{p}''(\boldsymbol{\nu})$ and
$\boldsymbol{p}'''(\boldsymbol{\nu})$, a compact notation to indicate
respectively the components (\ref{p_alpha}), (\ref{p'}), (\ref{p''})
and (\ref{p'''}), evaluated at
$\boldsymbol{\nu}=\boldsymbol{\nu}^\mathrm{sp}$.  When we set
$\boldsymbol{J}=\boldsymbol{0}$, since
$\left.\boldsymbol{\nu}^\mathrm{sp}\right|_{\boldsymbol{J}=\boldsymbol{0}}
= \boldsymbol{\Sigma}^{(1)}$ and
$\boldsymbol{p}^{(1)}=\boldsymbol{\Sigma}^{(1)}$, we have
\begin{eqnarray}
  \left.
    \boldsymbol{p}(\boldsymbol{\nu}^\mathrm{sp})
  \right|_{\boldsymbol{J}=\boldsymbol{0}} = 
  \boldsymbol{p}^{(1)},
  \\
  \left.
    \boldsymbol{p}'(\boldsymbol{\nu}^\mathrm{sp})
  \right|_{\boldsymbol{J}=\boldsymbol{0}} = 
  \boldsymbol{p}^{(2)},
  \\
  \left.
    \boldsymbol{p}''(\boldsymbol{\nu}^\mathrm{sp})
  \right|_{\boldsymbol{J}=\boldsymbol{0}} = 
  \boldsymbol{p}^{(3)},
  \\
  \left.
    \boldsymbol{p}'''(\boldsymbol{\nu}^\mathrm{sp})
  \right|_{\boldsymbol{J}=\boldsymbol{0}} = 
  \boldsymbol{p}^{(4)},
\end{eqnarray}
and so on. Notice that in the third-order derivatives (\ref{d3_phi})
there is only one term which contains $\boldsymbol{p}'''$, namely
\begin{eqnarray}
  \label{1termp4}
  \sum_{\alpha,\beta,\gamma,\delta\in\mathscr{H}} 
  \left( \nu_{\alpha_{1}}^{\mathrm{sp}}\delta_{\alpha_1\alpha} - 
    \nu_{\alpha_{1}}^{\mathrm{sp}}\nu_{\alpha}^{\mathrm{sp}}
    \chi_{\alpha_1\alpha} \right)
  \frac{p'''_{\alpha\beta\gamma\delta}
    \left(\boldsymbol{\nu}^{\mathrm{sp}}\right)} 
  {p_{\delta}\left(\boldsymbol{\nu}^{\mathrm{sp}}\right)}
  \nu_{\delta}^{\mathrm{sp}} 
  \frac{\partial\nu_{\beta}^{\mathrm{sp}}}{\partial J_{\alpha_2}}
  \frac{\partial\nu_{\gamma}^{\mathrm{sp}}}{\partial J_{\alpha_3}}.
\end{eqnarray} 
When this term is evaluated at $\boldsymbol{J}=\boldsymbol{0}$, the
factors
$\left. \nu_{\delta}^{\mathrm{sp}}\right|_{\boldsymbol{J}=\boldsymbol{0}}$
and $\left. p_{\delta}\left(\boldsymbol{\nu}^{\mathrm{sp}}\right)
\right|_{\boldsymbol{J}=\boldsymbol{0}}$ cancels each other out so
that, according to the sum rules (\ref{PPAR_SUM}), we have
\begin{eqnarray}
  \sum_{\delta\in\mathscr{H}} 
  \left.
    p'''_{\alpha\beta\gamma\delta}\left(\boldsymbol{\nu}^{\mathrm{sp}}\right)
  \right|_{\boldsymbol{J}=\boldsymbol{0}}
  =
  \sum_{\delta\in\mathscr{H}} 
  p^{(4)}_{\alpha\beta\gamma\delta} = 0.
\end{eqnarray} 
As a consequence, the system of equations which determines
$\boldsymbol{p}^{(3)}$ does not contain $\boldsymbol{p}^{(4)}$.  

The property is immediately extended to higher orders.  The system of
equations which determines
$p^{(4)}_{\alpha_1\alpha_2\alpha_3\alpha_4}$ depends on the fourth
order derivatives $\partial^4 \phi_{\mathrm{sp}}(\boldsymbol{J})/
\partial J_{\alpha_1}\partial J_{\alpha_2}\partial J_{\alpha_3}
\partial J_{\alpha_4}$ and these contain $\boldsymbol{p}$,
$\boldsymbol{p}'$, $\boldsymbol{p}''$, $\boldsymbol{p}'''$ and
$\boldsymbol{p}''''$.  The fourth order derivative
$\boldsymbol{p}''''$ may appear only in the term
\begin{eqnarray}
  \label{1termp5}
  \sum_{\alpha,\beta,\gamma,\delta,\varepsilon\in\mathscr{H}} 
  \left( \nu_{\alpha_{1}}^{\mathrm{sp}}\delta_{\alpha_1\alpha} - 
    \nu_{\alpha_{1}}^{\mathrm{sp}}\nu_{\alpha}^{\mathrm{sp}}
    \chi_{\alpha_1\alpha} \right)
  \frac{p''''_{\alpha\beta\gamma\delta\varepsilon}
    \left(\boldsymbol{\nu}^{\mathrm{sp}}\right)} 
  {p_{\delta}\left(\boldsymbol{\nu}^{\mathrm{sp}}\right)}
  \nu_{\delta}^{\mathrm{sp}} 
  \frac{\partial\nu_{\beta}^{\mathrm{sp}}}{\partial J_{\alpha_2}}
  \frac{\partial\nu_{\gamma}^{\mathrm{sp}}}{\partial J_{\alpha_3}}
  \frac{\partial\nu_{\varepsilon}^{\mathrm{sp}}}{\partial J_{\alpha_4}},
\end{eqnarray} 
obtained differentiating the factor
$p'''_{\alpha\beta\gamma\delta}\left(\boldsymbol{\nu}^{\mathrm{sp}}\right)$
of (\ref{1termp4}) with respect to $J_{\alpha_4}$.  When evaluated at
$\boldsymbol{J}=\boldsymbol{0}$, equation (\ref{1termp5}) vanishes and
we find that no $\boldsymbol{p}^{(5)}$ terms are contained in the
system of equations for $\boldsymbol{p}^{(4)}$.  Iterating, we
conclude that the property (ii) is valid at any higher order.

Now we focus on the property (i).  The system of equations
which determines $\boldsymbol{p}^{(3)}$ is linear in the unknown
tensor simply because the third-order derivatives $\partial^3
\phi_{\mathrm{sp}}(\boldsymbol{J})/
\partial J_{\alpha_1}\partial J_{\alpha_2}\partial J_{\alpha_3}$ are
linear in $\boldsymbol{p}''$, see equation (\ref{d3_phi}).  By further
differentiating (\ref{d3_phi}) with respect to $J_{\alpha_4}$ there is
no possibility to generate a term nonlinear in $\boldsymbol{p}'''$,
e.g. quadratic.  In fact, this would amount to have in $\partial^3
\phi_{\mathrm{sp}}(\boldsymbol{J})/
\partial J_{\alpha_1}\partial J_{\alpha_2}\partial J_{\alpha_3}$ a
term containing the product or the ratio of the components of
$\boldsymbol{p}'''$ and $\boldsymbol{p}''$.  However, we have seen
that the only term of (\ref{d3_phi}) containing $\boldsymbol{p}'''$ is
given by (\ref{1termp4}).  We conclude that the system of equations
which determines $\boldsymbol{p}^{(4)}$ is linear and, iterating, the
same holds at any higher order.

\section{Evaluation of the ground-state energy}
\label{E0.evaluation}
We have established that the ground state energy $E_0$ is the unique
solution of the scalar equation
\begin{eqnarray}
  \label{E0.eq0}
  \lim_{N\to\infty} \frac{1}{N} \log I_N(E_0) = 0,
  \qquad E_0 \leq V_\mathrm{min},
\end{eqnarray}
where
\begin{eqnarray}
  \label{I_N}
  I_{N}(E_0) = \int \mathrm{d}(N\boldsymbol{\nu})
  \mathcal{P}_N(N\boldsymbol{\nu})
  \mathrm{e}^{\left(N\boldsymbol{\nu},\boldsymbol{u}(E_0)\right)},
\end{eqnarray}
and
\begin{eqnarray}
  \label{u}
  \boldsymbol{u}^\mathrm{T}(E_{0}) =
  (\ldots -\log(-E_0+V) \ldots;
  \ldots \log T \ldots; \ldots \log \lambda \ldots).
\end{eqnarray}
In this section we want to determine $E_0$ when the probability
density $\mathcal{P}_N(N\boldsymbol{\nu})$ is given by the multinomial
perturbative scheme (\ref{OMEGA_pert}-\ref{p_alpha}).

For $N$ large we have already calculated the integral (\ref{I_N}).  In
fact, $I_N(E_0)$ coincides with the generating function
$Z_N(\boldsymbol{J})$ studied in section \ref{evaluation.arc} provided
we choose the source $\boldsymbol{J}=\boldsymbol{u}(E_0)$. Therefore,
equation (\ref{E0.eq0}) is equivalent to
\begin{eqnarray}
  \label{E0.eq1}
  \phi_{\mathrm{sp}}\left(\boldsymbol{u}\left(E_{0}\right)\right) = 0,
  \qquad E_0 \leq V_\mathrm{min},
\end{eqnarray}
where $
\phi_{\mathrm{sp}}\left(\boldsymbol{u}\left(E_{0}\right)\right)$ is
given by (\ref{phitilde_pert}) with
$\boldsymbol{J}=\boldsymbol{u}(E_0)$.  Of course, also the
saddle-point frequencies (\ref{nuspe_pert}) which appear in
(\ref{phitilde_pert}) must be evaluated with the same choice of
$\boldsymbol{J}$.  We conclude that the ground-state energy $E_0$ is
obtained, together with the saddle-point frequencies
$\boldsymbol{\nu}^{\mathrm{sp}}$, as the solution of the following
system of nonlinear coupled equations
\begin{eqnarray}\fl
  \label{E0.eq2}
  \sum_{V\in\mathscr{V}} 
  \frac{\widetilde{p}_{V}(\boldsymbol{\nu}^\mathrm{sp})}{-E_{0}+V}
  = 
  \frac{
    \mathrm{e}^{ \sum_{\alpha\in\mathscr{H}}\sum_{\beta\in\mathscr{H}}
      \frac{p'_{\alpha\beta}(\boldsymbol{\nu}^\mathrm{sp})}
      {p_{\beta}(\boldsymbol{\nu}^\mathrm{sp})}
      \nu_{\alpha}^\mathrm{sp} \nu_{\beta}^\mathrm{sp} 
    }}
  {\left(
      \sum_{T\in\mathscr{T}} 
      \widetilde{p}_{T}(\boldsymbol{\nu}^\mathrm{sp}) T
    \right)
    \left(
      \sum_{\lambda\in\mathscr{L}} 
      \widetilde{p}_{\lambda}(\boldsymbol{\nu}^\mathrm{sp}) \lambda
    \right)},
  \quad E_0 \leq V_\mathrm{min},
\end{eqnarray}
\begin{eqnarray}\fl
  \label{nusp.eq}
  \nu_{\alpha}^\mathrm{sp} = 
  \frac
  {\widetilde{p}_{\alpha}(\boldsymbol{\nu}^\mathrm{sp})
    ~\mathrm{e}^{u_{\alpha}(E_{0})} }
  {\sum_{\alpha'\in\mathscr{A}_{\alpha}}
    \widetilde{p}_{\alpha'}(\boldsymbol{\nu}^\mathrm{sp})
    ~\mathrm{e}^{u_{\alpha'}(E_{0})} },
  \qquad \alpha\in\mathscr{H},
\end{eqnarray}
where we have defined
\begin{eqnarray}
  \widetilde{p}_{\alpha}(\boldsymbol{\nu}) =
  p_{\alpha}(\boldsymbol{\nu})
  \mathrm{e}^{
    \sum_{\beta\in\mathscr{H}}
    \frac{p'_{\alpha\beta}(\boldsymbol{\nu})}
    {p_{\beta}(\boldsymbol{\nu})}
    \nu_{\beta} },
\end{eqnarray}
and we recall that
\begin{eqnarray*}\fl
  p_{\alpha}\left(\boldsymbol{\nu}\right)=
  \sum_{n=0}^{k_\mathrm{max}-1} \frac{1}{n!}
  \sum_{\beta_1\in\mathscr{H}}\dots\sum_{\beta_n\in\mathscr{H}}
  p^{(n+1)}_{\alpha\beta_1\dots\beta_n}
  \left(\nu_{\beta_1}-\Sigma^{(1)}_{\beta_1}\right)
  \dots
  \left(\nu_{\beta_n}-\Sigma^{(1)}_{\beta_n}\right),
\end{eqnarray*}
\begin{eqnarray*}\fl
  p'_{\alpha\beta}\left(\boldsymbol{\nu}\right) =
  \sum_{n=1}^{k_\mathrm{max}-1} \frac{1}{(n-1)!}
  \sum_{\gamma_1\in\mathscr{H}}\dots\sum_{\gamma_{n-1}\in\mathscr{H}}
  p^{(n+1)}_{\alpha\beta\gamma_1\dots\gamma_{n-1}}
  \left(\nu_{\gamma_1}-\Sigma^{(1)}_{\gamma_1}\right)
  \dots
  \left(\nu_{\gamma_{n-1}}-\Sigma^{(1)}_{\gamma_{n-1}}\right).
\end{eqnarray*}
Note that the above system of equations is valid at any perturbative
order. More precisely, for different choices of $k_\mathrm{max}$ only
the functions $p_{\alpha}\left(\boldsymbol{\nu}\right)$ and
$p'_{\alpha\beta}\left(\boldsymbol{\nu}\right)$ are to be modified
whereas the structure of equations (\ref{E0.eq2}-\ref{nusp.eq})
remains unchanged.

At the lowest perturbative order $k_\mathrm{max}=1$, we have
$\boldsymbol{p}(\boldsymbol{\nu})=\boldsymbol{\Sigma}^{(1)}$,
$\boldsymbol{p}'(\boldsymbol{\nu})=\boldsymbol{0}$ and
$\widetilde{\boldsymbol{p}}(\boldsymbol{\nu})=\boldsymbol{\Sigma}^{(1)}$.
It follows that equations (\ref{E0.eq2}) and (\ref{nusp.eq}) can be
solved separately.  Equation (\ref{E0.eq2}) reduces to
(\ref{E0multinomial}), i.e. $E_0$ is the solution of
\begin{eqnarray}
  \label{E0.eq3}
  \sum_{V\in\mathscr{V}} 
  \frac{\Sigma^{(1)}_{V}}{-E_{0}+V} = 
  \frac{1}
  {\left( \sum_{T\in\mathscr{T}} \Sigma^{(1)}_{T} T \right)
    \left( \sum_{\lambda\in\mathscr{L}} \Sigma^{(1)}_{\lambda} \lambda \right)},
  \qquad E_0 \leq V_\mathrm{min}.
\end{eqnarray}
This equation can be straightforwardly solved by bisection method.
Once $E_0$ is found, the saddle-point frequencies are given by
\begin{eqnarray}
  \label{nusp.V}
  \nu_V^\mathrm{sp} &=& 
  \frac
  {\Sigma_{V}^{(1)}(-E_{0}+V)^{-1}}
  {\sum_{V'\in\mathscr{V}} \Sigma_{V'}^{(1)}(-E_{0}+V')^{-1}},
  \qquad V\in\mathscr{V},
  \\
  \label{nusp.T}
  \nu_T^\mathrm{sp} &=& 
  \frac
  {\Sigma_{T}^{(1)} T}
  {\sum_{T'\in\mathscr{T}} \Sigma_{T'}^{(1)} T'},
  \qquad T\in\mathscr{T},
  \\
  \label{nusp.L}
  \nu_\lambda^\mathrm{sp} &=& 
  \frac
  {\Sigma_{\lambda}^{(1)} \lambda}
  {\sum_{\lambda'\in\mathscr{L}} \Sigma_{\lambda'}^{(1)} \lambda'},
  \qquad \lambda\in\mathscr{L}.
\end{eqnarray}

At higher perturbative orders $k_\mathrm{max}>1$, after the
perturbative parameters $\boldsymbol{p}^{(k)}$ for
$k=1,\dots,k_\mathrm{max}$ have been determined from the corresponding
exact cumulants $\boldsymbol{\Sigma}^{(1)}, \dots,
\boldsymbol{\Sigma}^{(k_\mathrm{max})}$, the system of equations
(\ref{E0.eq2}) and (\ref{nusp.eq}) can be solved numerically by a
globally convergent multidimensional Newton--Raphson method \cite{NR}.
Of course, we have to conjecture some initial value of
$(E_0,\boldsymbol{\nu}^\mathrm{sp})$ which is not too far from the
solution.  Actually, this may not represent a problem in the spirit of
our perturbative approach according to which the uncorrelated
multinomial probability density can roughly capture the features of
the system considered.  Therefore, we propose to use the solution of
(\ref{E0.eq3}) and the values (\ref{nusp.V}-\ref{nusp.L}) as an
initial guess for $(E_0,\boldsymbol{\nu}^\mathrm{sp})$.

\section{Numerical results for the Hubbard model}
\label{numerical.results}
In this section we apply the multinomial perturbative scheme to study
some example systems. We will focus on the Hubbard model with
pseudo-spin 1/2 hard-core bosons. The case of fermions will
be considered in a paper dedicated to the sign problem.

The Hubbard model is one of the simplest models displaying the real
word features of a many-body system. It plays essentially the same
role in the problem of the electron correlations as the Ising model in
the problem of spin correlations. The model describes interacting
particles, bosons or fermions, in a lattice, and the corresponding
Hamiltonian operator consists of two terms: a kinetic term allowing
for hopping of particles among the sites of the lattice and a
potential term consisting of an on-site interaction.  The model was
originally proposed by Hubbard \cite{Hubbard} to describe electrons in
solids and has since been the focus of particular interest as a model
for high-temperature superconductivity.  Recently a large interest has
been devoted also to the properties of the 2D Hubbard model on the
honeycomb lattice, as a basic model for the description of graphene
\cite{GM}.

Let us consider the first-neighbor uniform (FNU) Hubbard model
defined by the Hamiltonian
\begin{eqnarray}
  \label{FNUHubbard}
  \hat{H} = -\eta\sum_{(i,j)\in\Gamma}\sum_{\sigma=\uparrow\downarrow}
  \left(
    \hat{c}_{i\sigma}^{\dagger}\hat{c}_{j\sigma}^{\phantom{\dagger}}+
    \hat{c}_{j\sigma}^{\dagger}\hat{c}_{i\sigma}^{\phantom{\dagger}}
  \right)+
  \gamma\sum_{i\in\Lambda}
  \hat{c}_{i\uparrow}^{\dagger}\hat{c}_{i\uparrow}^{\phantom{\dagger}}
  \hat{c}_{i\downarrow}^{\dagger}\hat{c}_{i\downarrow}^{\phantom{\dagger}},
\end{eqnarray}
where $\Lambda \subset \mathbb{Z}^d$ is a finite $d$-dimensional
lattice with $|\Lambda|$ ordered sites, $\Gamma=\{(i,j):i<j\in\Lambda
\mbox{ and $i,j$ are first neighbor} \}$, and
$\hat{c}_{i\sigma}^{\phantom{\dagger}}$ a commuting destruction
operator at site $i\in\Lambda$ and spin index
$\sigma=\uparrow\downarrow$ with the property $\hat{c}_{i\sigma}^2=0$
(hard-core boson destruction operator).  Hopping strengths are uniform
through the lattice and described by the parameter $\eta>0$. Also the
on-site interactions are independent of the site index and their value
is fixed by the parameter $\gamma\geq 0$.  The system is described in
terms of Fock states labelled by the configurations
$\bm{n}=(n_{1\ua},n_{1\da},\dots,n_{|\Lambda|\ua},n_{|\Lambda|\da})$
with $n_{i\sigma}=0,1$.  In this $\bm{n}$-representation, the on-site
Hubbard interaction is the diagonal potential operator $\hat{V}$ with
matrix elements
\begin{eqnarray}
  V_{\bm{n},\bm{n}} = V_{\bm{n}} = \gamma \sum_{i\in\Lambda} n_{i\ua}n_{i\da},
  \label{V.Hubbard}
\end{eqnarray}
whereas the matrix elements of the hopping operator are
\begin{eqnarray}
  K_{\bm{n},\bm{n}'} = -\eta 
  \sum_{(i,j)\in\Gamma} \sum_{\sigma=\ua\da}
  \delta_{\bm{n},\bm{n}'\oplus\bm{\1}_{i\sigma}\oplus\bm{\1}_{j\sigma}},
  \label{K.Hubbard}
\end{eqnarray}
where 
 $\bm{\1}_{i\sigma}=(0,\dots,0,1_{i\sigma},0,\dots,0)$ and
$\oplus$ means $\mod$ 2 addition.  By comparing (\ref{K.Hubbard}) with
(\ref{K}) and observing that $\sum_{(i,j)\in\Gamma}
\sum_{\sigma=\ua\da}
\delta_{\bm{n},\bm{n}'\oplus\bm{\1}_{i\sigma}\oplus\bm{\1}_{j\sigma}}=0,1$,
the unit value being obtained only if $\bm{n}'\in A(\bm{n})$, where
\begin{eqnarray}
  A(\bm{n}) = \{
  \bm{n}\oplus\bm{\1}_{i\sigma}\oplus\bm{\1}_{j\sigma}:
  \mbox{$(i,j)\in\Gamma$, $\sigma=\ua\da$ with 
    $n_{i\sigma}\oplus n_{j\sigma}=1$} \}
  \label{An}
\end{eqnarray}
is the set of the configurations connected to $\bm{n}$ by the hopping
of one particle, we have
\begin{eqnarray}
  \eta_{\bm{n},\bm{n}'}=\eta, 
  \qquad
  \lambda_{\bm{n},\bm{n}'}=
  \left\{
    \begin{array}{ll}
      1, \quad &\bm{n}'\in A(\bm{n}),
      \\
      0, &\mbox{otherwise}.
    \end{array}
  \right.
\end{eqnarray}

In the rest of this section we will consider two dimensional square
lattices having $L_x \times L_y$ sites and containing $N_\mathrm{p}$ particles
per spin.  Periodic boundary conditions will be imposed.  At the
densities $N_\mathrm{p}/(L_xL_y) \leq 1/2$ took into consideration, the set of
all possible different values of the potential variables (\ref{V_k})
during an infinitely long random walk corresponds to the set of all
possible different values of $V_{\bm{n}}$, the matrix elements
(\ref{V.Hubbard}), over the configuration space.  It is
straightforward to see that
$\mathscr{V}=\{0,\gamma,2\gamma,\ldots,N_{p}\gamma\}$.  The set of all
possible different values of the hopping variables (\ref{T_k}) during
an infinitely long random walk, corresponds to the set of all possible
different values of $T_{\bm{n}} = \eta |A(\bm{n})|$ over the
configuration space, with $A(\bm{n})$ given by (\ref{An}).  The
determination of $\mathscr{T}$ depends both on the number of particles
and on the lattice size.  For instance, in a lattice $2\times 3$ we
have $\mathscr{T}=\{12\eta,16\eta,20\eta\}$ with $N_\mathrm{p}=3$ and
$\mathscr{T}=\{8\eta,10\eta,12\eta,14\eta,16\eta\}$ with $N_{p}=2$.
For the present model the set of all possible phase variables is
$\mathscr{L}=\{1\}$.

Once we have determined the sets $\mathscr{V}$, $\mathscr{T}$ and
$\mathscr{L}$, the asymptotic rescaled cumulants $\bm{\Sigma}^{(k)}$
associated with the considered Hubbard system can be measured as explained
in \cite{OP4}.  Note that the cumulants are unaffected by a change of
the parameters $\eta$ and $\gamma$ of the Hamiltonian
(\ref{FNUHubbard}), whereas the label sets $\mathscr{V}$,
$\mathscr{T}$ and $\mathscr{L}$ keep their cardinalities under the
same change. This implies, as already stated, that we can input the
cumulants measured for a particular value of $\eta$ and $\gamma$ 
into equations (\ref{E0.eq2}-\ref{nusp.eq}) to find the
ground-state energy $E_0$ as a function of $\eta$ and
$\gamma$. 
\begin{figure}
  \centering
  \includegraphics[width=10cm, bb=0 0 726 475]{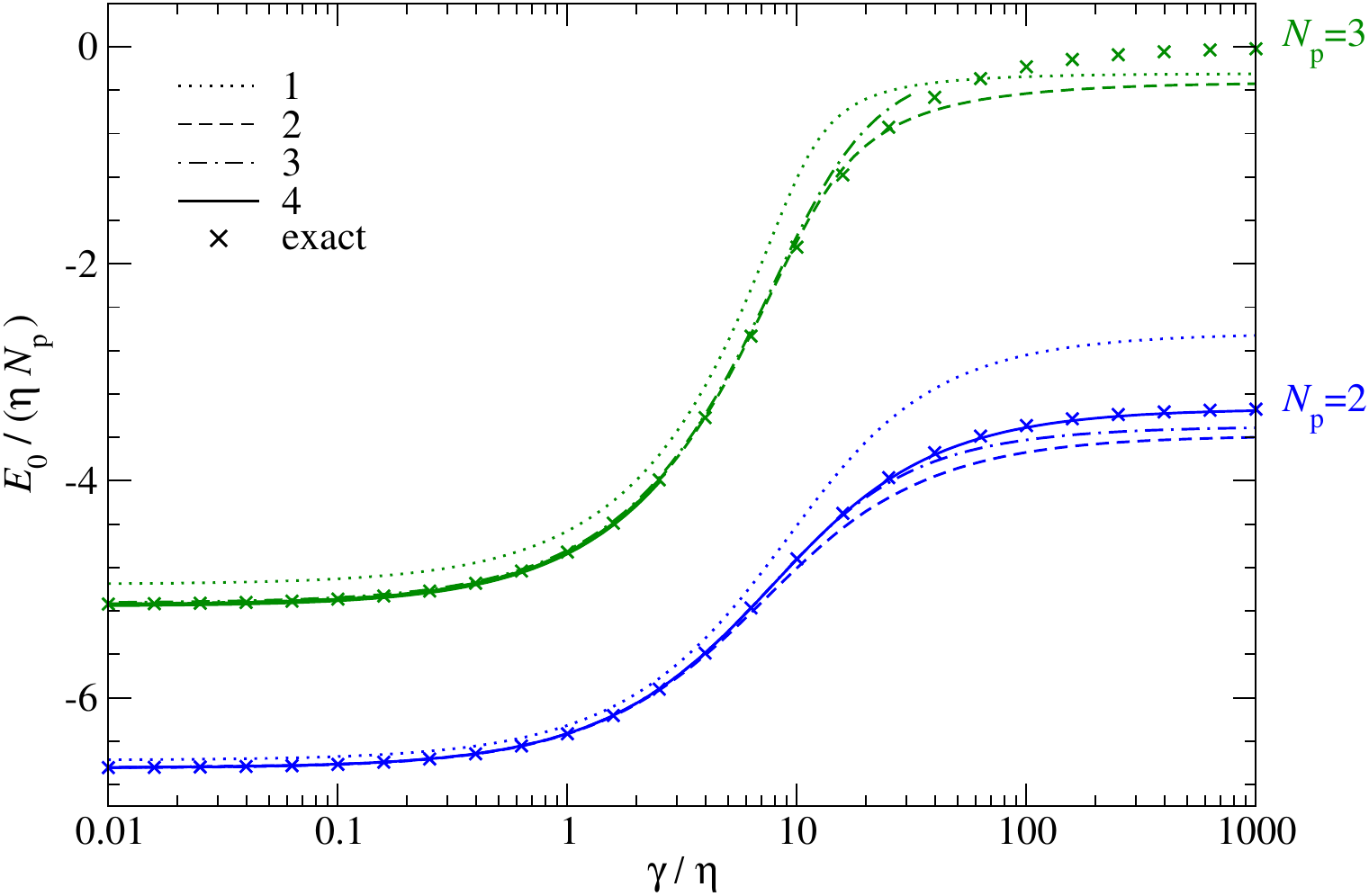}
  \caption{Ground-state energy per particle for the $2 \times 3$ FNU
    hard-core boson Hubbard model with periodic boundary conditions
    versus the interaction strength $\gamma$ with $N_\mathrm{p} = 2$ and
    $N_\mathrm{p}=3$ particles per spin. We compare the results from exact
    numerical diagonalization ($\times$) with those from present
    multinomial perturbative scheme by using cumulants up to order
    $k_\mathrm{max}=1,2,3,4$ (dotted, dashed, dot-dashed, solid lines,
    respectively).}
  \label{fig2}
\end{figure}

The results obtained by using cumulants up to order
$k_\mathrm{max}=1,2,3,4$ are shown in figure \ref{fig2} as a function
of the ratio $\gamma/\eta$ in the case of a lattice with $2\times 3$
sites and $N_\mathrm{p}=2,3$ particles per spin. In the same figure we also
depict the values of $E_0$ determined by exact numerical
diagonalization.  The curves obtained for $k_\mathrm{max}=1$ coincide
with the uncorrelated multinomial prediction and, as already noted,
their behavior is only qualitatively correct. Already at
$k_\mathrm{max}=2$, the quantitative agreement with the exact
ground-state energy becomes impressive at least for values of
$\gamma/\eta$ not too large.  By further increasing $k_\mathrm{max}$,
the quantitative agreement gradually improves in the whole range of
$\gamma/\eta$ which, note the horizontal log scale, goes from the
limit $\|\hat{V}\| \ll \|\hat{K}\|$ to the opposite one $\|\hat{V}\|
\gg \|\hat{K}\|$.  In the case with $N_\mathrm{p}=2$ particles per spin, the
curve $E_0(\gamma)$ obtained with $k_\mathrm{max}=4$ is
indistinguishable from the reported exact values. In the case with
$N_\mathrm{p}=3$ particles per spin, the nonlinear equations
(\ref{E0.eq2}-\ref{nusp.eq}) do not admit a solution for
$k_\mathrm{max}=3,4$ when $\gamma/\eta$ is larger than a threshold
value, namely $\gamma/\eta\simeq 31.6$ for $k_\mathrm{max}=3$ and
$\gamma/\eta\simeq 2.5$ for $k_\mathrm{max}=4$.  
As discussed at the beginning of section
\ref{multinomial.perturbative.scheme}, this means that the perturbative 
scheme is invalid at the order considered.

\begin{figure}[t]
  \centering
  \includegraphics[width=10cm, bb=0 0 726 475]{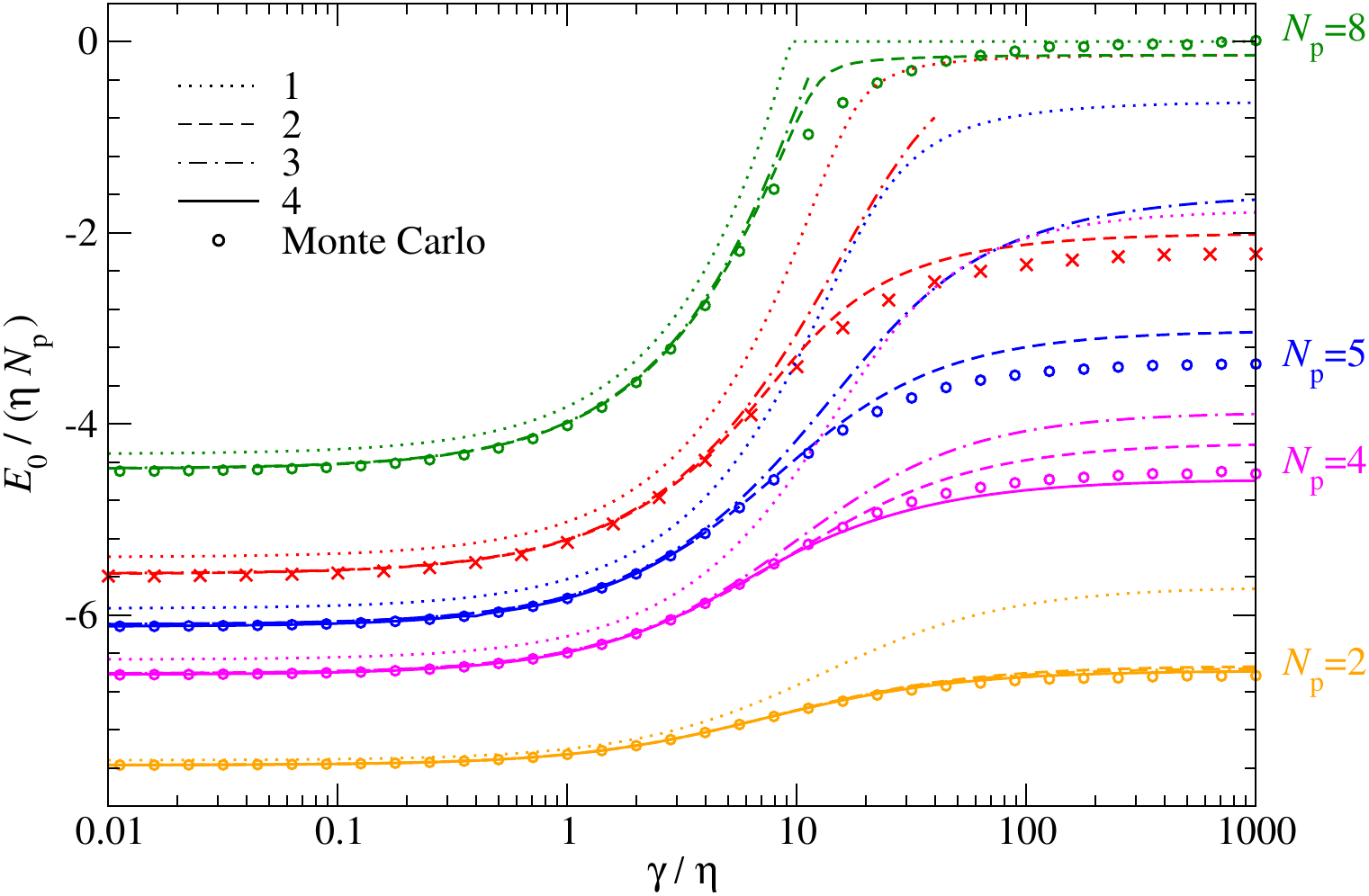}
  \caption{As in figure \ref{fig2} in the case of a $4 \times 4$ FNU
    hard-core boson Hubbard model with periodic boundary conditions
    and $N_\mathrm{p} = 2,4,5,8$ particles per spin. The data indicated by
    $\circ$ have been obtained by Monte Carlo simulations \cite{OP3}
    (the associated statistical errors increase with increasing $\gamma$
    and are of the order of the symbol size at $\gamma \simeq
    1000\eta$).}
  \label{fig3}
\end{figure}

\begin{figure}[t]
  \centering
  \includegraphics[width=10cm, bb=0 0 684
  475]{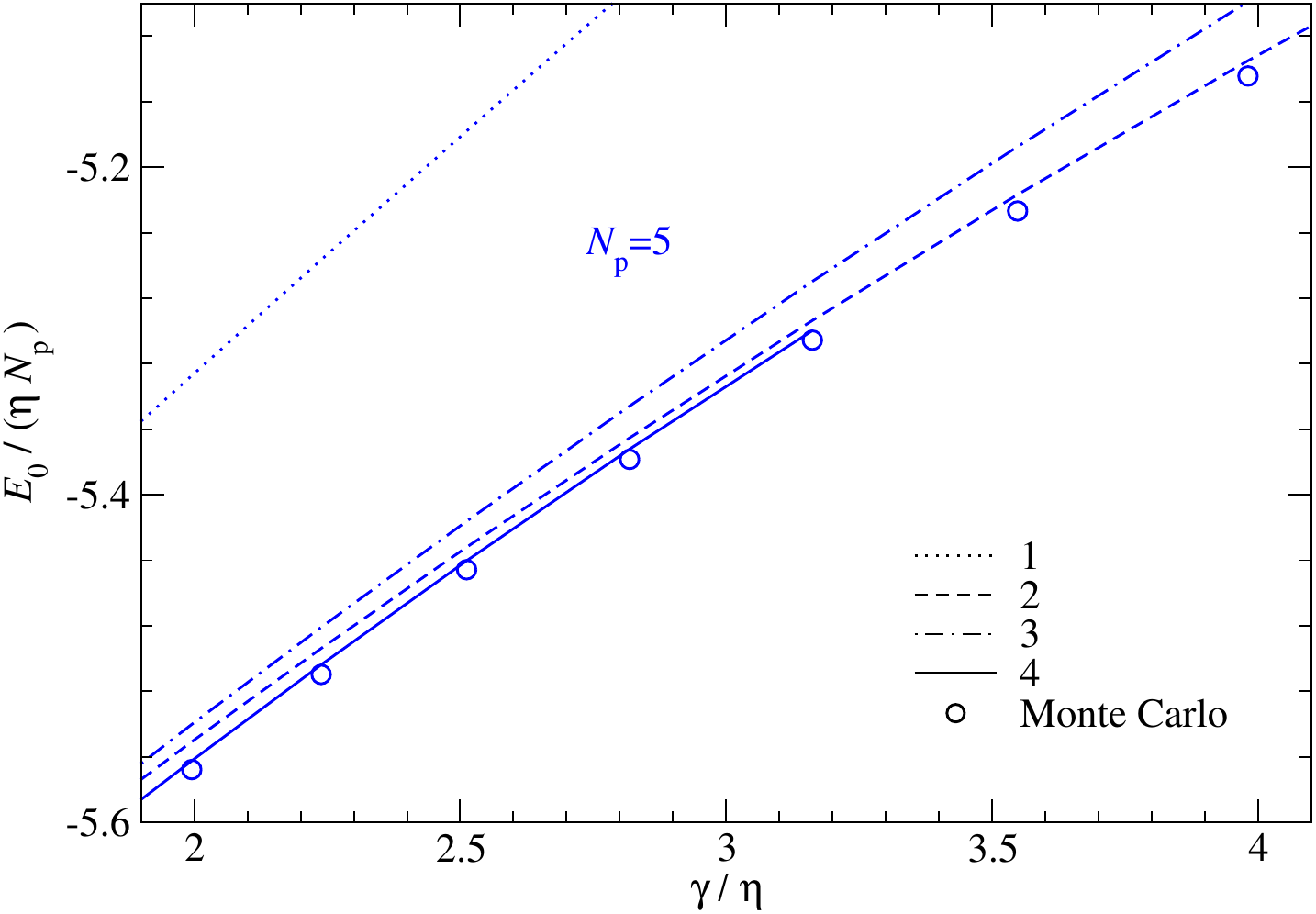}
  \caption{Enlargement of figure \ref{fig3}, case with $N_\mathrm{p}=5$.  The
    perturbative scheme with $k_\mathrm{max}=4$ admits a solution only
    for $\gamma\lesssim 3.2 \eta$ but in this range provides results
    more accurate than those obtained with $k_\mathrm{max}=1,2,3$.}
  \label{fig4}
\end{figure}

In figure \ref{fig3} we show the results obtained with a system of
larger size, namely a lattice $4\times 4$ with $N_\mathrm{p}=2,4,5,8$
particles.  In this case, the number of configurations is so large,
namely
\begin{eqnarray}
  M = \left( \frac{(L_xL_y)!}{N_\mathrm{p}! (L_xL_y-N_\mathrm{p})!} \right)^2,
\end{eqnarray}
that an exact numerical diagonalization of the Hamiltonian
(\ref{FNUHubbard}) is unfeasible. Therefore, we have compared the
values of $E_0$ predicted by the present multinomial perturbative
scheme with those obtained by a Monte Carlo simulation \cite{OP3}.
The conclusions that we reach from the analysis of figure \ref{fig3}
are similar to those we noticed after figure \ref{fig2}.  However, for
this larger system we see that the parity of the order
$k_\mathrm{max}$ may influence the quality of the approximation, a
fact which is not surprising.  For instance, in the case with $N_\mathrm{p}=5$
it is evident that the curve $E_0(\gamma)$ obtained with
$k_\mathrm{max}=3$ is not better than that obtained with
$k_\mathrm{max}=2$. However, as evidenced in the enlargement shown in
figure \ref{fig4}, the results obtained with $k_\mathrm{max}=4$ are
more accurate than those with $k_\mathrm{max}=2$, at least in the
$\gamma/\eta$ range where equations (\ref{E0.eq2}-\ref{nusp.eq}) admit
the $k_\mathrm{max}=4$ solution.

\section{Hubbard model with a magnetic field}
\label{magnetic.field}
In the examples considered in the previous section we have
$\mathscr{L}=\{1\}$, i.e. the phase variables play no role. The
situation is different in the case of fermions or for hard-core bosons
in the presence of a magnetic field. In order to illustrate how to
deal with the phase variables, in this section we will consider a
Hubbard model in a one-dimensional lattice with periodic boundary
conditions, namely a ring with $L$ sites, threaded by a line of
magnetic flux $\phi$. In the case of spin 1/2 fermions, this is a well
known model used to study electronic persistent currents, see
\cite{VKSDM} for a review. The model is free of sign problem in the
case of an even number of fermions per spin. In order to concentrate
on the effects due to the sole magnetic field, in the following we
will therefore assume $N_\mathrm{p}$, the number of particles per
spin, to be even.  This is equivalent to consider a system of
$N_\mathrm{p}+N_\mathrm{p}$ pseudo-spin 1/2 hard-core bosons.

The Hamiltonian of the system is
\begin{eqnarray}
  \label{HubbardRing}
  \hat{H} = -\eta\sum_{j=1}^{L}\sum_{\sigma=\uparrow\downarrow}
  \left(
    \E^{-\I\theta}
    \hat{c}_{j+1\sigma}^{\dagger}\hat{c}_{j\sigma}^{\phantom{\dagger}}+
    \E^{\I\theta}
    \hat{c}_{j\sigma}^{\dagger}\hat{c}_{j+1\sigma}^{\phantom{\dagger}}
  \right)+
  \gamma\sum_{j=1}^{L}
  \hat{c}_{j\uparrow}^{\dagger}\hat{c}_{j\uparrow}^{\phantom{\dagger}}
  \hat{c}_{j\downarrow}^{\dagger}\hat{c}_{j\downarrow}^{\phantom{\dagger}},
\end{eqnarray}
where site correspondence $j\pm L = j$ is assumed and
$\E^{\pm\I\theta}$ are the Peierls phase factors with
$\theta=2\pi\phi/(L \phi_0)$, $\phi_0=h/(2e)$ being the magnetic flux
quantum.  The spectrum of the Hamiltonian (\ref{HubbardRing}) can be
determined exactly in terms of the Bethe ansatz \cite{LW}, however,
when the Fock dimension $(L!/(N_\mathrm{p}!(L-N_\mathrm{p})!))^2$ is
not too large, a numerical diagonalization represents a simpler
alternative. The ground state energy $E_0$ as well as all the excited
eigenenergies of $\hat{H}$ are periodic functions of the flux $\phi$
with period $\phi_0$.  In the non-interacting case $\gamma=0$, the
ground state energy $E_0^{(0)}$ has the simple expression
\begin{eqnarray}
  \label{E00ring}
  E_0^{(0)}(\phi) = - 4\eta\frac{\sin(\pi N_\mathrm{p}/L)}{\sin(\pi/L)}
  \cos\left(\frac{2\pi}{L}\frac{\phi}{\phi_0}\right),
  \qquad \phi \in [-\phi_0/2,\phi_0/2].
\end{eqnarray}

The sets of the potential, hopping and phase variables which apply to
the present model are found out at once. We have
$\mathscr{V}=\{0,\gamma,2\gamma,\ldots,N_\mathrm{p}\gamma\}$,
$\mathscr{T}=\{N_\mathrm{p}\eta,(N_\mathrm{p}+2)\eta,(N_\mathrm{p}+4)\eta,
\dots,2N_\mathrm{p}\eta\}$ and
$\mathscr{L}=\{\E^{\I\theta},\E^{-\I\theta}\}$.  These data, together
with the asymptotic rescaled cumulants $\bm{\Sigma}^{(k)}$ measured up
to some order $k\leq k_\mathrm{max}$, are input into equations
(\ref{E0.eq2}-\ref{nusp.eq}) to determine the ground-state energy
$E_0$.  Let us start considering the non-interacting case $\gamma=0$
at the lowest perturbative order $k_\mathrm{max}=1$.  According to
equation (\ref{E0.eq3}) and considering that $\mathscr{V}=\{0\}$ and
$\Sigma_{\lambda=\E^{\I\theta}}^{(1)} =
\Sigma_{\lambda=\E^{-\I\theta}}^{(1)} = 1/2$ (for each forward
movement of a particle in the ring there is another possible backward
jump) we have
\begin{eqnarray}
  \label{E00ring1}
  E_0^{(0)} = - \left( \sum_{T\in\mathscr{T}} \Sigma^{(1)}_{T} T \right)
  \cos\left(\frac{2\pi}{L}\frac{\phi}{\phi_0}\right).
\end{eqnarray}
Compared to the exact expression (\ref{E00ring}) this is a very
promising result.  However, equations (\ref{nusp.V}), (\ref{nusp.T})
and (\ref{nusp.L}) show that, whereas $\nu_{V=0}^\mathrm{sp}=1$ and
$0<\nu_{T}^\mathrm{sp}<1$ for $T\in\mathscr{T}$, as expected, the
saddle-point frequencies associated with the phase variables
$\lambda=\e^{\pm\I\theta}$ are complex conjugated, namely
\begin{eqnarray}
  \nu_\pm^\mathrm{sp} = \frac{1}{2} \pm \frac{\I}{2} \tan\theta.
\end{eqnarray}
To simplify the notation, hereafter we use the subscripts $\pm$
instead of $\lambda=\E^{\pm\I\theta}$.  The situation does not change
for $\gamma\neq 0$ or at higher perturbative orders. From equation
(\ref{nusp.eq}) we see that any $\nu_\alpha^\mathrm{sp}$ lies outside
the real unit simplex. What is the meaning of these complex
frequencies?  Do they imply an unphysical complex solution for $E_0$?
\begin{figure}
  \centering
  \includegraphics[width=10cm, bb=0 0 1201 836]{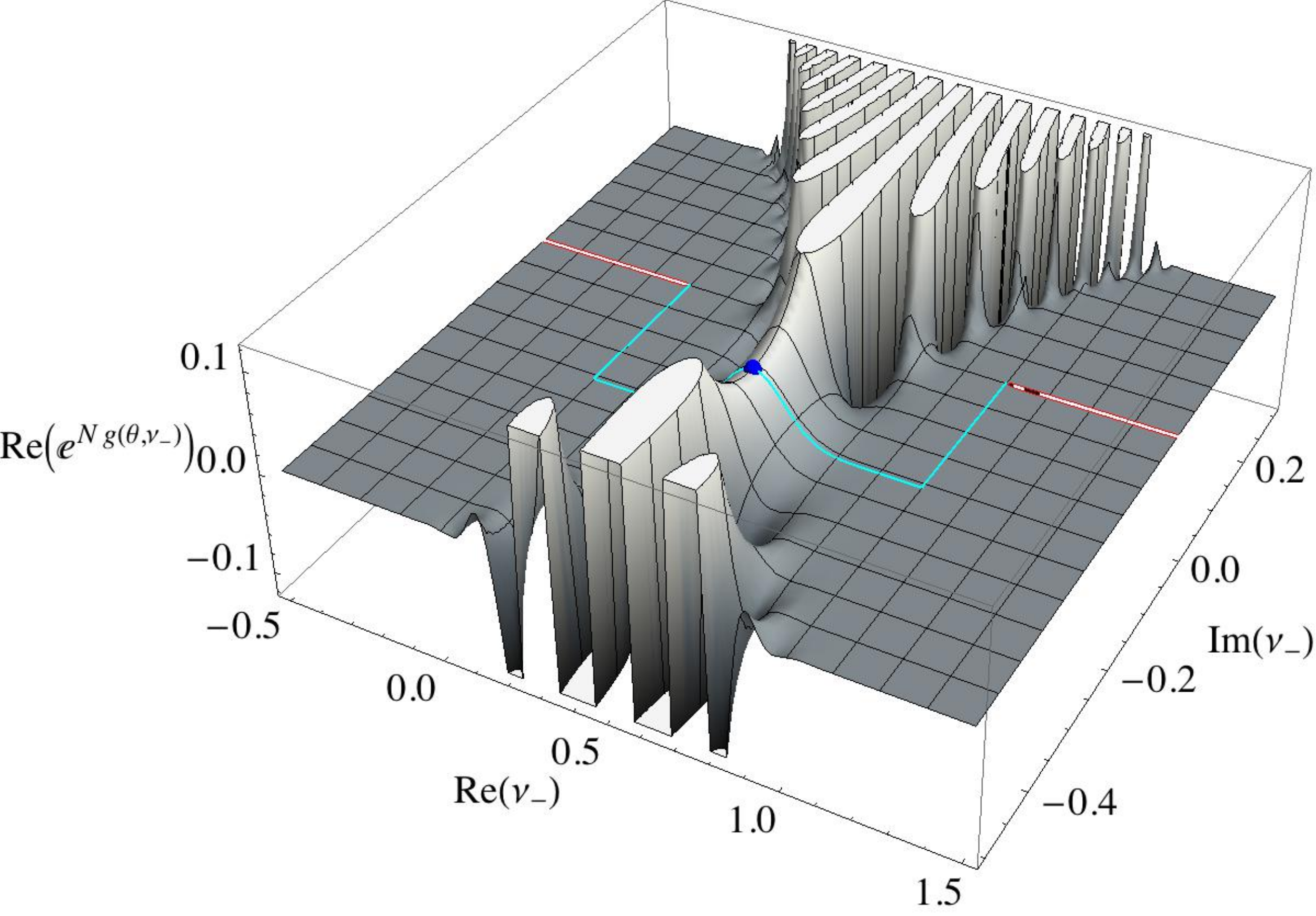}
  \includegraphics[width=10cm, bb=0 0 1201 836]{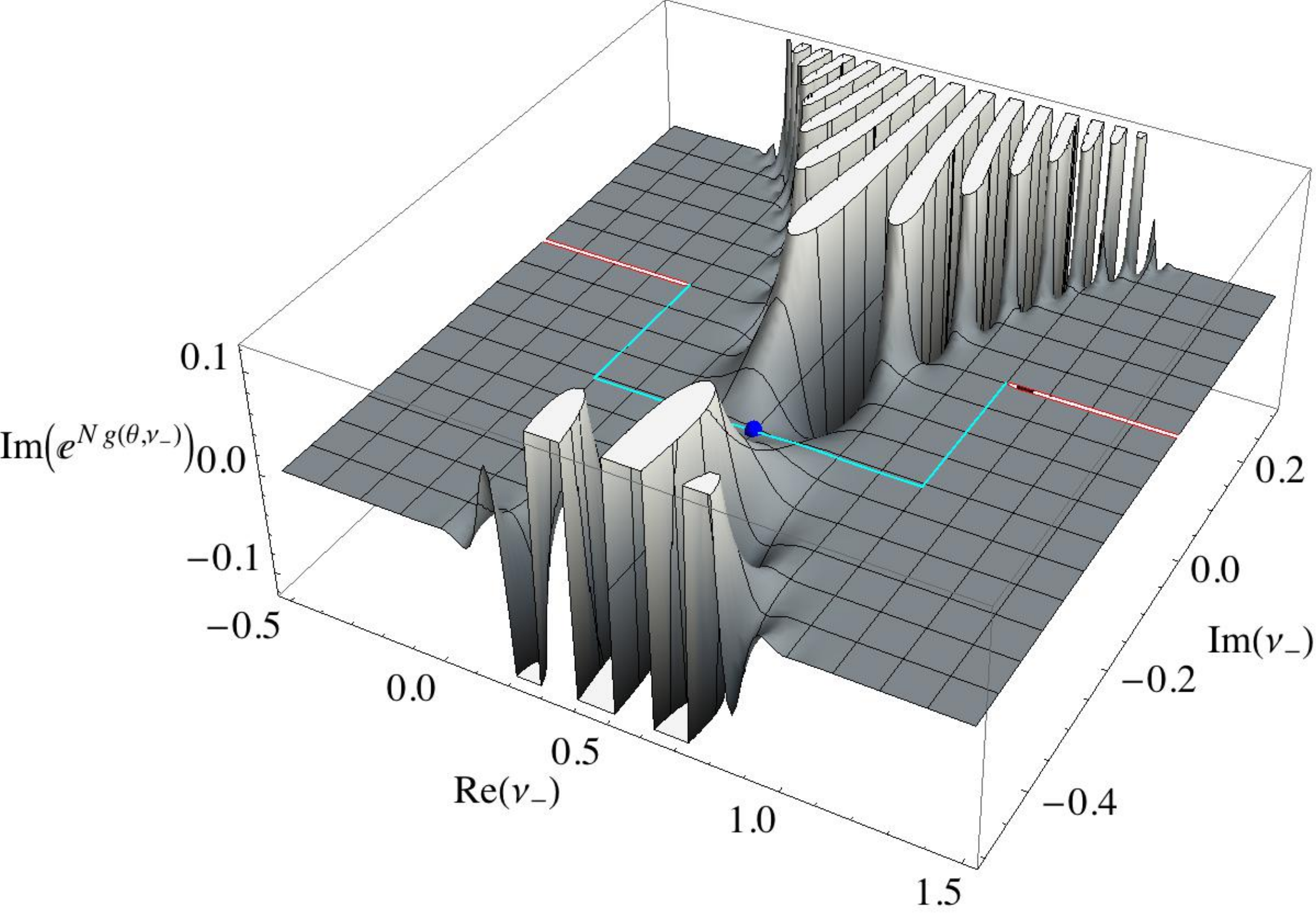}
  \caption{Real and imaginary parts of principal branch of $\exp(N
    g(\theta,\nu_-))$ as a function of the complex variable $\nu_-$
    for $N=35$ and $\theta=\pi/8$.  A contour from $\nu_-=0$ to
    $\nu_-=1$ is shown which goes through the saddle point
    $\nu_-^\mathrm{sp}=(1-\I \tan \theta)/2$ in the direction of the
    steepest descent. The branch cuts $(-\infty, 0]$ and $[1,\infty)$
    along the real axis are also indicated.}
  \label{fig5}
\end{figure}\noindent

To answer the above questions we re-examine the derivation of the
fundamental equations (\ref{E0.eq2}-\ref{nusp.eq}) in the specific
case of two phase variables. Let us start to consider the asymptotic
evaluation of the integral (\ref{I_N}) at the lowest pertubative order
$k_\mathrm{max}=1$.  The integral $I_N$ coincides with the generating
function (\ref{Z_N}) provided we choose the source
$\bm{J}=\bm{u}(E_0)$, which in the present case reads
\begin{eqnarray}
  \label{umag}
  \boldsymbol{u}^\mathrm{T}(E_{0}) =
  (\ldots -\log(-E_0+V) \ldots;
  \ldots \log T \ldots;\I\theta,-\I\theta).
\end{eqnarray}
For $k_\mathrm{max}=1$, the integral factorizes
\begin{eqnarray}
  \label{factorization}
  I_N = I_{N,\mathscr{V}} ~I_{N,\mathscr{T}} ~I_{N,\mathscr{L}},
\end{eqnarray}
where for $\mathscr{A}=\mathscr{V},\mathscr{T},\mathscr{L}$, up to
inessential constants, we have
\begin{eqnarray}
  I_{N,\mathscr{A}} = \int \prod_{\alpha\in\mathscr{A}}\D{\nu_\alpha}
  ~\E^{N 
    \sum_{\alpha\in\mathscr{A}} \nu_{\alpha}
    \left[  
      \log\left(\frac{p^{(1)}_{\alpha}}{\nu_{\alpha}}
      \right) + u_{\alpha} \right] }
  \delta\left( \sum_{\alpha\in\mathscr{A}} \nu_{\alpha}-1 \right),
\end{eqnarray}
For $\mathscr{A}=\mathscr{V},\mathscr{T}$ the functions to be
integrated are real so that $I_{N,\mathscr{V}}$ and
$I_{N,\mathscr{T}}$ can be evaluated asymptotically by the Laplace
method as explained before. The corresponding saddle-points
frequencies $\nu_{V}^\mathrm{sp}$, $V\in \mathscr{V}$, and
$\nu_{T}^\mathrm{sp}$, $T\in \mathscr{T}$, are real and lie in the
unit simplex.  In the case of $I_{N,\mathscr{L}}$ we use the Dirac
$\delta$ to eliminate the frequency $\nu_+=1-\nu_-$ and obtain
\begin{eqnarray}
  \label{I_NL}
  I_{N,\mathscr{L}} =  \int_0^1 \D{\nu_-}  
  ~\E^{N g(\theta,\nu_-)},
\end{eqnarray}
where
\begin{eqnarray}
  \label{g}
  g(\theta,\nu_-) = 
  (1-\nu_-) \log\left(\frac{1/2}{1-\nu_-}\right) +
  \nu_- \log\left(\frac{1/2}{\nu_-}\right)
  +\I\theta(1-2\nu_-).
\end{eqnarray}
Due to the factor $\E^{\I N \theta (1-2\nu_-)}$, for $N$ large the
integral (\ref{I_NL}) suffers from wild cancellations hard to
estimate. However, $\exp(N g(\theta,\nu_-))$ thought of as a function
of the complex variable $\nu_-$ is analytic in the whole complex plane
except the branch cuts $(-\infty, 0]$ and $[1,\infty)$ along the real
axis (we consider the principal branch). Thus we can evaluate
(\ref{I_NL}) by deforming the integration contour in the complex
plane. Any contour going from $\nu_-=0$ to $\nu_-=1$ and passing
through the saddle point
$\nu_-^\mathrm{sp}(\theta)=(1-\I\tan\theta)/2$, solution of the
equation $\D{g(\theta,\nu_-)}/\D{\nu_-}=0$, in the direction of the
steepest descent provides the asymptotic logarithm equality
\begin{eqnarray}
  I_{N,\mathscr{L}} \simeq \E^{N g(\theta,\nu_-^\mathrm{sp}(\theta))}
  = \E^{N \log(\cos\theta)}. 
\end{eqnarray}
An example of the steepest descent contour is shown in figure
\ref{fig5}.  Note that, despite the complex nature of the saddle point
$\nu_-^\mathrm{sp}$, the asymptotic result of the integration is real
as required. It follows that the corresponding equation for the
ground-state energy, obtained as $\lim_{N\to\infty} N^{-1}\log
I_N(E_0)=0$, with $E_0 \leq V_\mathrm{min}$, explicitly gives
\begin{eqnarray}
  \sum_{V\in\mathscr{V}} 
  \frac{\Sigma^{(1)}_{V}}{-E_{0}+V} = 
  \frac{1}
  {\left( \sum_{T\in\mathscr{T}} \Sigma^{(1)}_{T} T \right)
    \cos\theta},
  \qquad E_0 \leq V_\mathrm{min}.
\end{eqnarray}
This equation always admits one and only one real solution $E_0$.

At higher perturbative orders the situation is more complicated.  The
factorization (\ref{factorization}) does not apply and the oscillating
factor $\E^{\I N \theta (1-2\nu_-)}$ affects, via the correlations
induced by $\bm{p}(\bm{\nu})$, the evaluation of the integrals over
all frequencies $\nu_\alpha$, $\alpha\in\mathscr{H}$.  Once again,
however, the asymptotics of $I_N$ is correctly estimated by the value
of its integrand function at the complex saddle point
$\bm{\nu}^\mathrm{sp}$ determined by equation (\ref{nusp.eq}).
Despite $\bm{\nu}^\mathrm{sp}\in\mathbb{C}^{|\mathscr{H}|}$, we expect
the asymptotic value of $I_N$ to be real and equation (\ref{E0.eq2})
to admit a real solution $E_0$.  A mathematical justification for the
saddle-point method in $\mathbb{C}^{|\mathscr{H}|}$ is given by
theorem 2.8 of \cite{Sjostrand}.
\begin{figure}
  \centering
  \includegraphics[width=10cm, bb=1 1 713 498]{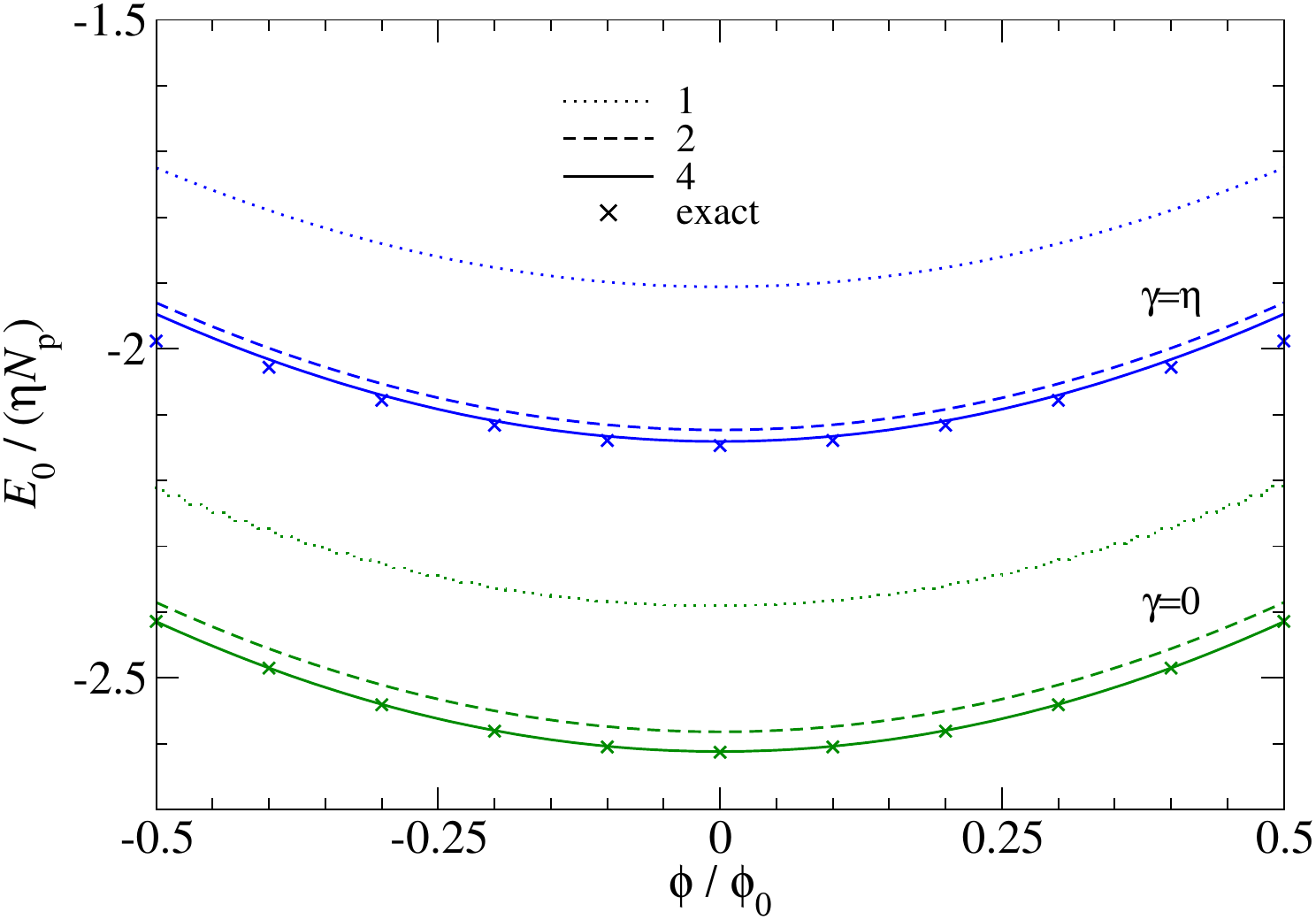}
  \caption{Ground-state energy per particle for the FNU pseudo-spin
    1/2 hard-core boson Hubbard model in a ring threaded by a magnetic
    flux $\phi$. The ring has $L=8$ sites and the number of particles
    per spin is $N_\mathrm{p}=4$.  Exact values of $E_0$ ($\times$)
    are compared with the results from present multinomial
    perturbative scheme by using cumulants up to order
    $k_\mathrm{max}=1,2,4$ (dotted, dashed, solid lines, respectively)
    for two different values of the interaction strength $\gamma$.}
  \label{fig6}
\end{figure}

We have checked the scenario depicted above by numerical simulations
on the model described by the Hamiltonian (\ref{HubbardRing}).  In
figure \ref{fig6} we show the behavior of $E_0(\phi)$ determined at
perturbative orders $k_\mathrm{max}=1,2,4$ in comparison with the
exact values of the ground-state energy obtained by numerical
diagonalization of $\hat{H}$. In all cases the solution of the system
of equations (\ref{E0.eq2}-\ref{nusp.eq}) in terms of complex unknowns
$(E_0,\bm{\nu}^\mathrm{sp})$ provides a ground-state energy which is
real within the statistical errors associated with the input
cumulants.  The agreement with the exact values of $E_0(\phi)$
increases on increasing the perturbative order in the whole range of
the magnetic flux.  At $\gamma=0$ the solution for $k_\mathrm{max}=4$
is practically exact.  When hopping and interaction have equal
strengths, i.e. for $\gamma=\eta$, the $k_\mathrm{max}=4$ solution is
in excellent agreement with $E_0(\phi)$ at small fluxes. At the flux
edges $\phi=\pm \phi_0/2$ a residual error of about $2\%$ is observed.
Note that for systems of larger size we have smaller maximum phases
$\theta=\pi/L$ and therefore we expect a better performance of our
approach already at small perturbative orders. Of course at large
sizes the measurement of the input cumulant is statistically heavier.

\section{Conclusions}
\label{conclusions}
In the framework of the probabilistic approach previously developed by
us to study the ground-state properties of many-body quantum systems,
we have introduced a multinomial perturbative scheme which has the
following characteristics.  At any perturbative order, the probability
distribution of the potential, hopping and phase multiplicities, whose
knowledge would allow for an exact solution of the problem, is
approximated by a multinomial-like distribution with infinitely many
statistical moments. By increasing the perturbative order, an
increasing number of cumulants of the distribution is made to coincide
with the corresponding exact cumulants of the system.

We have tested the proposed perturbative scheme in the case of Hubbard
models with pseudo-spin 1/2 hard-core bosons in
two-dimensional lattices and in a ring threaded by a magnetic flux.

For the two-dimensional lattices, we find, already at 
second perturbative order, a ground-state energy in good quantitative
agreement with the exact one for any value of the ratio $\gamma/\eta$,
$\gamma$ and $\eta$ being the strengths of the interaction and hopping
terms of the Hamiltonian of the system.  The agreement improves at
higher perturbative orders. At orders $\geq 3$, however, the scheme
may not always be consistent, i.e. in some systems a solution for the
ground-state energy is found only for values of $\gamma/\eta$ smaller
than a threshold.  As a matter of fact, we observe that in all our
test cases at, or near, the \sfrac{1}{4} particle filling, which is a
case of remarkable physical interest, the solution of the perturbative
method turns out to exist for all values of the interaction parameter
up to the largest explored perturbative order, $k_\mathrm{max}=4$,
where it provides stunning results.

The ring-shaped one-dimensional lattice with an orthogonal magnetic
field is a well known model to study electronic persistent currents
and, remarkably, presents a phase problem. For this model we discuss
in detail how our approach handles the phase problem and allow to find
the correct behavior of the ground-state energy as a function of the
threading flux $\phi$.  As in the previous phase-problem--free cases,
the quantitative agreement of $E_0(\phi)$ with the corresponding exact
values increases on increasing the perturbative order both for
non-interacting or interacting systems.

The limits, merits and scaling properties of our approach can be
summarized as follows.

The main uncertainty is that we do not know \textit{a priori} if our
perturbative scheme is meaningful at any order. For the systems
considered here, the second perturbative order always provides a
fairly accurate ground-state energy. Sometimes, at third and fourth
order the nonlinear system of equations which must be solved to find
$E_0$ does not admit a solution. An increased statistical accuracy of
the input data used to define the coefficients of these equations
and/or more accurate numerical methods to solve the system of
equations (\ref{E0.eq2}-\ref{nusp.eq}) could relieve this problem.

In our approach, the perturbative probability distribution at order
$k$ is built up from the knowledge of the first $k$ connected
statistical moments of the potential, hopping and phase multiplicities
of the system. These cumulants, more precisely the associated
asymptotic rescaled cumulants, are \textit{measured} by Monte Carlo
simulations as explained in \cite{OP4}. We use initial configurations
randomly distributed according to the invariant measure of the Markov
chain which provides their evolution. Thus, in a sense, ours is a
perfect simulation \cite{Thonnes}. Moreover, the mentioned Markov
chain has a finite correlation length which grows slowly, at least for
the cases studied, with the size of the system. This implies that
sampling cumulants of relatively high order is statistically reliable
also for large size systems. Our statistical accuracy, however, could
be increased by faster unbiased estimators based on the umbral
calculus \cite{NS}. In the present paper, the highest cumulant order
considered is 4 merely because the perturbative coefficients
$\bm{p}^{(k)}$ have been explicitly calculated only up to $k=4$.

It is difficult to precisely assess the scaling of the computational
costs of our method with the size $S$ of the system
considered. Unquestionably, the cardinality of the set $\mathscr{H}$
grows only linearly with $S$ so that the evaluation of the cumulants
of order $k$ can be safely bounded by $S^k$. However, from the limited
data at our disposal it is rash to figure out the behavior of
$k_\mathrm{max}(\varepsilon,S)$, the maximum order $k$ needed to
calculate $E_0$ at size $S$ with error $\epsilon$.

The cumulants input into equations (\ref{E0.eq2}-\ref{nusp.eq}) are
independent of the parameters $\gamma$ and $\eta$, namely the
strengths of the interaction and hopping terms of the Hamiltonian of
the system.  Once the probability distribution is determined at the
chosen approximation, the ground-state energy $E_0$ can be found by
solving numerically a small system of nonlinear equations.  The latter
job has a computational cost negligible with respect to the
determination of the cumulants, which, in turn, has a cost roughly
equivalent to a direct Monte Carlo evaluation of $E_0$.  Thus, the
advantage of our approach in comparison to a direct Monte Carlo
simulation is remarkable.  Different Monte Carlo runs are needed to
evaluate $E_0$ for different values of $\eta$ and/or $\gamma$, whereas
in our approach we have to solve each time a small system of nonlinear
equations and, \textit{una tantum}, calculate the cumulants.

Another advantage of our approach is that no extra efforts are
required to evaluate generic ground-state correlation functions.  The
key point is, again, the analytical dependence of our equations
(\ref{E0.eq2}-\ref{nusp.eq}), and, therefore, of its solution $E_0$,
on $\gamma$ and $\eta$ as well as on any other parameter entering the
Hamiltonian of the system.  In fact, the quantum expectation of an
observable $\hat{O}$ in the ground state of $\hat{H}$ is reconduced,
via the Hellman-Feynman theorem, to the ability to take the
derivative with respect to the parameter $\xi$ of the ground-state
energy of the ancillary Hamiltonian $\hat{H}+\xi\hat{O}$.

The present perturbative probabilistic approach is particularly
promising for systems affected by the so called sign problem, for
which unbiased Monte Carlo simulations of $E_0$ are impractical.  In
fact, the statistical evaluation of the cumulants of the potential,
hopping and phase multiplicities is \textit{unaffected} by sign/phase
problems.  Oscillations and cancellations remain confined in the
expression of the perturbative probability distribution and can be
tackled by complex analysis techniques.  Here, we have provided an
example of this strategy in a somewhat softer phase problem.  We plan
to discuss the case of fermions in a future paper.

% \section*{Acknowledgments}

\appendix
\setcounter{section}{0}
\section{Solution of nonsymmetric algebraic Riccati equations}
\label{calculation.p2}
In section \ref{equation.p2} we have seen that the parameters
$\boldsymbol{p}^{(2)}$, more precisely the associated reduced matrix
$\hat{\boldsymbol{p}}^{(2)}$, are determined by the NARE (\ref{SOL_p2M_fin}).
In general, the NAREs are defined as the quadratic matrix equations of the kind
\begin{eqnarray}
  \label{eq_NARE}
  \boldsymbol{X}\boldsymbol{C}\boldsymbol{X} - 
  \boldsymbol{A}\boldsymbol{X} - \boldsymbol{X}\boldsymbol{D} +
  \boldsymbol{B} = \boldsymbol{0},
\end{eqnarray}
where we assume that the unknown $\boldsymbol{X}$, as well as the
coefficients $\boldsymbol{A}$, $\boldsymbol{B}$, $\boldsymbol{C}$ and
$\boldsymbol{D}$ are quadratic matrices of finite size.
In this section we illustrate two numerical methods developed to solve
equation (\ref{eq_NARE}). The first one is an iterative method based
on a fixed-point technique \cite{Guo2011}, whereas the second one is
a direct method based on the Sch\"ur decomposition \cite{BIMP}.

Equations (\ref{eq_NARE}) play an important role in the study 
of the stochastic fluid
models and have been extensively studied. In general, a NARE admits
more than one solution. In most stochastic fluid models, the
coefficients $\boldsymbol{A}$, $\boldsymbol{B}$, $\boldsymbol{C}$ and
$\boldsymbol{D}$ form a super matrix
\begin{eqnarray}
  \label{eq_H}
  \boldsymbol{H} = 
  \left(
    \begin{array}{cc}
      \boldsymbol{D} & -\boldsymbol{C} \\
      -\boldsymbol{B} & \boldsymbol{A}
    \end{array}
  \right).
\end{eqnarray}
with the property to be a so called $M$-matrix \cite{BIMP}. It can be
proved that in this case the Sch\"ur method provides the minimal
non negative solution of the NARE, which is, there, the solution of
physical interest.

In our context, $\boldsymbol{H}$ is not a $M$-matrix and it is not
clear which solution of the NARE (\ref{SOL_p2M_fin}) has to be
considered. We propose to consider the unique solution given by the
Sch\"ur method. This solution coincides with that obtained by the
iterative method in which $\boldsymbol{X}$ is chosen at the zeroth iteration 
as the solution of (\ref{eq_NARE}) with
$\boldsymbol{C}=\boldsymbol{0}$. Since in our case $\boldsymbol{X}$
represents the matrix of the perturbative parameters
$\hat{\boldsymbol{p}}^{(2)}$, which we expect to be small, the above
proposed solution seems the most natural one.

\subsection{Iterative method}
In \cite{Guo2011} a class of fixed-point methods 
is considered to solve equation (\ref{eq_NARE}). 
These fixed-point iterations are based on a suitable
splitting of the matrices $\boldsymbol{A}$ and $\boldsymbol{D}$, that
is $\boldsymbol{A}=\boldsymbol{A}_{1}-\boldsymbol{A}_{2}$ and
$\boldsymbol{D}=\boldsymbol{D}_{1}-\boldsymbol{D}_{2}$, and have the form
\begin{eqnarray}
  \label{eq_NAREite}
  \boldsymbol{A}_{1}\boldsymbol{X}_{k+1} + 
  \boldsymbol{X}_{k+1}\boldsymbol{D}_{1} =
  \boldsymbol{X}_{k}\boldsymbol{C}\boldsymbol{X}_{k} + 
  \boldsymbol{A}_{2}\boldsymbol{X}_{k} + \boldsymbol{X}_{k}\boldsymbol{D}_{2} +
  \boldsymbol{B},
\end{eqnarray}
with $k=0,1,2,\dots$ and $\boldsymbol{X}_{0}=\boldsymbol{0}$.
In particular, for 
$\boldsymbol{A}_{1}=\boldsymbol{A}$ and
$\boldsymbol{D}_{1}=\boldsymbol{D}$ we have
\begin{eqnarray}
  \label{eq_ite}
  \boldsymbol{A}\boldsymbol{X}_{k+1} + \boldsymbol{X}_{k+1}\boldsymbol{D} =
  \boldsymbol{X}_{k}\boldsymbol{C}\boldsymbol{X}_{k} + 
  \boldsymbol{B},
  \qquad \boldsymbol{X}_{0}=\boldsymbol{0}.
\end{eqnarray}
Note that finding $\boldsymbol{X}_{k+1}$ in terms
of $\boldsymbol{X}_{k}$ at $k$th iteration implies to solve a Sylvester 
equation. This can be accomplished by vectorization, namely 
\begin{eqnarray}
  \label{eq_ite2}
  \mathrm{vec}\left(\boldsymbol{A}\boldsymbol{X}_{k+1}\right) + 
  \mathrm{vec}\left(\boldsymbol{X}_{k+1}\boldsymbol{D}\right) =
  \mathrm{vec}\left(\boldsymbol{X}_{k}\boldsymbol{C}\boldsymbol{X}_{k} + 
    \boldsymbol{B}\right).
\end{eqnarray}
Using the properties of the $\mathrm{vec}$ operator, in
particular
\begin{eqnarray}
  \mathrm{vec}\left(\boldsymbol{A}\boldsymbol{X}_{k+1}\right) = 
  \left(\boldsymbol{I}\otimes\boldsymbol{A}\right)
  \mathrm{vec}\left(\boldsymbol{X}_{k+1}\right),
\end{eqnarray} 
\begin{eqnarray}
  \mathrm{vec}\left(\boldsymbol{X}_{k+1}\boldsymbol{D}\right) =
  \left(\boldsymbol{D}^{T}\otimes\boldsymbol{I}\right)
  \mathrm{vec}\left(\boldsymbol{X}_{k+1}\right),
\end{eqnarray}
where $\otimes$ indicates the Kronecker product and $^T$ the transpose,
equation (\ref{eq_ite2}) is rewritten as
\begin{eqnarray}
  \label{eq_ite3}
  \left[\left(\boldsymbol{I}\otimes\boldsymbol{A}\right) +
    \left(\boldsymbol{D}^{T}\otimes\boldsymbol{I}\right)\right]
  \mathrm{vec}\left(\boldsymbol{X}_{k+1}\right)=
  \mathrm{vec}\left(\boldsymbol{X}_{k}\boldsymbol{C}\boldsymbol{X}_{k} + 
    \boldsymbol{B}\right).
\end{eqnarray}
This is a linear matrix equation which can be solved by 
standard methods, e.g. LU-factorization \cite{NR}.

The convergence of the full class of iterative schemes
(\ref{eq_NAREite}) to a solution $\boldsymbol{X}$ of (\ref{eq_NARE})
is ensured by a theorem \cite{Guo2011}. In this class, the iterative
scheme (\ref{eq_ite}) is the most expensive from a computational point
of view, but, on the other hand, it has the highest (linear)
convergence speed.

\subsection{Sch\"ur method}
In the following we discuss a different approach to solve equation
(\ref{eq_NARE}), based on the ordered Sch\"ur decomposition. This
approach was conceived by Laub \cite{Laub} for a symmetric algebraic
Riccati equation and extended by Guo \cite{Guo} to the study of NAREs.

Let us rewrite the matrix $\boldsymbol{H}$ associated with the
coefficients of (\ref{eq_NARE}) as
\begin{eqnarray}
  \boldsymbol{H} =
  \left(
    \begin{array}{cc}
      \boldsymbol{D} & -\boldsymbol{C} \\
      -\boldsymbol{B} & \boldsymbol{A}
    \end{array}
  \right)= 
  \left(
    \begin{array}{cc}
      \boldsymbol{H}_{11} & \boldsymbol{H}_{12} \\
      \boldsymbol{H}_{21} & \boldsymbol{H}_{22}
    \end{array}
  \right).
\end{eqnarray}
Note that $\boldsymbol{H}$ is real in our case.  We look for an
orthogonal transformation
\begin{eqnarray}
  \boldsymbol{U} = 
  \left(
    \begin{array}{cc}
      \boldsymbol{U}_{11} & \boldsymbol{U}_{12} \\
      \boldsymbol{U}_{21} & \boldsymbol{U}_{22}
    \end{array}
  \right),
\end{eqnarray}
which leaves $\boldsymbol{H}$ in a semi-ordered real Sch\"ur form,
\begin{eqnarray}
  \label{schur.decomposition}
  \boldsymbol{U}^{T}\boldsymbol{H}\boldsymbol{U} = \boldsymbol{S} =
  \left(
    \begin{array}{cc}
      \boldsymbol{S}_{11} & \boldsymbol{S}_{12} \\
      \boldsymbol{0}           & \boldsymbol{S}_{22}
    \end{array}
  \right),
\end{eqnarray}
in which $\boldsymbol{S}_{11}$ and $\boldsymbol{S}_{22}$ contain only
blocks, denoted $\boldsymbol{s}_{ij}$, $i,j=1,2,\dots$, of size 1 or
2.  The eigenvalues of the $2\times2$ diagonal blocks
$\boldsymbol{s}_{ii}$ provide the complex conjugated eigenvalues of
$\boldsymbol{H}$ whereas the $1\times 1$ blocks are the real
eigenvalues of $\boldsymbol{H}$.  The diagonal blocks are semi-ordered
in the sense that if $\boldsymbol{s}_{ii}$, $\boldsymbol{s}_{jj}$ and
$\boldsymbol{s}_{kk}$ have eigenvalues with positive, null and
negative real parts, respectively, then $i<j<k$. It is possible to
show that the matrix $\boldsymbol{U}_{11}$ is invertible\footnotemark[2]
\footnotetext[2]{See theorem 4 of \cite{Guo}}
and that
\begin{eqnarray}
  \label{sol_schur}
  \boldsymbol{X} = \boldsymbol{U}_{21}\boldsymbol{U}_{11}^{-1}
\end{eqnarray}
solves (\ref{eq_NARE}). Note that the semi-ordered decomposition
(\ref{schur.decomposition}) is unique and so is the solution
(\ref{sol_schur}).  We used the subroutines of LAPACK library
\cite{LAPACK} to numerically implement the Sch\"ur method.

\section{Equations for the perturbative parameters: fourth order}
\label{equation.p4}
The perturbative parameters $\boldsymbol{p}^{(4)}$ are determined by
the system of equations
\begin{eqnarray}
  \label{SYSTEM_P4}
  \arc{\nu_{\alpha_1}\nu_{\alpha_2}\nu_{\alpha_3}\nu_{\alpha_4}}
  (\boldsymbol{p}^{(1)},\boldsymbol{p}^{(2)},\boldsymbol{p}^{(3)},
  \boldsymbol{p}^{(4)}) = 
  \Sigma_{\alpha_1\alpha_2\alpha_3,\alpha_4}^{(4)},
\end{eqnarray}
with $\alpha_1,\alpha_2,\alpha_3,\alpha_4\in\mathscr{H}$.  By using
(\ref{arc-der}) and taking the derivative of (\ref{d3_phi}) with
respect to $J_{\alpha_4}$, the above system can be cast in the form
\begin{eqnarray}\fl
  \sum_{\alpha\in\mathscr{H}}
  \sum_{\beta\in\mathscr{H}}
  \sum_{\gamma\in\mathscr{H}}
  \sum_{\delta\in\mathscr{H}}
  \left( 
    \Lambda^{(2,0)}_{{\alpha}_{1}\alpha}
    \Sigma^{(2)}_{\beta{\alpha}_{2}}
    \Sigma^{(2)}_{\gamma{\alpha}_{3}}
    \Sigma^{(2)}_{\delta{\alpha}_{4}} +
    \Sigma^{(2,0)}_{{\alpha}_{1}\alpha}
    \Lambda^{(2)}_{\beta{\alpha}_{2}}
    \Sigma^{(2)}_{\gamma{\alpha}_{3}} 
    \Sigma^{(2)}_{\delta{\alpha}_{4}} 
  \right.
  \nonumber \\+
  \left.
    \Sigma^{(2,0)}_{{\alpha}_{1}\alpha}
    \Sigma^{(2)}_{\beta{\alpha}_{2}}
    \Lambda^{(2)}_{\gamma{\alpha}_{3}}
    \Sigma^{(2)}_{\delta{\alpha}_{4}} +
    \Sigma^{(2,0)}_{{\alpha}_{1}\alpha} 
    \Sigma^{(2)}_{\beta{\alpha}_{2}}
    \Sigma^{(2)}_{\gamma{\alpha}_{3}}
    \Lambda^{(2)}_{\delta{\alpha}_{4}}
  \right) p^{(4)}_{\alpha\beta\gamma\delta} =
  \Delta_{{\alpha}_1{\alpha}_{2}{\alpha}_{3}{\alpha}_{4}},
\end{eqnarray}
where $\mathbf{\Sigma}^{(2)}$ is the asymptotic rescaled cumulant of
order 2 and the matrices $\mathbf{\Sigma}^{(2,0)}$,
$\mathbf{\Lambda}^{(2)}$ and $\mathbf{\Lambda}^{(2,0)}$ are defined by
(\ref{Sigma20}), (\ref{Lambda2}) and (\ref{Lambda20}), respectively.
The tensor $\boldsymbol{\Delta}$ has components
$\alpha_1,\alpha_2,\alpha_3,\alpha_4\in\mathscr{H}$ given by
\begin{eqnarray}\fl
  \Delta_{\alpha_{1}\alpha_{2}\alpha_{3}\alpha_{4}} =
  \Sigma^{(4)}_{\alpha_{1}\alpha_{2}\alpha_{3}\alpha_{4}} - 
  \Sigma^{(4,0)}_{\alpha_{1}\alpha_{2}\alpha_{3}\alpha_{4}}
  \nonumber \\  -
  \sum_{\alpha\in\mathscr{H}}\sum_{\beta\in\mathscr{H}}
  \left(
    \Sigma^{(2,0)}_{\alpha_{1}\alpha} M^{(2)}_{\alpha\beta}
    \Sigma^{(4)}_{\beta\alpha_{2}\alpha_{3}\alpha_{4}} +
    \Sigma^{(4,0)}_{\alpha_{1}\alpha\alpha_{3}\alpha_{4}}
    M^{(2)}_{\alpha\beta} \Sigma^{(2)}_{\beta\alpha_{2}}
  \right)
  \nonumber \\ -
  \sum_{\alpha\in\mathscr{H}}\sum_{\beta\in\mathscr{H}}
  \Sigma^{(3,0)}_{\alpha_{1}\alpha\alpha_{3}}
  \left(
    M^{(3)}_{\alpha\beta\alpha_{4}} \Sigma^{(2)}_{\beta\alpha_{2}} +
    M^{(2)}_{\alpha\beta} \Sigma^{(3)}_{\beta\alpha_{2}\alpha_{4}}
  \right)
  \nonumber \\ -
  \sum_{\alpha\in\mathscr{H}}\sum_{\beta\in\mathscr{H}}
  \Sigma^{(3,0)}_{\alpha_{1}\alpha\alpha_{4}}
  \left(
    M^{(3)}_{\alpha\beta\alpha_{3}} \Sigma^{(2)}_{\beta\alpha_{2}} +
    M^{(2)}_{\alpha\beta} \Sigma^{(3)}_{\beta\alpha_{2}\alpha_{3}}
  \right)
  \nonumber \\ -
  \sum_{\alpha\in\mathscr{H}}\sum_{\beta\in\mathscr{H}}
  \Sigma^{(2,0)}_{\alpha_{1}\alpha}
  \left(
    M^{(3)}_{\alpha\beta\alpha_{3}} \Sigma^{(3)}_{\beta\alpha_{2}\alpha_{4}} + 
    M^{(3)}_{\alpha\beta\alpha_{4}} \Sigma^{(3)}_{\beta\alpha_{2}\alpha_{3}}
  \right)
  \nonumber \\ -
  \sum_{\alpha\in\mathscr{H}}\sum_{\beta\in\mathscr{H}}
  \sum_{\gamma\in\mathscr{H}}
  \Sigma^{(2,0)}_{\alpha_{1}\alpha}
  F^{(3)}_{\alpha\beta\gamma}
  \Sigma^{(2)}_{\beta\alpha_{2}}
  \Sigma^{(3)}_{\gamma\alpha_{3}\alpha_{4}}
  \nonumber \\ +
  \sum_{\alpha\in\mathscr{H}}\sum_{\beta\in\mathscr{H}}
  \sum_{\gamma\in\mathscr{H}}\sum_{\delta\in\mathscr{H}}
  \Sigma^{(2,0)}_{\alpha_{1}\alpha} 
  F^{(4)}_{\alpha\beta\gamma\delta}
  \Sigma^{(2)}_{\beta\alpha_{2}}
  \Sigma^{(2)}_{\gamma\alpha_{3}}
  \Sigma^{(2)}_{\delta\alpha_{4}},
\end{eqnarray}
where
\begin{eqnarray}\fl
  \Sigma^{(4,0)}_{\alpha\beta\gamma\delta} =
  \Sigma^{(3)}_{\beta\gamma\delta}\delta_{\alpha\beta} -
  \left(
    \Sigma^{(3)}_{\alpha\gamma\delta}\Sigma^{(1)}_{\beta} +
    \Sigma^{(3)}_{\beta\gamma\delta}\Sigma^{(1)}_{\alpha} +
    \Sigma^{(2)}_{\alpha\delta}\Sigma^{(2)}_{\beta\gamma} +
    \Sigma^{(2)}_{\alpha\gamma}\Sigma^{(2)}_{\beta\delta}    
  \right) \chi_{\alpha\beta},
\end{eqnarray}
\begin{eqnarray}\fl
  M^{(2)}_{\alpha\beta} = 
  \left.M_{\alpha\beta}\left(\boldsymbol{\nu}^{\mathrm{sp}}\right)
  \right|_{\boldsymbol{J}=\boldsymbol{0}} =
  \frac{p^{(2)}_{\alpha\beta}}{\Sigma^{(1)}_{\alpha}} +
  \frac{p^{(2)}_{\alpha\beta}}{\Sigma^{(1)}_{\beta}} - 
  \sum_{\gamma\in\mathscr{H}}
  \frac{p^{(2)}_{\alpha\gamma} p^{(2)}_{\gamma\beta}}
  {\Sigma^{(1)}_{\gamma}},
\end{eqnarray}
\begin{eqnarray}\fl
  M^{(3)}_{\alpha\beta\gamma} = 
  \left.\frac{\partial M_{\alpha\beta}(\boldsymbol{\nu}^{\mathrm{sp}})}
    {\partial J_{\gamma}}\right|_{\boldsymbol{J}=\boldsymbol{0}} =
  \sum_{\delta\in\mathscr{H}} 
  F^{(3)}_{\alpha\beta\delta} \Sigma^{(2)}_{\delta\gamma},
\end{eqnarray}
\begin{eqnarray}\fl
F^{(3)}_{\alpha\beta\gamma} =
    \frac{p^{(3)}_{\alpha\beta\gamma}}{\Sigma^{(1)}_{\alpha}} + 
    \frac{p^{(3)}_{\alpha\beta\gamma}}{\Sigma^{(1)}_{\beta}} + 
    \frac{p^{(3)}_{\alpha\beta\gamma}}{\Sigma^{(1)}_{\gamma}} - 
    \frac{p^{(2)}_{\alpha\beta} p^{(2)}_{\alpha\gamma}}
    {{\Sigma^{(1)}_{\alpha}}^2} - 
    \frac{p^{(2)}_{\alpha\beta} p^{(2)}_{\beta\gamma}}
    {{\Sigma^{(1)}_{\beta}}^2} -
    \frac{p^{(2)}_{\alpha\gamma} p^{(2)}_{\gamma\beta}}
    {{\Sigma^{(1)}_{\gamma}}^2}
  \nonumber \\ -
    \sum_{\delta\in\mathscr{H}}
    \left(
      \frac{p^{(3)}_{\alpha\beta\delta} p^{(2)}_{\delta\gamma}}
      {\Sigma^{(1)}_{\delta}} + 
      \frac{p^{(3)}_{\alpha\delta\gamma} p^{(2)}_{\delta\beta}}
      {\Sigma^{(1)}_{\delta}} +
      \frac{p^{(3)}_{\delta\beta\gamma} p^{(2)}_{\alpha\delta}}
      {\Sigma^{(1)}_{\delta}} -     
      2\frac{p^{(2)}_{\alpha\delta} p^{(2)}_{\delta\beta} 
        p^{(2)}_{\delta\gamma}}
      {{\Sigma^{(1)}_{\delta}}^2}
    \right), 
\end{eqnarray}
\begin{eqnarray}\fl
  F^{(4)}_{\alpha\beta\gamma\delta}  =
  \frac{
    p_{\alpha\beta\gamma}^{(3)} p_{\alpha\delta}^{(2)} +
    p_{\alpha\beta\delta}^{(3)} p_{\alpha\gamma}^{(2)} +
    p_{\alpha\gamma\delta}^{(3)} p_{\alpha\beta}^{(2)}}
  {{\Sigma_{\alpha}^{(1)}}^2} + 
  \frac{
    p_{\alpha\beta\gamma}^{(3)} p_{\beta\delta}^{(2)} +
    p_{\alpha\beta\delta}^{(3)} p_{\beta\gamma}^{(2)} +
    p_{\beta\gamma\delta}^{(3)} p_{\alpha\beta}^{(2)}}
  {{\Sigma_{\beta}^{(1)}}^2}
  \nonumber \\ +
  \frac{
    p_{\alpha\beta\gamma}^{(3)} p_{\gamma\delta}^{(2)} +
    p_{\alpha\gamma\delta}^{(3)} p_{\gamma\beta}^{(2)} +
    p_{\beta\gamma\delta}^{(3)} p_{\alpha\gamma}^{(2)}}
  {{\Sigma_{\gamma}^{(1)}}^2} +
  \frac{
    p_{\alpha\beta\delta}^{(3)} p_{\delta\gamma}^{(2)} +
    p_{\alpha\delta\gamma}^{(3)} p_{\delta\beta}^{(2)} +
    p_{\delta\beta\gamma}^{(3)} p_{\alpha\delta}^{(2)}}
  {{\Sigma_{\delta}^{(1)}}^2}
  \nonumber \\ -
  2\left(
    \frac{
      p_{\alpha\beta}^{(2)} p_{\alpha\gamma}^{(2)} p_{\alpha\delta}^{(2)}}
    {{\Sigma_{\alpha}^{(1)}}^3} + 
    \frac{
      p_{\alpha\beta}^{(2)} p_{\beta\gamma}^{(2)} p_{\beta\delta}^{(2)}}
    {{\Sigma_{\beta}^{(1)}}^3} +
    \frac{
      p_{\alpha\gamma}^{(2)} p_{\gamma\beta}^{(2)} p_{\gamma\delta}^{(2)}}
    {{\Sigma_{\gamma}^{(1)}}^3} + 
    \frac{
      p_{\alpha\delta}^{(2)} p_{\delta\beta}^{(2)} p_{\delta\gamma}^{(2)}}
    {{\Sigma_{\delta}^{(1)}}^3}
  \right)
    \nonumber \\ + 
    \sum_{\varepsilon\in\mathscr{H}}
    \left[
      \frac{
        p_{\alpha\beta\varepsilon}^{(3)}p_{\varepsilon\gamma\delta}^{(3)} +
        p_{\alpha\varepsilon\gamma}^{(3)}p_{\varepsilon\beta\delta}^{(3)} +
        p_{\varepsilon\beta\gamma}^{(3)}p_{\varepsilon\alpha\delta}^{(3)}
      }
      {\Sigma_{\varepsilon}^{(1)}}
    \right.
    \nonumber\\ -
    2\left(\frac{
      p_{\alpha\beta\varepsilon}^{(3)}
      p_{\varepsilon\gamma}^{(2)} p_{\varepsilon\delta}^{(2)} + 
      p_{\alpha\varepsilon\gamma}^{(3)}
      p_{\varepsilon\beta}^{(2)} p_{\varepsilon\delta}^{(2)} +
      p_{\varepsilon\beta\gamma}^{(3)}
      p_{\alpha\varepsilon}^{(2)} p_{\varepsilon\delta}^{(2)}
    }{{\Sigma_{\varepsilon}^{(1)}}^{2}}\right.
    \nonumber\\ +
    \left.\frac{
      p_{\alpha\varepsilon\delta}^{(3)}
      p_{\varepsilon\beta}^{(2)} p_{\varepsilon\gamma}^{(2)} +
      p_{\varepsilon\beta\delta}^{(3)}
      p_{\varepsilon\alpha}^{(2)} p_{\varepsilon\gamma}^{(2)} +
      p_{\varepsilon\gamma\delta}^{(3)}
      p_{\varepsilon\alpha}^{(2)} p_{\varepsilon\beta}^{(2)}
    }{{\Sigma_{\varepsilon}^{(1)}}^{2}}\right) +
  \left.
    6 \frac{p_{\varepsilon\alpha}^{(2)} p_{\varepsilon\beta}^{(2)}
      p_{\varepsilon\gamma}^{(2)} p_{\varepsilon\delta}^{(2)}}
    {{\Sigma^{(1)}_{\varepsilon}}^3}
  \right] .
\end{eqnarray}
To find $\boldsymbol{p}^{(4)}$, we first determine the reduced tensor 
$\hat{\boldsymbol{p}}^{(4)}$ which is the solution of the linear system 
\begin{eqnarray}\fl
  \sum_{\alpha\in\hat{\mathscr{H}}}
  \sum_{\beta\in\hat{\mathscr{H}}}
  \sum_{\gamma\in\hat{\mathscr{H}}}
  \sum_{\delta\in\hat{\mathscr{H}}}
  \left(
    \widetilde{\Lambda}_{{\alpha}_{1}\alpha}^{(2,0)}
    \widetilde{\Sigma}_{\beta{\alpha}_{2}}^{(2)}
    \widetilde{\Sigma}_{\gamma{\alpha}_{3}}^{(2)}
    \widetilde{\Sigma}_{\delta{\alpha}_{4}}^{(2)} +
    \widetilde{\Sigma}_{{\alpha}_{1}\alpha}^{(2,0)}
    \widetilde{\Lambda}_{\beta{\alpha}_{2}}^{(2)}
    \widetilde{\Sigma}_{\gamma{\alpha}_{3}}^{(2)} 
    \widetilde{\Sigma}_{\delta{\alpha}_{4}}^{(2)} 
  \right.
  \nonumber \\ +
  \left.
    \widetilde{\Sigma}_{{\alpha}_{1}\alpha}^{(2,0)}
    \widetilde{\Sigma}_{\beta{\alpha}_{2}}^{(2)}
    \widetilde{\Lambda}_{\gamma{\alpha}_{3}}^{(2)}
    \widetilde{\Sigma}_{\delta{\alpha}_{4}}^{(2)} +
    \widetilde{\Sigma}_{{\alpha}_{1}\alpha}^{(2,0)}
    \widetilde{\Sigma}_{\beta{\alpha}_{2}}^{(2)}
    \widetilde{\Sigma}_{\gamma{\alpha}_{3}}^{(2)}
    \widetilde{\Lambda}_{\delta{\alpha}_{4}}^{(2)}
  \right)
  \hat{p}_{\alpha\beta\gamma\delta}^{(4)} =
  \Delta_{{\alpha}_1{\alpha}_{2}{\alpha}_{3}{\alpha}_{4}},
\end{eqnarray}
with $\alpha_1,\alpha_2,\alpha_3,\alpha_4 \in \hat{\mathscr{H}}$. 
The matrices
$\widetilde{\boldsymbol{\Sigma}}^{(2)}$,
$\widetilde{\boldsymbol{\Sigma}}^{(2,0)}$,
$\widetilde{\boldsymbol{\Lambda}}^{(2)}$ and 
$\widetilde{\boldsymbol{\Lambda}}^{(2,0)}$ are defined
by (\ref{Sigma2.tilde}), (\ref{Sigma20.tilde}), (\ref{Lambda2.tilde}) and 
(\ref{Lambda20.tilde}), respectively.
The complete fourth-order perturbative parameter
$\boldsymbol{p}^{(4)}$ is then recovered using the sum rules
(\ref{PPAR_SUM}) for $k=4$.

\end{document}